\documentclass[prb,twocolumn,superscriptaddress,showpacs,floatfix]{revtex4}
\pdfoutput=1
\usepackage{amsmath}
\usepackage{bm}
\usepackage{graphicx}
\usepackage{amssymb}
\usepackage{datetime}
\usepackage{sidecap}
\begin{document}
\newcommand{\figdir}{.}
\newcommand{\figwidth}{0.9\columnwidth}
\newcommand{\ffigwidth}{0.4\columnwidth}
%
\title{Multifractal finite-size-scaling and universality at the Anderson transition}

\author{Alberto Rodriguez}
\email[Corresponding author:\\]{argon@usal.es}
\affiliation{
Department of Physics and Centre for Scientific Computing, University of Warwick, Coventry, CV4 7AL, United Kingdom}
\affiliation{
Departamento de F\'isica Fundamental, Universidad de Salamanca, 37008 Salamanca, Spain}
\author{Louella J.~Vasquez}
\affiliation{
Institute of Advanced Study, Complexity Science Centre and Department of Statistics, University of Warwick, Coventry, CV4 7AL, United Kingdom}
\author{Keith Slevin}
\affiliation{
Department of Physics, Graduate School of Science, Osaka University, 1-1 Machikaneyama, Toyonaka, Osaka 560-0043, Japan}
\author{Rudolf A.~R\"omer}
\affiliation{
Department of Physics and Centre for Scientific Computing, University of Warwick, Coventry, CV4 7AL, United Kingdom}
\date{$Revision: 1.122 $, compiled \today, \currenttime}
%
\begin{abstract}
We describe a new multifractal finite size scaling (MFSS) procedure and its application to the Anderson localization-delocalization transition.
MFSS permits the simultaneous estimation of the critical parameters and the multifractal exponents.
Simulations of system sizes up to $L^3=120^3$ and involving nearly $10^6$ independent wavefunctions have yielded unprecedented precision for
the critical disorder $W_c=16.530 (16.524,16.536)$ and the critical exponent $\nu=1.590 (1.579,1.602)$.
We find that the multifractal exponents $\Delta_q$ exhibit a previously predicted symmetry relation and we confirm the non-parabolic nature of their spectrum.
We explain in detail the MFSS procedure first introduced in our Letter [Phys.\ Rev.\ Lett.\ 105, 046403 (2010)] and, in addition, we show how to take account
of correlations in the simulation data.
The MFSS procedure is applicable to any continuous phase transition exhibiting multifractal fluctuations in the vicinity of the critical point.
\end{abstract}
%
\maketitle

\section{Introduction}
\label{sec-introduction}

One of the most fascinating aspects of the Anderson localization-delocalization transition is the occurrence of multifractal fluctuations of the wavefunction
intensity at the critical point.\cite{EveM08,MorKMG02,HasSWI08,FaeSPL09}
While, strictly speaking, the fluctuations are truly multifractal only at the critical point, where the correlation length $\xi$ diverges, multifractal fluctuations nevertheless
persist on either side of the transition on length scales less than the correlation length.\cite{Cuevas2007}
The persistence of the fluctuations can be clearly seen in Fig.~\ref{fig-states}, where the wavefunction intensities for some typical critical, metallic and localized wavefunctions are plotted.
In this paper we show how to exploit this persistence by combining multifractal analysis with finite size scaling to arrive at a very powerful method
for the quantitative analysis of the Anderson transition, or any continuous phase transition that exhibits multifractal fluctuations: Multifractal Finite Size Scaling (MFSS).

Below we describe in detail the MFSS procedure and we demonstrate its potential
by employing it to make a more comprehensive analysis of the Anderson transition in three dimensions than given in our Letter.\cite{RodVSR10}
As we describe, the MFSS procedure permits the simultaneous estimation of the usual critical parameters, such as the location of the critical point and the critical exponent,
and the multifractal exponents.
Moreover, MFSS offers the opportunity to examine the consistency of the estimates of the critical parameters against the choice of multifractal exponent used in the scaling analysis.

The organization of the paper is as follows. In Section \ref{sec-model}, we describe briefly the Anderson model of a disordered systems and the numerical simulation of the Anderson transition.
In Section \ref{sec-gmfe}, we define the generalized multifractal exponents (GMFEs) used in MFSS, and derive the corresponding scaling laws.
The relation between the GMFEs and the scaling properties of the probability density function (PDF) of wavefunction intensities is discussed in Section \ref{sec-pdf}.
In Section \ref{sec-datacor} we demonstrate the necessity of avoiding correlations in the wavefunctions.
In Sections \ref{sec-fixlambda} and \ref{sec-3dfss} we present results from standard and multifractal FSS, including estimates for the critical parameters and  multifractal exponents.
Finally, details of how to account for the inevitable correlations in different coarse grainings of the same simulation data, as well as how to check the stability of the scaling fits, are collected in the Appendices.
%
\section{The eigenstates of the Anderson model}
\label{sec-model}
We consider the three-dimensional (3D) Anderson Hamiltonian in site basis,
\begin{equation}
\mathcal{H}=\sum_{i} \varepsilon_i~\vert i\rangle\langle i\vert + \sum_{\langle i,j\rangle}~\vert i\rangle\langle j\vert,
\label{eq:anderson_H1}
\end{equation}
where site $i=(x,y,z)$ is the position of an electron in a cubic lattice of volume $L^3$ --- where $L$ is measured in terms of the lattice constant  ---, $\langle i,j\rangle$ denote nearest neighbors, periodic boundary conditions are assumed and $\varepsilon_i$ are random on-site energies uniformly distributed in the interval $[-W/2,W/2]$. The $L^3\times L^3$ Hamiltonian is diagonalized in the vicinity of the band centre $E=0$ for different degrees of disorder $W$, close to the critical value $W_c\sim 16.5$ where the localization-delocalization transition occurs. The eigenstates $\Psi=\sum_{i} \psi_i |i\rangle$ are numerically obtained using the {\sc Jadamilu} library.\cite{SchBR06,BolN07,VasRR08}

\begin{table}
\caption{Average number of uncorrelated wavefunctions $\langle \mathcal{N}\rangle$ considered for each disorder $W$ for each system size $L$. The maximum and minimum numbers of states for a given $W$ are shown in brackets for each $L$. A total of 17 disorder values in the interval $[15,18]$ were considered.}
\begin{tabular}{cc}
\hline\hline
       $L$ & $\langle \mathcal{N} \rangle$ $(\mathcal{N}_{\rm max}, \mathcal{N}_{\rm min})$ \\
\hline
20 &  5138 (5006, 5374)  \\
30 &  5079 (5011, 5143)  \\
40 &  5168 (5012, 5351)  \\
50 &  5042 (5005, 5125)  \\
60 &  5027 (5009, 5082)  \\
70 &  5032 (5010, 5058)  \\
80 &  5028 (5013, 5048)  \\
90 &  5083 (5006, 5328)  \\
100 &  5024 (5020, 5041) \\
110 &  4331 (4214, 4589) \\
120 &  3103 (3000, 3757) \\
\hline\hline
\end{tabular}
\label{tab-states}
\end{table}
\begin{figure*}
 \includegraphics[width=.31\textwidth]{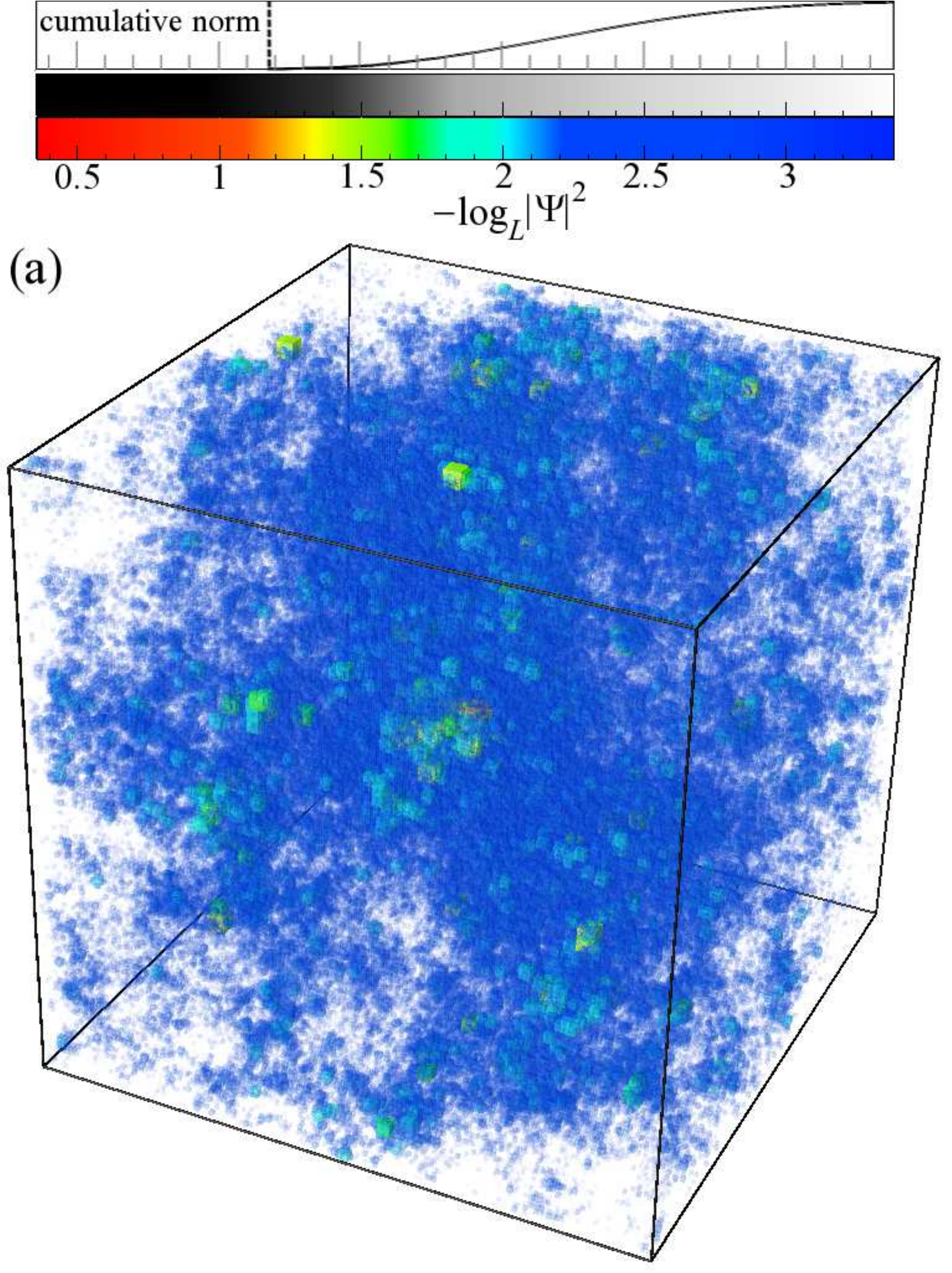}
 \includegraphics[width=.31\textwidth]{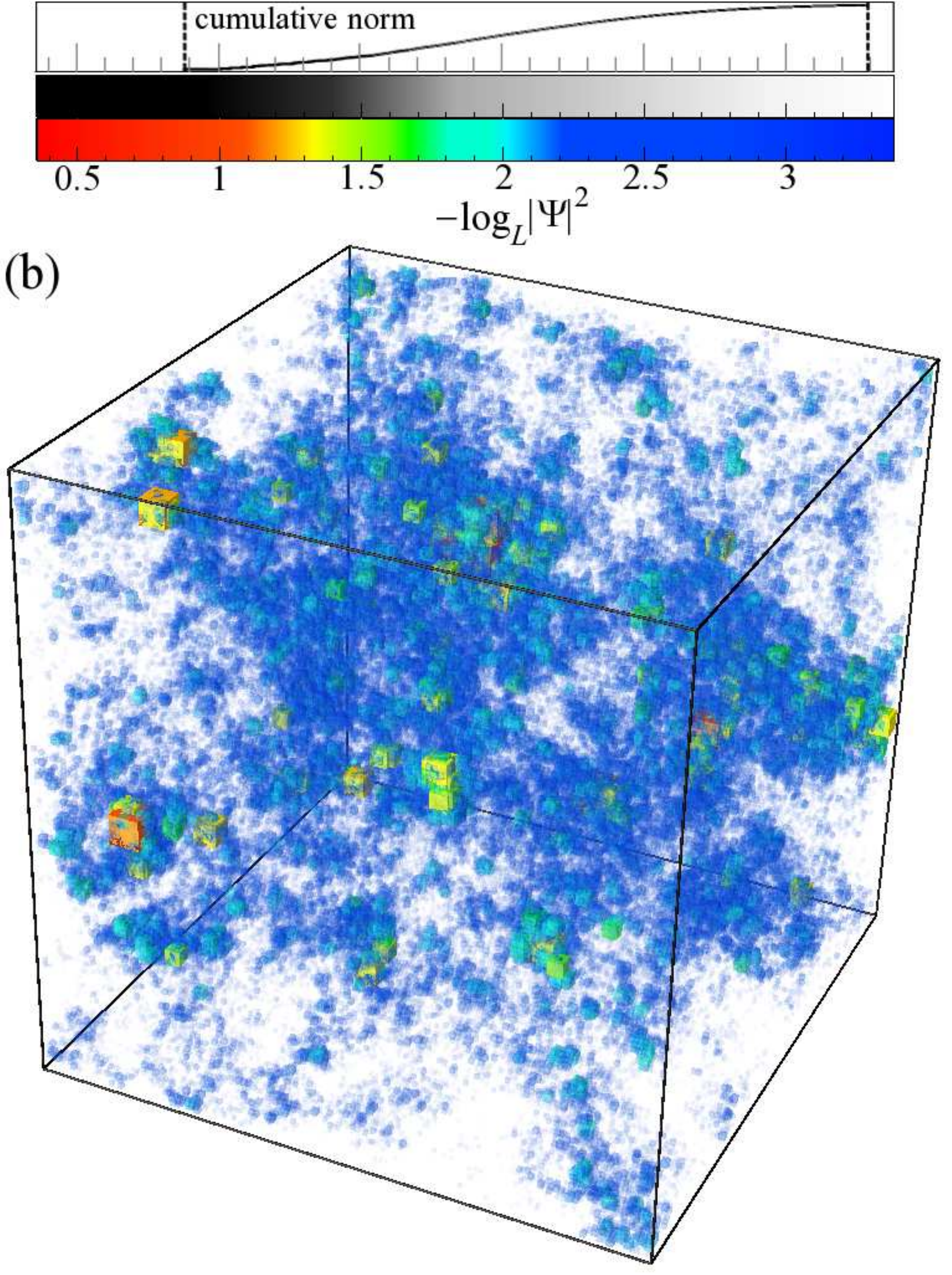}
 \includegraphics[width=.31\textwidth]{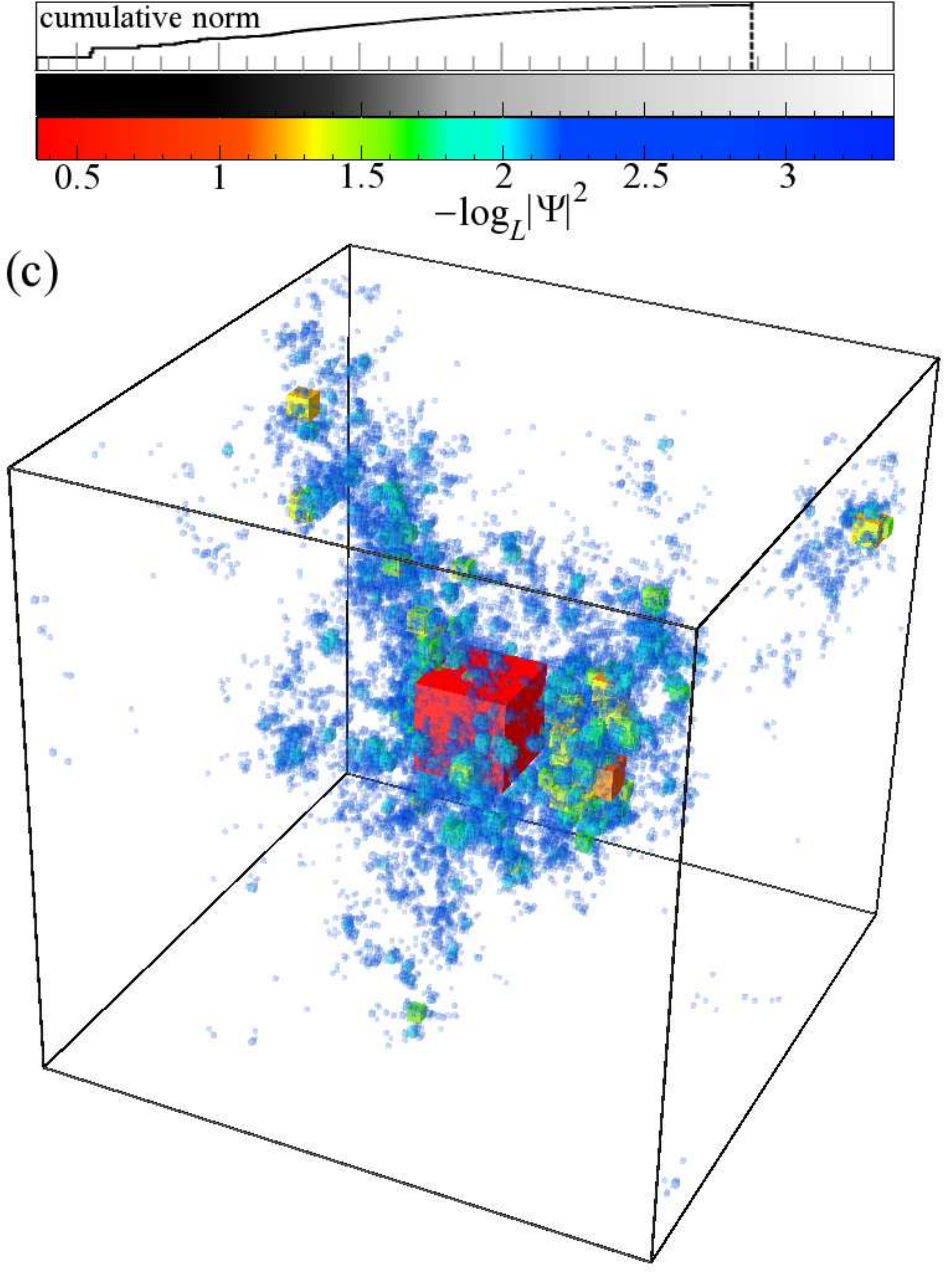}
\caption{(color online) Wavefunctions for the 3D Anderson model near the band center ($E=0$) for a system of size $L^3=120^3$.
From left to right, we move from the metallic phase through the transition to the insulating phase, as the degree of disorder increases: (a) $W=15$, (b) $W=16.5$, (c) $W=18$. Sites contributing to $98$\% of the wavefunction's norm (from large to small values) are shown as cubes whose volume is proportional to $|\psi_i|^2$. The color and opacity of the cubes is chosen according to the value of $-\log_L|\psi_i|^2$. The top plot shows the cumulative norm of the wavefunction as a function of the cut-off value considered for $-\log_L|\psi_i|^2$. Vertical dashed lines mark the minimum and maximum values of $|\psi_i|^2$ occurring in the wavefunction plots, which are (a) $9.4\times 10^{-8}\leqslant |\psi_i|^2\leqslant 0.0035$, (b) $1.4\times 10^{-7}\leqslant |\psi_i|^2\leqslant 0.015$ and (c) $1.0\times 10^{-6}\leqslant |\psi_i|^2\leqslant 0.18$. The opacity and color scales are indicated below the cumulative norm plot.}
\label{fig-states}
\end{figure*}
We have considered only a single eigenstate per sample (disorder realization), namely, the eigenstate with energy closest to $E=0$. This is costly in terms of computing time, but as we shall show later, absolutely essential to avoid the strong correlations that exist between eigenstates of the same sample.
System sizes range from $20^3$ to $120^3$, and disorder values are in the interval  $15\leqslant W \leqslant 18$. For each size and disorder combination, we have taken at least $5000(3000)$ samples for $L\leqslant100(>100)$, for a total of $\sim904 000$ wavefunctions. The average number of states considered for each $L,W$ pair is indicated in Table \ref{tab-states}.
In Fig.~\ref{fig-states} we show some wavefunctions for $L=120$ around the critical point.
\section{Scaling Laws for generalized multifractal exponents around the critical point}
\label{sec-gmfe}
\subsection{Multifractality at the critical point}

It is known that the eigenstates of the 3D Anderson model \eqref{eq:anderson_H1} exhibit multifractal fluctuations at the critical point.\cite{Aok83,Jan94a,EveM08}
Here, we recapitulate briefly the basics of multifractal analysis.

To analyze the multifractal properties of wavefunctions in $d$ dimensions of a system of size $L$, we coarse grain the wavefunction intensity on a scale $l<L$.
The system is partitioned into $(L/l)^d$ boxes of volume $l^d$.
A probability
\begin{equation}
\mu_k \equiv\sum_{j \in \text{box }k} |\psi_j|^2,
\label{eq-muk}
\end{equation}
is defined for each box $k$.
It is more convenient to work, not directly with the box probability $\mu$, but with
a related random variable $\alpha$, defined by
\begin{equation}
\alpha \equiv \frac{\ln \mu }{\ln \lambda}.
\end{equation}
Here, $\lambda$ is the ratio of the box size $l$ to the system size $L$
\begin{equation}
\lambda \equiv \frac{l}{L}.
\end{equation}
Multifractality  means that, if we count the number of boxes $N(\alpha)$ for which the value of the random variable $\alpha$
falls in a given small interval $[\alpha,\alpha+\Delta \alpha ]$, this number scales with $\lambda$ as
\begin{equation}
N(\alpha) \sim \lambda^{-f(\alpha)},
\end{equation}
in the limit that $\lambda \rightarrow 0$, i.e.\ that these boxes form a fractal with a fractal dimension $f( \alpha)$
that depends on $\alpha$.
The set of all fractal dimensions $f(\alpha)$ is known as the multifractal spectrum.

Generalized inverse participation ratios (GIPR) or $q$-moments are obtained by summing over the boxes
\begin{equation}
R_q\equiv\sum_k \mu_k^q.
\end{equation}
For later use, it is also convenient to define
\begin{equation}
S_q\equiv \frac{d R_q}{dq} =\sum_k \mu_k^q \ln \mu_k.
\end{equation}
As a consequence of multifractality it can be shown that at the critical point the GIPRs obey the scaling law
\begin{equation}
 \langle R_q \rangle \sim \lambda^{\tau_q},
 \label{eq-iprscal}
\end{equation}
in the limit that $\lambda \rightarrow 0$.
Here, the brackets denote an \emph{ensemble} average.
The mass exponents $\tau_q$ depend non-linearly on $q$.
They are conveniently expressed in terms of anomalous scaling exponents $\Delta_q$,
\begin{equation}
\tau_q=d(q-1) +\Delta_q.
\end{equation}
The multifractal spectrum and the exponents $\tau_q$ are related via a Legendre transformation,
\begin{equation}\label{Legendre}
    \alpha_q=d\tau_q/dq, \qquad f_q=q\alpha_q-\tau_q,
\end{equation}
which defines singularity strengths $\alpha_q$ and a singularity spectrum $f_q$.

Multifractal exponents can also be defined from the scaling law corresponding to the geometric or \emph{typical} average of the GIPRs
\begin{equation}
    \exp\langle \ln R_q \rangle \sim \lambda^{\tau^\text{typ}_q}.
    \label{eq-iprscaltyp}
\end{equation}
The relation between the typical and ensemble averaged multifractal exponents is now well understood, \cite{FosRL09,EveM08} and
the multifractal properties of the 3D Anderson transition have been thoroughly studied using this standard formalism.\cite{SchG91,GruS95,MilRs97,VasRR08,RodVR08,RodVR08a}

\subsection{Multifractal behavior in the vicinity of the critical point}

To extend multifractal analysis beyond the critical point we define a generalized mass exponent
\begin{equation}
    \widetilde{\tau}_q(W,L,l)\equiv \ln \langle R_q \rangle/\ln \lambda.
\end{equation}
Here, the tilde is used to emphasize that this equation applies throughout the critical region not just at the critical point.
This generalized mass exponent becomes the usual mass exponent $\tau_q$ at the critical point $W_{\rm c}$ in the limit $\lambda\rightarrow 0$.

Next we proceed to suggest a finite size scaling law for the GIPRs and from that derive a scaling law for these generalized mass exponents.
Close to the transition, we may suppose that the GIPRs are determined by the ratios of the length scales $l$ and $L$ to the
localization (correlation) length in the insulating (metallic) phase $\xi$.
This can be justified using renormalization group arguments,\cite{Car96} and is the basis for the scaling theory of localization.\cite{AbrALR79,LeeR85}
This leads to the following scaling law for the GIPRs\cite{YakO98}
\begin{equation}
 \langle R_q \rangle (W,L,l)= \lambda^{\tau_q} \mathcal{R}_q\left( L/\xi,l/\xi\right).
 \label{eq-iprscalgen}
\end{equation}
At the critical point $W_c$ the correlation length has a power law divergence
\begin{equation}
\xi \propto |W-W_c|^{-\nu},
\end{equation}
described by a critical exponent $\nu$.
Thus, the scaling correctly reproduces the invariance of the GIPRs with $\lambda$ exhibited by Eq.\ \eqref{eq-iprscal} at the critical point.

The scaling law for the GIPRs can be rearranged as follows to give a scaling law for the generalized mass exponents,
\begin{equation}
   \widetilde{\tau}_q(W,L,l) = \tau_q + \frac{q(q-1)}{\ln \lambda} {\mathcal{T}_q}\left( L/\xi,l/\xi\right).
   \label{eq-tauq}
\end{equation}
The function ${\mathcal{T}_q}$ is related to the original $\mathcal{R}_q$.
The factor $q(q-1)$ has been explicitly included so that
\begin{equation}
\widetilde{\tau}_0=\tau_0 = -d, \qquad \widetilde{\tau}_1=\tau_1 = 0.
\end{equation}
The generalized anomalous scaling exponents
\begin{equation}
\widetilde{\Delta}_q\equiv \widetilde{\tau}_q-d(q-1),
\end{equation}
will then obey
\begin{equation}
   \widetilde{\Delta}_q(W,L,l) = \Delta_q + \frac{q(q-1)}{\ln \lambda} {\mathcal{T}_q}\left( L/\xi,l/\xi\right).
   \label{eq-deltaq}
\end{equation}

By exact analogy with  Eq.~\eqref{Legendre} we may define generalized singularity strengths
\begin{equation}\label{eq-def-alphaq}
\widetilde{\alpha}_q \equiv d \widetilde{\tau}_q/dq = \big\langle S_q \big\rangle/ \left(\langle R_q \rangle \ln \lambda\right).
\end{equation}
The scaling law for these quantities follows immediately from
Eq.~\eqref{eq-tauq},
\begin{equation}
 \widetilde{\alpha}_q(W,L,l) = \alpha_q + \frac{1}{\ln \lambda} \mathcal{A}_q \left( L/\xi,l/\xi\right).
   \label{eq-alphaq}
\end{equation}
We may also define a generalized singularity spectrum
\begin{equation}\label{gensingspec}
\widetilde{f}_q \equiv q\widetilde{\alpha}_q - \widetilde{\tau}_q,
\end{equation}
obeying a corresponding scaling law,
\begin{equation}
 \widetilde{f}_q(W,L,l) = f_q + \frac{q}{\ln \lambda} \mathcal{F}_q \left( L/\xi,l/\xi\right).
 \label{eq-falphaq}
\end{equation}
The scaling law for the GIPRs can be expressed in the entirely equivalent form
\begin{equation}
 \langle R_q \rangle (W,L,\lambda)= \lambda^{\tau_q} \mathcal{R}_q\left( L/\xi,\lambda \right).
\end{equation}
This remark applies equally well to the scaling laws for other quantities.
Writing the scaling laws in this way immediately suggests a standard FSS analysis by fitting the
disorder and system size dependence \emph{at fixed} $\lambda$.
As we show below this is indeed possible and works well. It does not, however, permit the estimation of the various multifractal exponents.

A much more exciting application of the above scaling laws \eqref{eq-tauq}--\eqref{eq-falphaq} is to fit the variation with disorder, system size \emph{and} box size.
This allows not only the estimation of the usual critical parameters $W_c$ and $\nu$ but also the simultaneous determination of a multifractal exponent for a particular $q$.
Moreover, the use of different moments $q$ of the wavefunctions and different averages (ensemble, typical) provides a test of the stability of the estimates for the critical parameters, as these should be average- and $q$-independent.

\subsection{Averaging and the $\lambda\rightarrow 0$ limit}
\label{sec-gmfe-lambda}
%
\begin{table*}
 \caption{Definitions of GMFEs for ensemble and typical averages.
 For a particular $(W,L,l)$, $R_q$ and $S_q$ are calculated for each wavefunction and the ensemble average $\langle \cdots \rangle$ taken over samples.
 Formulae for error estimation are also given, where $\sigma$ stands for the standard deviation.
 Note that $\langle S_q \rangle$ and $\langle R_q\rangle$ are highly correlated and their covariance must be
 taken into account in the error estimation.
 Definitions of the ensemble average exponents in terms of the PDF $\mathcal{P}(\alpha)$ are also given.}
 \label{tab-gme}
 \begin{tabular}{lll}
 \hline\hline
 GMFE & Error estimation & Definition using the PDF \\\hline
 $\widetilde{\Delta}_q=\dfrac{\ln \langle R_q\rangle }{\ln \lambda}-d(q-1)$ &  $\sigma_{\widetilde{\Delta}_q}= \dfrac{\sigma_{\langle R_q\rangle}}{\langle R_q\rangle \ln \lambda}$
 & $\widetilde{\Delta}_q=-dq + \dfrac{1}{\ln \lambda} \ln \int_0^\infty \lambda^{q\alpha} \mathcal{P}(\alpha) d\alpha$\\
 $\widetilde{\Delta}^\text{typ}_q=\dfrac{\langle \ln  R_q \rangle}{\ln \lambda}-d(q-1)$ &  $\sigma_{\widetilde{\Delta}^\text{typ}_q}= \dfrac{\sigma_{\langle \ln R_q\rangle}}{\ln \lambda}$ & \\
 $\widetilde{\alpha}_q = \dfrac{\langle S_q \rangle}{\langle R_q \rangle \ln \lambda}$ &  $\sigma_{\widetilde{\alpha}_q}= \dfrac{1}{\ln \lambda}
   \sqrt{\dfrac{\sigma_{\langle S_q \rangle}^2}{\langle R_q\rangle^2} + \dfrac{\langle S_q \rangle^2 \sigma_{\langle R_q \rangle}^2 }{\langle R_q \rangle^4} -\dfrac{2 \langle S_q \rangle}{\langle R_q\rangle^3} \textrm{cov}(\langle S_q\rangle,\langle R_q\rangle) }$ &
   $\widetilde{\alpha}_q = \dfrac{\int_0^\infty \alpha \lambda^{q\alpha} \mathcal{P}(\alpha) d\alpha}{\int_0^\infty \lambda^{q\alpha} \mathcal{P}(\alpha) d\alpha} $\\
 $\widetilde{\alpha}^\text{typ}_q = \dfrac{1}{\ln \lambda}\left\langle \dfrac{S_q}{R_q}\right\rangle$ &  $\sigma_{\widetilde{\alpha}^\text{typ}_q}= \dfrac{\sigma_{\langle S_q/R_q \rangle} }{\ln \lambda}$ & \\
 \hline
 \end{tabular}
\end{table*}
\begin{figure*}
 \includegraphics[width=.3\textwidth]{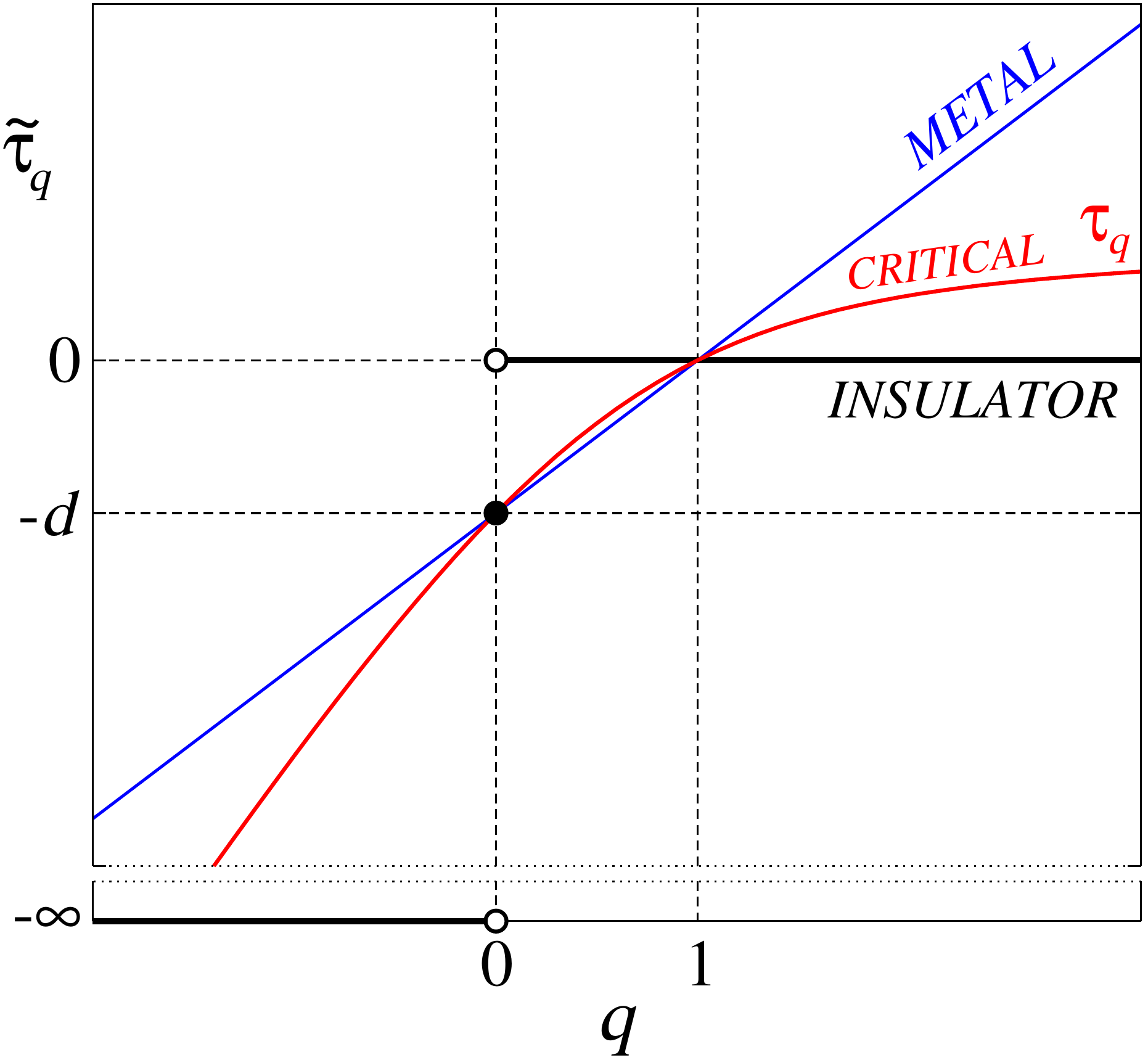}\hfill
 \includegraphics[width=.301\textwidth]{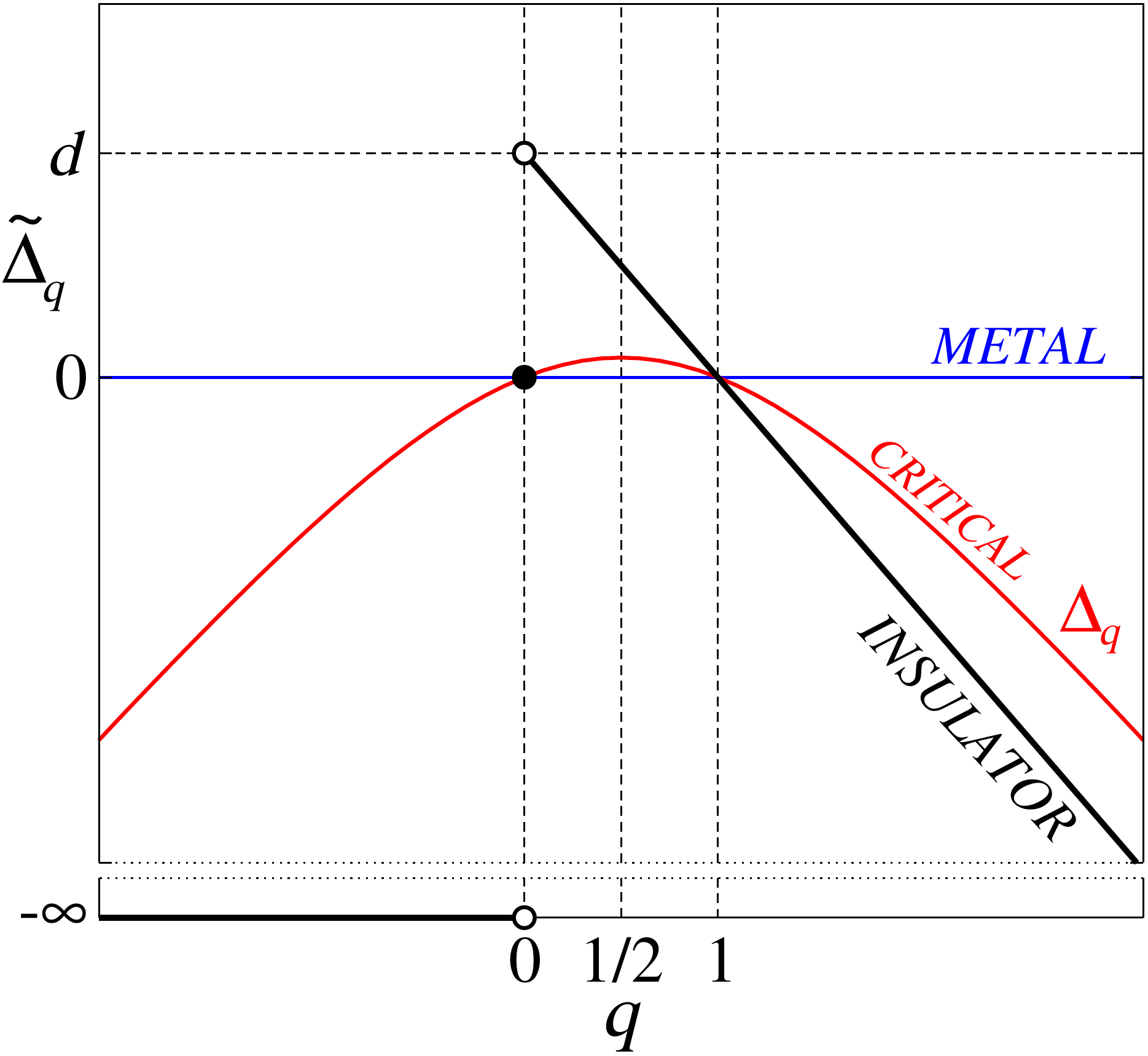}\hfill
 \includegraphics[width=.301\textwidth]{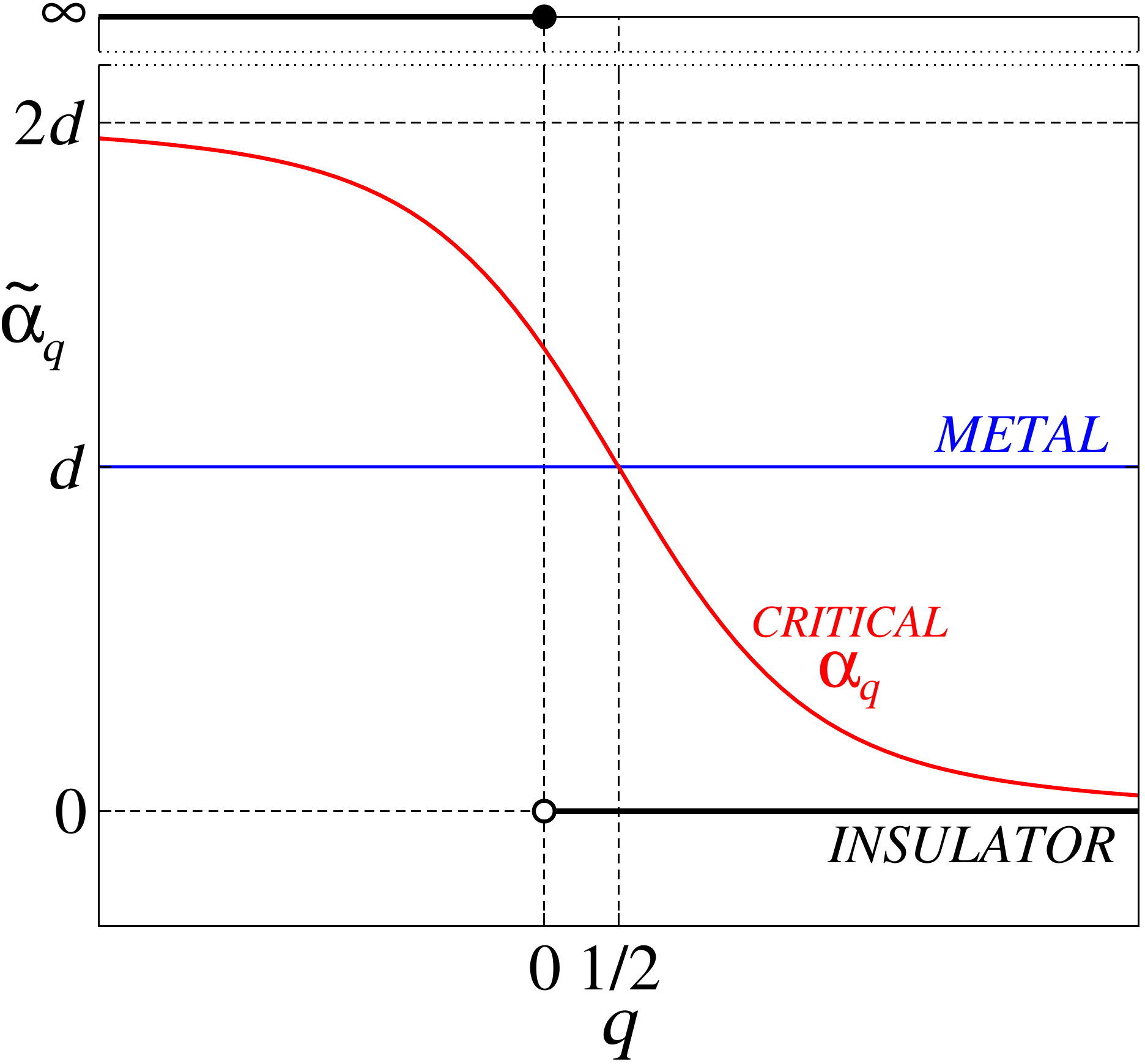}
 \caption{(color online) Schematic phase diagrams for the GMFEs in the limit $\lambda\rightarrow0$ for the metallic ($W<W_c$), critical $(W=W_c)$ and insulating ($W>W_c$) regimes.
 The metallic and insulating limits can also be reached as $L\rightarrow\infty$ at fixed $\lambda$.
 However, the multifractal exponents are obtained only at the critical point as $\lambda\rightarrow 0$.}
 \label{fig-MEphases}
\end{figure*}

We emphasize that the generalized multifractal exponents (GMFEs) $\widetilde{\tau}_q,\widetilde{\Delta}_q,\widetilde{\alpha}_q$ are equal to the corresponding \emph{scale invariant} multifractal exponents
$\tau_q,\Delta_q,\alpha_q$ only at the critical point $W=W_c$ in the limit $\lambda\rightarrow0$.

Recalling the trivial scaling of $R_0$ and $R_1$, we see that GMFEs $\widetilde{\tau}_0=-d$ and $\widetilde{\tau}_1=0$, independent of $W$, $L$, and $l$.
This is equivalent to  $\widetilde{\Delta}_0=\widetilde{\Delta}_1=0$.

To understand better the behavior of the GMFEs as functions of $q$, which is depicted schematically in Fig.~\ref{fig-MEphases}, we discuss the expected behaviors in the thermodynamic
limit in the metallic and insulating phases, and at the critical point.

In the metallic phase, for sufficiently large system size or sufficiently small disorder, the states are homogeneously extended and $\mu_k\rightarrow\lambda^d$.
It follows that $R_q\rightarrow\lambda^{d(q-1)}$, $\widetilde{\tau}_q\rightarrow d(q-1)$ and $\widetilde{\Delta}_q\rightarrow 0$ for all $q$.

In the insulating phase, for sufficiently large system sizes or large enough disorder, the wavefunctions approach an extremely localized state and $\mu_k\rightarrow \delta_{k,k_0}$.
Then, for $q>0$, $R_q \rightarrow 1$, $\widetilde{\tau}_q\rightarrow 0$ and $\widetilde{\Delta}_q\rightarrow -d(q-1)$.
While for $q<0$, the moments diverge, $\widetilde{\tau}_q\rightarrow -\infty$ and $\widetilde{\Delta}_q\rightarrow -\infty$. We emphasize that the insulating limit has been confirmed by
analytical calculations for 3D exponentially localized states (with finite localization lengths).\cite{note-gmfelimit}
We note that the metallic and insulating limits can be reached either as $\lambda\rightarrow0$ or as $L\rightarrow\infty$ for any fixed $\lambda$.

At the critical point, we recover the multifractal exponents as $\lambda\rightarrow 0$.
These are known to obey a symmetry relation,\cite{MirFME06}
\begin{equation}\label{deltaq_symrel}
\Delta_q=\Delta_{1-q}
\end{equation}
or equivalently
\begin{equation}\label{alphaq_symrel}
\alpha_q+\alpha_{1-q}=2d.
\end{equation}
This symmetry has been studied for Anderson transitions in different systems and dimensionality,\cite{MilSNE07,MilE07,EveMM08a,ObuSFGL08,ObuSFGL07,FyoOR09,RodVR08} and has also been experimentally measured.\cite{FaeSPL09}
The corresponding limits  for $\widetilde{\alpha}_q$ can be obtained following similar reasonings and calculations.\cite{note-gmfelimit}

We may also define $\widetilde{\tau}^\text{typ}_q,\widetilde{\Delta}^\text{typ}_q,\widetilde{\alpha}^\text{typ}_q$, in terms of the \emph{typical} average
for the moments $R_q$ (see Eq.~\eqref{eq-iprscaltyp}).
The scaling laws have the same form as for the ensemble average exponents.
For the benefit of the reader, we collect in Table \ref{tab-gme} the definition of the exponents $\widetilde{\Delta}_q$ and $\widetilde{\alpha}_q$ for both cases.

%
%
\section{Scaling for the PDF of wavefunction intensities around the critical point}
\label{sec-pdf}
It is also possible, and indeed, sometimes more convenient, to work directly with the probability density function (PDF) $\mathcal{P}(\alpha; W,L,l)\equiv \mathcal{P}(\alpha)$ of $\alpha$.\cite{RodVR09}
At the critical point the distribution is multifractal
\begin{equation}
 \mathcal{P}(\alpha; W = W_c, L, l)\underset{\lambda\rightarrow 0}{\propto} \sqrt{|\ln \lambda|}\,\lambda^{d-f(\alpha)},
 \label{eq-pdf}
\end{equation}
and exhibits scale invariance, provided that $\lambda$ is held fixed.
Equation \eqref{eq-pdf} can be used to estimate the multifractal spectrum directly from numerically calculated histograms of $\alpha$ values.\cite{RodVR09}
When we broaden attention to the critical regime, we find that, when $\lambda$ is sufficiently small, we may approximate the
PDF using the generalized multifractal spectrum \eqref{gensingspec}
\begin{equation}
\mathcal{P}(\alpha;W,L,l) \propto \sqrt{|\ln \lambda|}\,\lambda^{d - \widetilde{f}(\alpha;W,L,l)}.
\end{equation}
This relation could be the basis for an alternative definition of the generalized singularity spectrum from the PDF. While there will be quantitative differences with
\eqref{gensingspec} we would expect the results of scaling analysis to be unchanged.

The GMFEs can be obtained from the PDF.
We consider $\widetilde{\alpha}_0$ as an example.
Setting $q=0$ in Eq.\ \eqref{eq-def-alphaq} we have
\begin{equation}
\widetilde{\alpha}_0 = \frac{ \left< \sum_k \ln \mu_k \right>}{\lambda^{-d} \ln \lambda}.
\end{equation}
The result of averaging is the same for all boxes, so
\begin{equation}
\widetilde{\alpha}_0 = \langle \alpha \rangle = \int_0^\infty \alpha \mathcal{P}(\alpha) d\alpha.
\end{equation}
Thus $\widetilde{\alpha}_0$ corresponds to the mean value of the PDF.
Expressions for general $q$ are given in Table \ref{tab-gme}.

\begin{figure}[tb]
 \includegraphics[width=\figwidth]{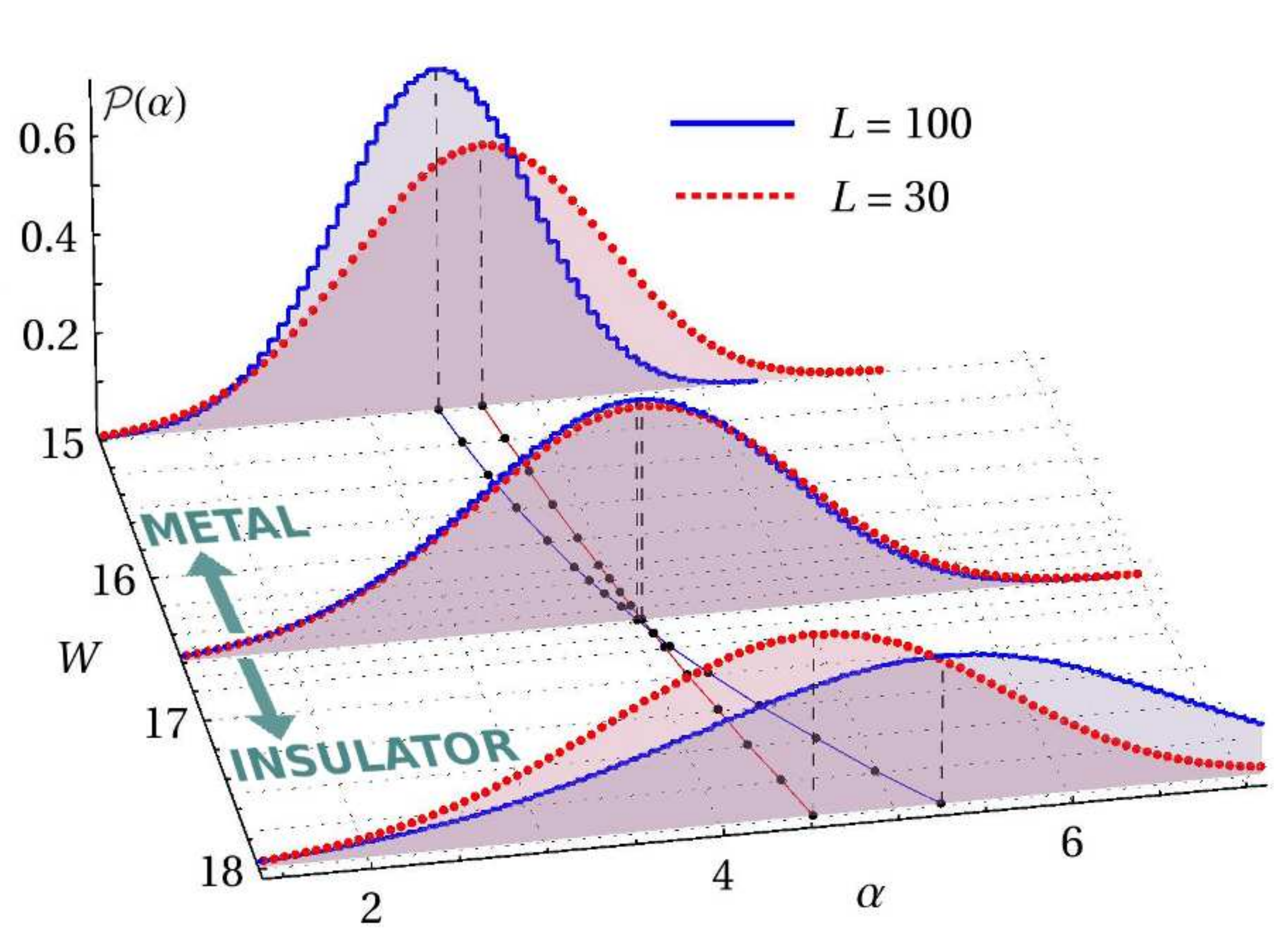}
 \caption{(color online) Reproduced from Ref.~\onlinecite{RodVSR10}. Evolution of the wavefunction intensity distribution $\mathcal{P}(\alpha;W,L,l)$ as a function of disorder $W$ across the Anderson transition,
 at fixed $\lambda=0.1$ for two system sizes $L$. Each distribution was computed with $10^4$ wavefunctions. The data points ($\bullet$) and solid lines on the bottom plane mark the trajectories of the maximum.
   For clarity, distributions are shown at $W=15$, $16.6$ and $18$ only.}
 \label{fig-pdf}
\end{figure}

The scaling of $\mathcal{P}(\alpha)$ with system size for fixed $\lambda$ in the vicinity of the transition is shown in Fig.~\ref{fig-pdf}. This figure shows clearly that
the scaling of the distribution of wavefunction intensities, or the distribution of a related quantity such as the local density of states (LDOS), could be used to characterize the
Anderson transition.\cite{MirF94}
Indeed, in Ref.~\onlinecite{RodVSR10} the critical parameters were successfully estimated from a scaling analysis of the system size and disorder dependence of the maximum of the PDF of $\alpha$ at fixed $\lambda$.
A similar procedure might be applied to experimental LDOS data obtained using STM techniques,\cite{HasSWI08,RicRMZ10,MorKMG02,KraCWC10,MilKRR10} or by direct imaging of ultracold atom systems.\cite{BilJZB08,RoaDFF08}

\section{Remarks on correlations amongst wavefunctions}
\label{sec-datacor}

In exact diagonalization studies of disordered systems, it is common practice to average over eigenstates located within a small energy window.
This is because the initialization and diagonalization of very large matrices is computationally demanding. Time can be saved by generating several eigenstates for the \emph{same} sample.
There is, however, a price to be paid.
Eigenstates of the same sample are correlated, since they are solutions of the Schr\"{o}dinger equation with the same potential.
We have found that these correlations distort the statistical analysis.
In particular, the error estimation becomes unreliable and the precision of critical parameters is overestimated, i.e.\ the error bars are erroneously small.

\begin{figure}[tb]
 \includegraphics[width=\columnwidth]{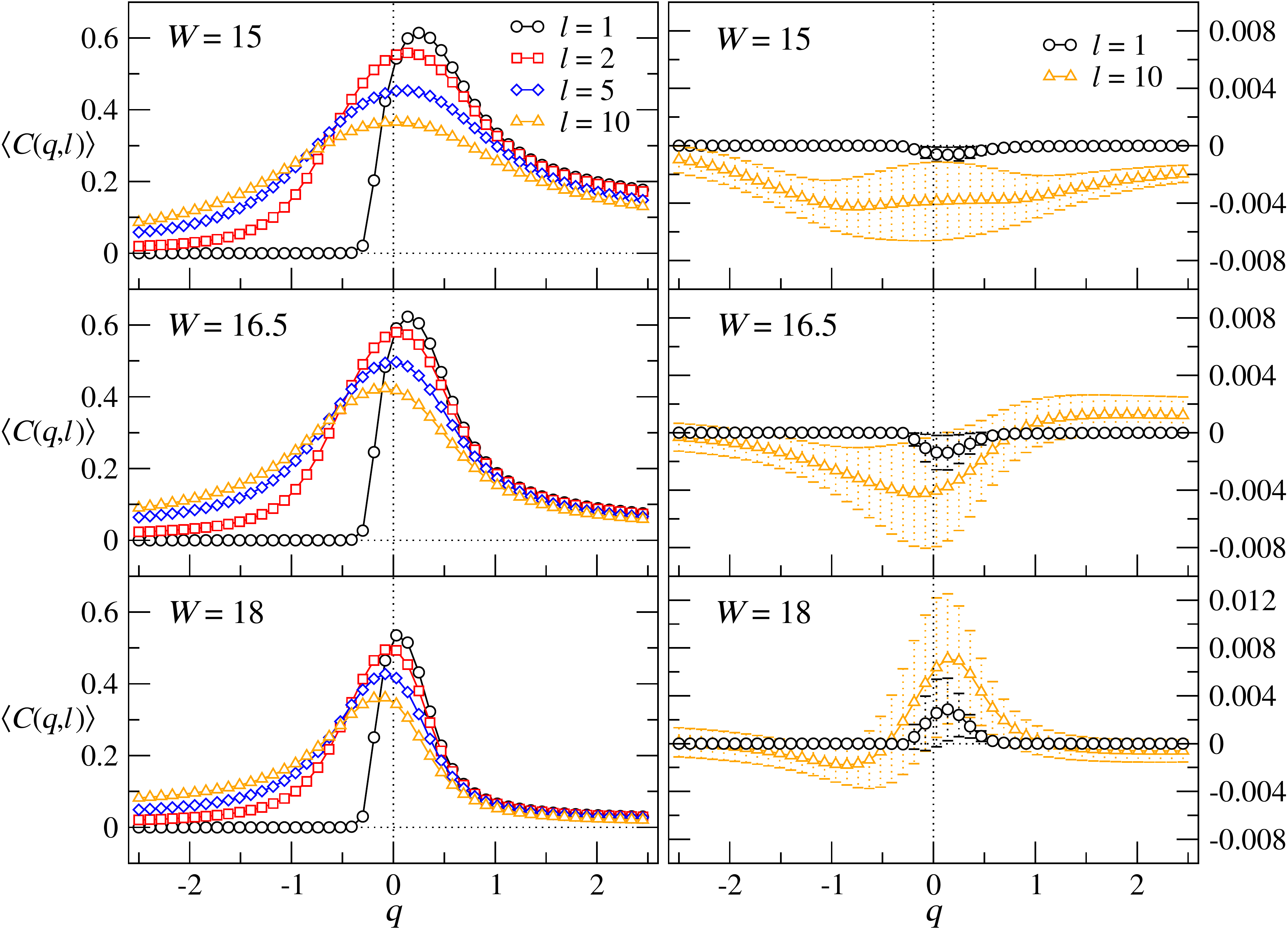}
 \caption{(color online) Averaged correlation (over $1000$ samples) between eigenstates as a function of $q$, for $L^3=100^3$ and varying box-size $l$ and disorder $W$.
 Data on the left were obtained using pairs of eigenstates from the same sample.
 Data on the right, for eigenstates of two different samples. Note the different scales for the ordinate axis.
 Whenever not shown, errors are smaller than the symbol size.}
\label{fig-5corr}
\end{figure}

To quantify the correlations amongst wavefunctions from the same sample, we have studied the average correlation between two eigenstates with energies $E$, $E'$, close to $E=0$, such that $|E-E'|\ll1$.
The correlation is defined as,
\begin{equation}
 \mathcal{C}(q,l)=\frac{\textrm{cov}(\mu^{q},\mu^{\prime q})}{\sigma(\mu^{q}) \sigma(\mu^{\prime q})}.
 \label{eq-corr}
\end{equation}
Here, $\mu^q$ is the $q$-th power of the box probability Eq.~\eqref{eq-muk} for the eigenstate with energy $E$, and the prime indicates the same quantity calculated for the
eigenstate with energy $E^{\prime}$.
The covariance is calculated using \eqref{eq-covestimate}, where the sum is over boxes. The covariance is normalized to the product of the standard deviations,
where again the sum is over boxes.
The correlation can thus be calculated for two eigenfunctions from the sample, or two eigenfunctions from a pair of samples.
A further ensemble average over $1000$ samples, or pairs of samples, as appropriate, was taken to arrive at Fig.~\ref{fig-5corr}.
On the left is the correlation $\mathcal{C}(q,l)$ obtained using pairs of eigenstates of the same sample.
On the right is the correlation obtained using eigenstates of two different samples.
A maximum correlation (anticorrelation) corresponds to $\mathcal{C}(q,l)=1$ $(-1)$, whereas statistical independence implies a vanishing $\mathcal{C}(q,l)$.
The correlation is shown as a function of the moment $q$, for different degrees of coarse-graining $l$ and for three values of the disorder $W=15,16.5,18$.
(Note that, in the limit $q\rightarrow 0$, the numerator and the denominator in \eqref{eq-corr} both go to zero.
The finite solution of this indetermination is not necessarily unity, as one may naively expect.)
Near the critical point, the correlation between eigenstates of the same sample is high, while for eigenstates of different samples correlation is, as expected, absent.
We conclude that it is not safe to include in the MFSS analysis more than one state from a given sample, as these correlations render the subsequent statistical analysis unreliable.

On the left of Fig.~\ref{fig-5corr} we observe how the correlation varies with $q$.
For large positive $q$, large wavefunction amplitudes dominate the $q$-th moment of the box probability, while for large
negative $q$ small amplitudes dominate. In both cases the correlation is reduced.
We think that this occurs because only a small number of boxes contribute significantly to the extreme values of the distribution.
We have checked that the overall shape of the correlations in the insulating regime agrees with calculations for 1D exponentially localized states.
Another feature is that for negative $q$ correlations are absent for box size $l=1$ but are restored when the wavefunction is coarse-grained.
This may indicate that very small wavefunction amplitudes are affected by random noise, which averages away when the wavefunction is coarse-grained.
We have verified that similar behavior occurs for wavefunctions calculated using other numerical libraries, e.g.\ {\sc LAPACK}.\cite{AndBBB87}
We strongly recommend the use of coarse-graining to evaluate negative moments, even for \mbox{1-D} models, such as power-law random banded matrices\cite{MirFME06} and others.\cite{FyoOR09}

We emphasize that in this work --- with the exception of the data on the left of Fig.\ \ref{fig-5corr} --- we have used only one eigenfunction per sample.

\section{Single-parameter scaling at fixed $\lambda$}
\label{sec-fixlambda}

We study first the scaling of the GMFEs $\widetilde{\Delta}_q$, $\widetilde{\Delta}_q^\text{typ}$, $\widetilde{\alpha}_q$, $\widetilde{\alpha}_q^\text{typ}$ at fixed $\lambda\equiv l/L$.\cite{note-scalinglaw}
This simplifies the scaling laws \eqref{eq-deltaq} and \eqref{eq-alphaq}, which become one parameter functions,
\begin{equation}
 \Gamma_q(W,L)= \mathcal{G}_q(L/\xi),
 \label{eq-gen}
\end{equation}
where $\Gamma_q$ denotes any of the above mentioned exponents.
Since $\lambda$ is fixed, the multifractal exponents cannot be estimated, as they merge with the zero-th order term in the expansion of the scaling functions $\mathcal{T}_q$ and $\mathcal{A}_q$ (see Eqs.~\eqref{eq-deltaq} and \eqref{eq-alphaq}). Nevertheless the critical disorder $W_c$ and the critical exponent $\nu$ can be estimated.
We demonstrate the consistency of the estimates of these parameters for different $q$ values, and their independence on the type of average (typical or ensemble) considered.

\subsection{Expansion in relevant and irrelevant scaling variables}
\label{sec-fixlambda-expansion}

In order to fit data for the GMFEs, we follow the standard procedure of Ref.\ \onlinecite{SleO99a} and include two
kinds of corrections to scaling,
(i) nonlinearities of the $W$ dependence of the scaling variables,
and (ii) an irrelevant scaling variable that accounts for a shift with $L$ of the apparent critical disorder at which the $\Gamma_q(W,L)$ curves cross.
After expanding to first order in the irrelevant scaling variable,
the scaling functions take the form
\begin{equation}
  \mathcal{G}_q(\varrho L^{1/\nu}, \eta L^{y})= {\mathcal G}_q^0(\varrho L^{1/\nu}) +\eta L^{y} {\mathcal G}_q^1(\varrho L^{1/\nu}).
  \label{eq-Gexpansion}
\end{equation}
Here, $\varrho$ and $\eta$ are the relevant and irrelevant scaling variables, respectively.
The irrelevant component is expected to vanish for large $L$, so $y<0$.
Both the scaling functions are Taylor-expanded
\begin{equation}
 \mathcal{G}_q^k(\varrho L^{1/\nu})= \sum_{j=0}^{n_k} a_{kj}\varrho^j L^{j/\nu}, \quad \text{for } k=0, 1.
\end{equation}
The scaling variables are expanded in terms of $w\equiv(W-W_c)$
up to order $m_{\varrho}$ and $m_{\eta}$, respectively,
\begin{equation}
  \varrho(w)=w+\sum_{m=2}^{m_{\varrho}} b_m w^m,\quad
   \eta(w)=1+\sum_{m=1}^{m_{\eta}} c_m w^m.
   \label{eq-fieldex}
\end{equation}
The fitting function is characterized by the expansion orders $n_0, n_1, m_{\varrho}, m_{\eta}$.
The total number of free parameters to be determined in the fit is
$N_P=n_0+ n_1+m_{\varrho}+ m_{\eta}+4$ (including $\nu$, $y$ and $W_c$).

The localization (correlation) length, up to a constant of proportionality, is $\xi=|\varrho(w)|^{-\nu}$.
After subtraction of corrections to scaling
\begin{equation}
\Gamma_q^\text{corr}\equiv\Gamma_q(W,L)- \eta L^{y} {\mathcal G}_q^1(\varrho L^{1/\nu}),
\end{equation}
and the data for the GMFEs should fall on the single-parameter curves
\begin{equation}
\Gamma_q^\text{corr} = \mathcal{G}_q^0(\pm (L/\xi)^{1/\nu}).
\end{equation}

\subsection{Numerical procedure at fixed $\lambda$}
\label{sec-fixlambda-numerics}

When performing FSS, the aim is to identify a \emph{stable} expansion of the scaling function that fits the numerical data.
The best fit is found by minimizing the $\chi^2$ statistic over the parameter space.
The validity of the fit is decided by the $p$-value or goodness-of-fit.
We take $p\geqslant 0.1$ as the threshold for an acceptable fit.
As a rule of thumb the expansion orders $n_0, n_1, m_{\varrho}, m_{\eta}$ are kept as low as possible while giving acceptable and stable fits.
Once a stable fit has been found, the precision of the estimates of the critical parameters is estimated by Monte Carlo simulation, i.e.\ by fitting a large set of synthetic data sets
generated by adding appropriately scaled random normal errors to an ideal data set generated from the best-fit model.
A detailed description of the FSS procedure, with some examples, is given in Appendix \ref{app-fss}.

We performed a detailed FSS analysis for $\widetilde{\Delta}_q$ and $\widetilde{\alpha}_q$ for both ensemble and typical averages for $13$ different values of $q\in[-1,2]$ at a fixed value of $\lambda=0.1$, i.e.\ the box-size is always $l=L/10$.
The GMFEs were obtained for system-sizes ranging between $L=20$ and $L=120$ and for $17$ values of the disorder $W\in[15,18]$.
The average number of independent wavefunctions involved in the calculation for each $L$, $W$ is indicated in Table \ref{tab-states}.

\subsection{Results for $\lambda=0.1$}
\begin{figure*}
 \includegraphics[width=.33\textwidth]{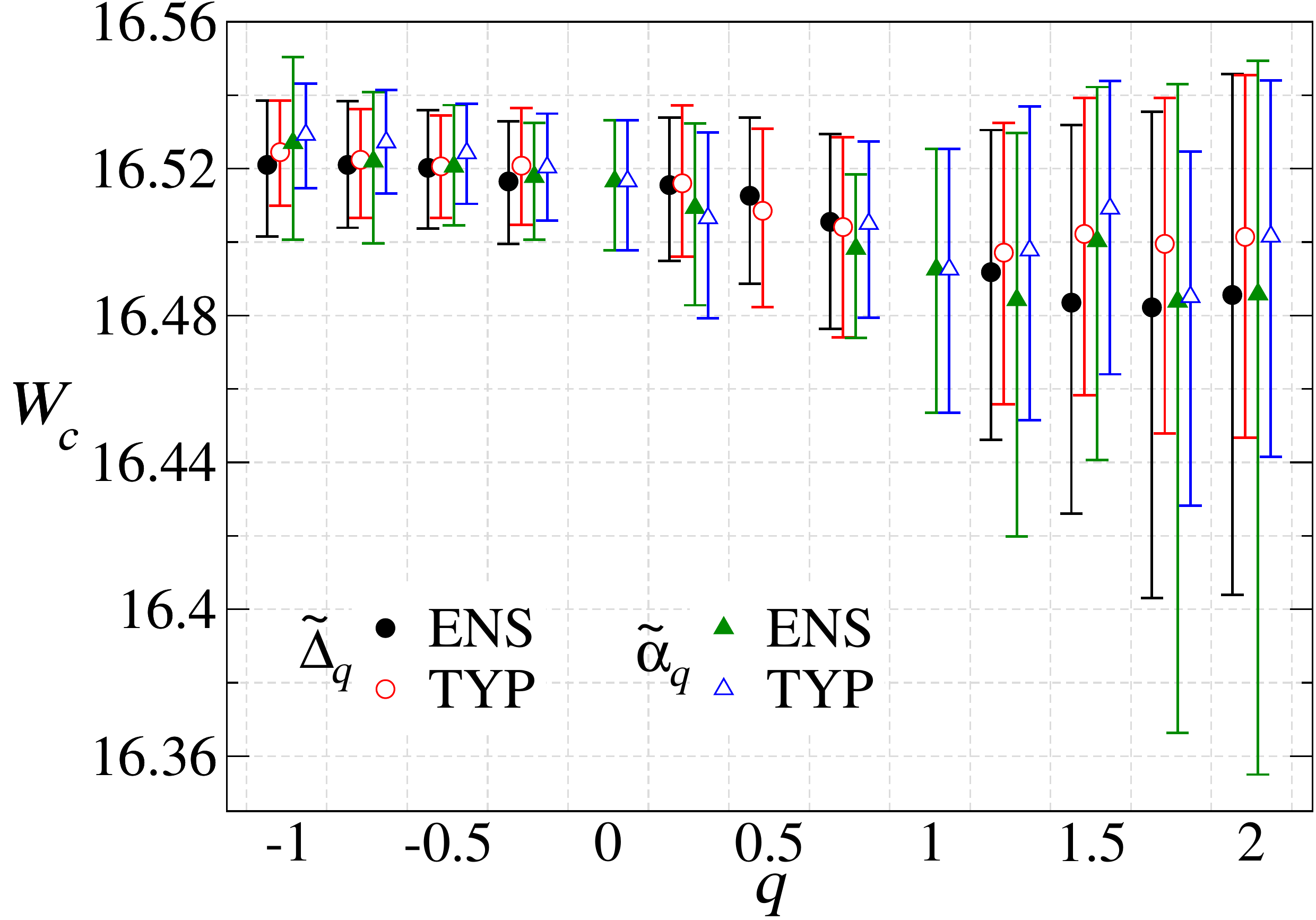}\hfill
 \includegraphics[width=.315\textwidth]{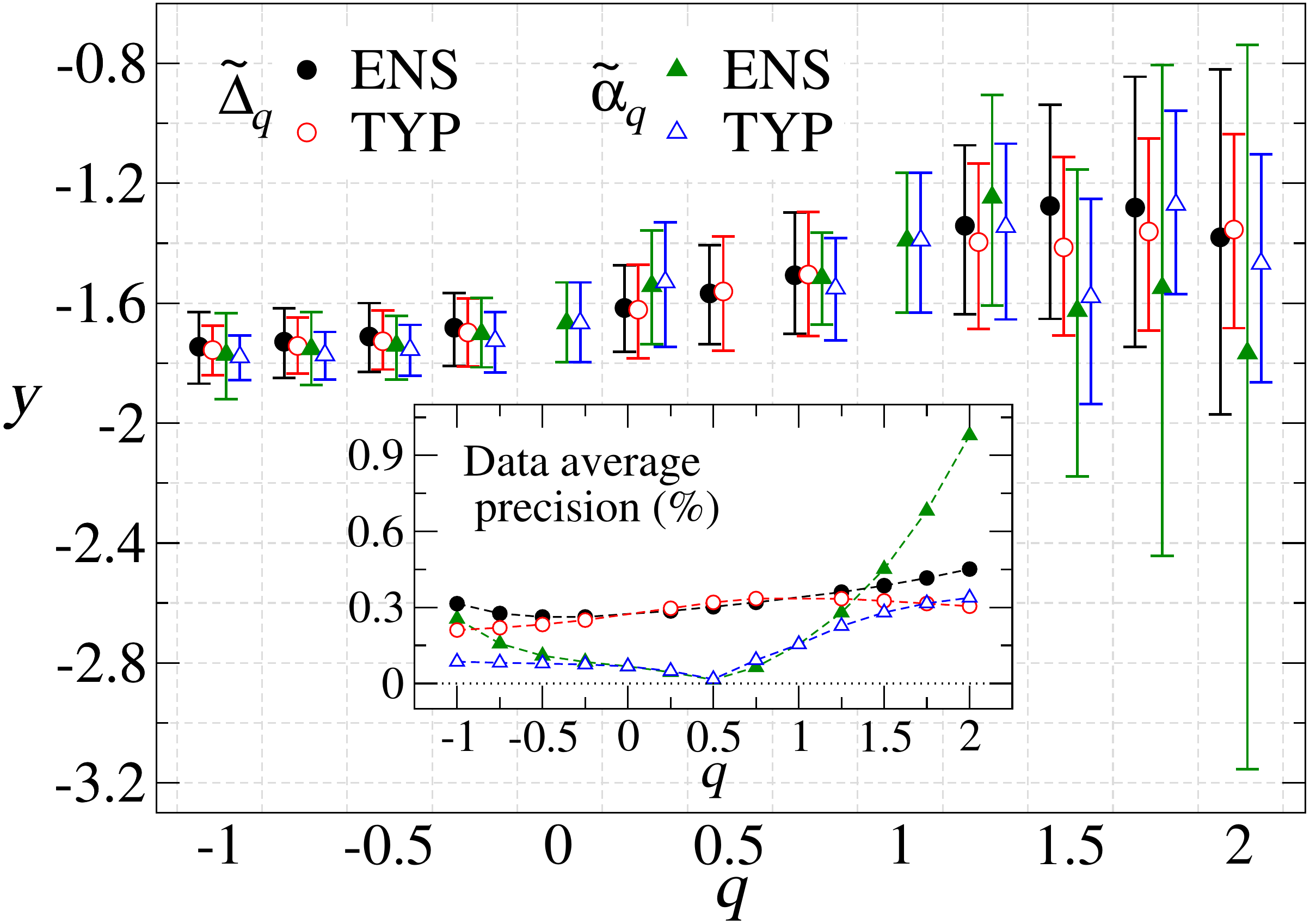}\hfill
 \includegraphics[width=.31\textwidth]{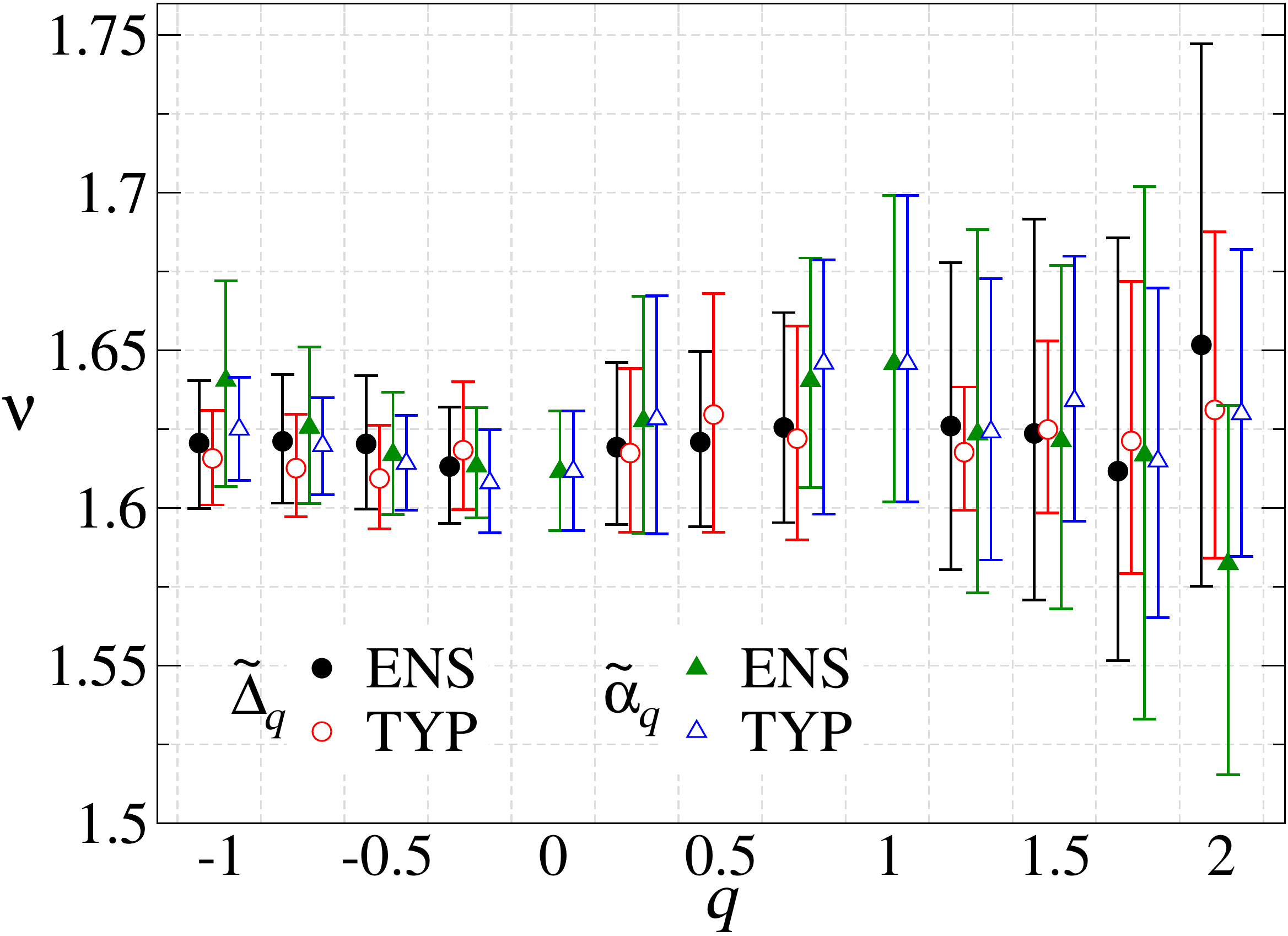}
 \caption{(color online) The estimates of the critical disorder $W_c$, critical exponent $\nu$,  and irrelevant exponent $y$, as functions of $q$, obtained from single-parameter FSS at fixed $\lambda=0.1$. Results for different averages and different exponents have been slightly offset in the $q$ direction. Error bars are 95\% confidence intervals. The corresponding values are listed in Table \ref{tab-results0.1}. The inset in the center plot shows the average data precision versus $q$ for the different GMFEs considered. }
 \label{fig-CPvsq0.1}
\end{figure*}

%
\begin{table*}
\caption{The estimates of the critical parameters, together with 95\% confidence intervals,  from single-parameter FSS at fixed $\lambda=0.1$.
Results for both ensemble (ENS) and typical (TYP) averages, for different values of $q\in[-1,2]$, are given.
Notice that for $\widetilde{\alpha}_0$, and also for $\widetilde{\alpha}_1$, the typical and ensemble averages coincide.
The number of data is $N_D$ and its average percentage precision is given in parentheses.
The number of free parameters in the fit is $N_P$, $\chi^2$ is the value of the chi-squared statistic for the best fit, and $p$ is the goodness of fit probability.
The orders of the expansions are specified in the last column. The system sizes used are $L\in[20, 120]$, and the range of disorder is $W\in\left[15,18\right]$.%
}
\begin{tabular}{ccccccccccc}
\hline\hline
   & $q$ & Average & $\nu$  & $W_c$ & $-y$ & $N_D\text{(prec.\%)}$ & $N_P$ & $\chi ^2$ & $p$ & $n_0\,n_1\,m_{\rho }\,m_{\eta }$ \\
\hline
 $\widetilde{\Delta}_q$ & $-1$ & ENS & $1.621(1.600,1.640)$ & $16.521(16.502,16.539)$ & $1.75(1.63,1.87)$ & $187\,(0.31)$ & 10 & 181 & 0.39 & 3 1 2 0 \\
 &        & TYP & $1.616(1.601,1.631)$ & $16.525(16.510,16.539)$ & $1.76(1.67,1.84)$ & $187\,(0.21)$ & 10 & 177 & 0.49 &  3 1 2 0  \\
 &$-0.75$ & ENS &  $1.621(1.601,1.642)$ & $16.521(16.504,16.538)$ & $1.73(1.62,1.85)$ & $187\,(0.28)$ & 11 & 181 & 0.38 & 4 1 2 0 \\
 &        & TYP & $1.613(1.597,1.630)$ & $16.522(16.507,16.536)$ & $1.74(1.65,1.84)$ & $187\,(0.22)$ & 10 & 177 & 0.49 & 3 1 2 0 \\
 &$-0.5$  & ENS &  $1.620(1.600,1.642)$ & $16.520(16.504,16.536)$ & $1.71(1.60,1.83)$ & $187\,(0.26)$ & 11 & 180 & 0.40 & 4 1 2 0 \\
 &        & TYP & $1.609(1.593,1.626)$ & $16.521(16.507,16.535)$ & $1.73(1.62,1.82)$ & $187\,(0.23)$ & 10 & 178 & 0.47 & 3 1 2 0 \\
 &$-0.25$ & ENS & $1.613(1.595,1.632)$ & $16.517(16.500,16.533)$ & $1.68(1.57,1.81)$ & $187\,(0.26)$ & 11 & 176 & 0.49 & 3 2 2 0 \\
 &        & TYP & $1.618(1.599,1.640)$ & $16.521(16.505,16.537)$ & $1.70(1.58,1.81)$ & $187\,(0.25)$ & 11 & 173 & 0.56 & 4 1 2 0 \\
 &$0.25$  & ENS & $1.619(1.595,1.646)$ & $16.516(16.495,16.534)$ & $1.62(1.47,1.76)$ & $187\,(0.29)$ & 12 & 168 & 0.63 & 5 1 2 0 \\
 &        & TYP & $1.617(1.592,1.644)$ & $16.516(16.496,16.537)$ & $1.62(1.47,1.78)$ & $187\,(0.30)$ & 12 & 171 & 0.58 & 5 1 2 0 \\
 &$0.5$   & ENS & $1.621(1.594,1.650)$ & $16.513(16.489,16.534)$ & $1.57(1.41,1.74)$ & $187\,(0.30)$ & 12 & 167 & 0.65 & 5 1 2 0 \\
 &        & TYP & $1.630(1.592,1.668)$ & $16.509(16.482,16.531)$ & $1.56(1.38,1.76)$ & $187\,(0.32)$ & 12 & 182 & 0.33 & 4 1 3 0 \\
 &$0.75$  & ENS & $1.626(1.595,1.662)$ & $16.506(16.476,16.529)$ & $1.51(1.30,1.70)$ & $187\,(0.32)$ & 11 & 174 & 0.53 & 4 1 2 0 \\
 &        & TYP & $1.622(1.590,1.658)$ & $16.504(16.474,16.529)$ & $1.51(1.29,1.71)$ & $187\,(0.33)$ & 11 & 178 & 0.45 & 4 1 2 0 \\
 &$1.25$  & ENS & $1.626(1.580,1.678)$ & $16.492(16.446,16.531)$ & $1.34(1.07,1.64)$ & $187\,(0.36)$ & 11 & 172 & 0.56 & 4 1 2 0 \\
 &        & TYP & $1.618(1.599,1.638)$ & $16.497(16.456,16.532)$ & $1.40(1.13,1.69)$ & $187\,(0.33)$ & 11 & 167 & 0.64 & 5 0 2 0 \\
 &$1.5$   & ENS & $1.624(1.571,1.692)$ & $16.484(16.426,16.532)$ & $1.28(0.93,1.65)$ & $187\,(0.39)$ & 11 & 175 & 0.51 & 4 1 2 0 \\
 &        & TYP & $1.625(1.598,1.653)$ & $16.502(16.458,16.539)$ & $1.41(1.11,1.71)$ & $187\,(0.33)$ & 11 & 168 & 0.65 & 4 0 3 0 \\
 &$1.75$  & ENS & $1.612(1.552,1.686)$ & $16.482(16.403,16.536)$ & $1.28(0.84,1.75)$ & $187\,(0.42)$ & 11 & 174 & 0.52 & 4 1 2 0 \\
 &        & TYP & $1.621(1.579,1.672)$ & $16.500(16.448,16.539)$ & $1.36(1.05,1.69)$ & $187\,(0.32)$ & 11 & 170 & 0.61 & 4 1 2 0 \\
 &$2$     & ENS & $1.652(1.575,1.747)$ & $16.486(16.404,16.546)$ & $1.38(0.82,1.97)$ & $187\,(0.45)$ & 12 & 166 & 0.67 &  4 1 3 0 \\
 &        & TYP & $1.631(1.584,1.688)$ & $16.501(16.447,16.546)$ & $1.35(1.04,1.68)$ & $187\,(0.30)$ & 12 & 168 & 0.64 & 4 2 2 0 \\
     \hline
 $\widetilde{\alpha}_q$ &$-1$    & ENS & $1.640(1.607,1.672)$ & $16.527(16.501,16.550)$ & $1.77(1.63,1.92)$ & $187\,(0.25)$ & 10 & 179 & 0.45 & 3 1 2 0 \\
 &        & TYP & $1.625(1.609,1.641)$ & $16.529(16.515,16.543)$ & $1.78(1.71,1.86)$ & $187\,(0.09)$ & 10 & 178 & 0.47 & 3 1 2 0 \\
 &$-0.75$ & ENS & $1.626(1.601,1.651)$ & $16.522(16.500,16.541)$ & $1.75(1.63,1.87)$ & $187\,(0.16)$ & 10 & 182 & 0.38 & 3 1 2 0 \\
 &        & TYP & $1.620(1.604,1.635)$ & $16.527(16.513,16.541)$ & $1.77(1.70,1.85)$ & $187\,(0.08)$ & 10 & 175 & 0.52 & 3 1 2 0 \\
 &$-0.5$  & ENS & $1.617(1.598,1.637)$ & $16.521(16.505,16.537)$ & $1.74(1.64,1.86)$ & $187\,(0.11)$ & 10 & 184 & 0.34 & 3 1 2 0 \\
 &        & TYP & $1.614(1.599,1.629)$ & $16.524(16.510,16.538)$ & $1.76(1.67,1.84)$ & $187\,(0.08)$ & 10 & 176 & 0.50 & 3 1 2 0 \\
 &$-0.25$ & ENS & $1.613(1.597,1.632)$ & $16.518(16.501,16.533)$ & $1.70(1.58,1.81)$ & $187\,(0.08)$ & 11 & 179 & 0.43 & 3 2 2 0 \\
 &        & TYP & $1.608(1.592,1.625)$ & $16.520(16.506,16.535)$ & $1.73(1.63,1.83)$ & $187\,(0.07)$ & 10 & 179 & 0.45 & 3 1 2 0 \\
 &$0$     & ENS/TYP & $1.612(1.593,1.631)$ & $16.517(16.498,16.533)$ & $1.67(1.53,1.80)$ & $187\,(0.07)$ & 10 & 175 & 0.53 & 3 1 2 0 \\
 &$0.25$  & ENS & $1.628(1.592,1.667)$ & $16.509(16.483,16.532)$ & $1.54(1.36,1.74)$ & $187\,(0.04)$ & 12 & 178 & 0.43 & 4 1 3 0 \\
 &        & TYP & $1.628(1.592,1.667)$ & $16.507(16.479,16.530)$ & $1.53(1.33,1.75)$ & $187\,(0.05)$ & 12 & 187 & 0.25 & 4 1 3 0 \\
 &$0.75$  & ENS & $1.640(1.607,1.679)$ & $16.498(16.474,16.519)$ & $1.52(1.37,1.67)$ & $187\,(0.06)$ & 13 & 168 & 0.62 & 4 2 3 0 \\
 &        & TYP & $1.646(1.598,1.679)$ & $16.505(16.479,16.527)$ & $1.55(1.38,1.72)$ & $187\,(0.09)$ & 14 & 168 & 0.59 & 5 2 3 0 \\
 &$1$     & ENS/TYP & $1.646(1.602,1.699)$ & $16.493(16.454,16.525)$ & $1.39(1.17,1.63)$ & $187\,(0.15)$ & 13 & 168 & 0.61 & 4 3 2 0 \\
 &$1.25$  & ENS & $1.623(1.573,1.688)$ & $16.484(16.420,16.530)$ & $1.25(0.90,1.61)$ & $187\,(0.28)$ & 11 & 176 & 0.48 & 4 1 2 0 \\
 &        & TYP & $1.624(1.583,1.673)$ & $16.498(16.452,16.537)$ & $1.35(1.07,1.65)$ & $187\,(0.23)$ & 11 & 171 & 0.59 & 4 1 2 0 \\
 &$1.5$   & ENS & $1.621(1.568,1.677)$ & $16.500(16.441,16.542)$ & $1.63(1.15,2.18)$ & $187\,(0.45)$ & 11 & 183 & 0.35 & 5 1 1 0 \\
 &        & TYP & $1.634(1.596,1.680)$ & $16.509(16.464,16.544)$ & $1.58(1.25,1.94)$ & $187\,(0.28)$ & 11 & 194 & 0.17 & 5 1 1 0 \\
 &$1.75$  & ENS & $1.617(1.533,1.702)$ & $16.484(16.366,16.543)$ & $1.55(0.81,2.44)$ & $187\,(0.68)$ & 10 & 175 & 0.54 & 4 1 1 0 \\
 &        & TYP & $1.615(1.565,1.670)$ & $16.485(16.428,16.525)$ & $1.27(0.96,1.57)$ & $187\,(0.32)$ & 10 & 172 & 0.58 & 3 2 1 0 \\
 &$2$     & ENS & $1.582(1.515,1.632)$ & $16.486(16.355,16.549)$ & $1.77(0.74,3.15)$ & $187\,(0.98)$ & 9 & 169 & 0.68 & 4 0 1 0 \\
 &        & TYP & $1.630(1.585,1.682)$ & $16.502(16.442,16.544)$ & $1.47(1.10,1.86)$ & $187\,(0.34)$ & 10 & 187 & 0.28 & 4 1 1 0 \\
\hline
\hline
\end{tabular}
\label{tab-results0.1}
\end{table*}
%
The details of the fits are listed in Table \ref{tab-results0.1}.\cite{note-fits} %
In Fig.~\ref{fig-CPvsq0.1} we plot the estimates of $W_c$, $\nu$, and $y$, along with their $95\%$ confidence intervals, as functions of $q$.
\begin{SCfigure*}
 \parbox{.33\textwidth}{\includegraphics[width=.33\textwidth]{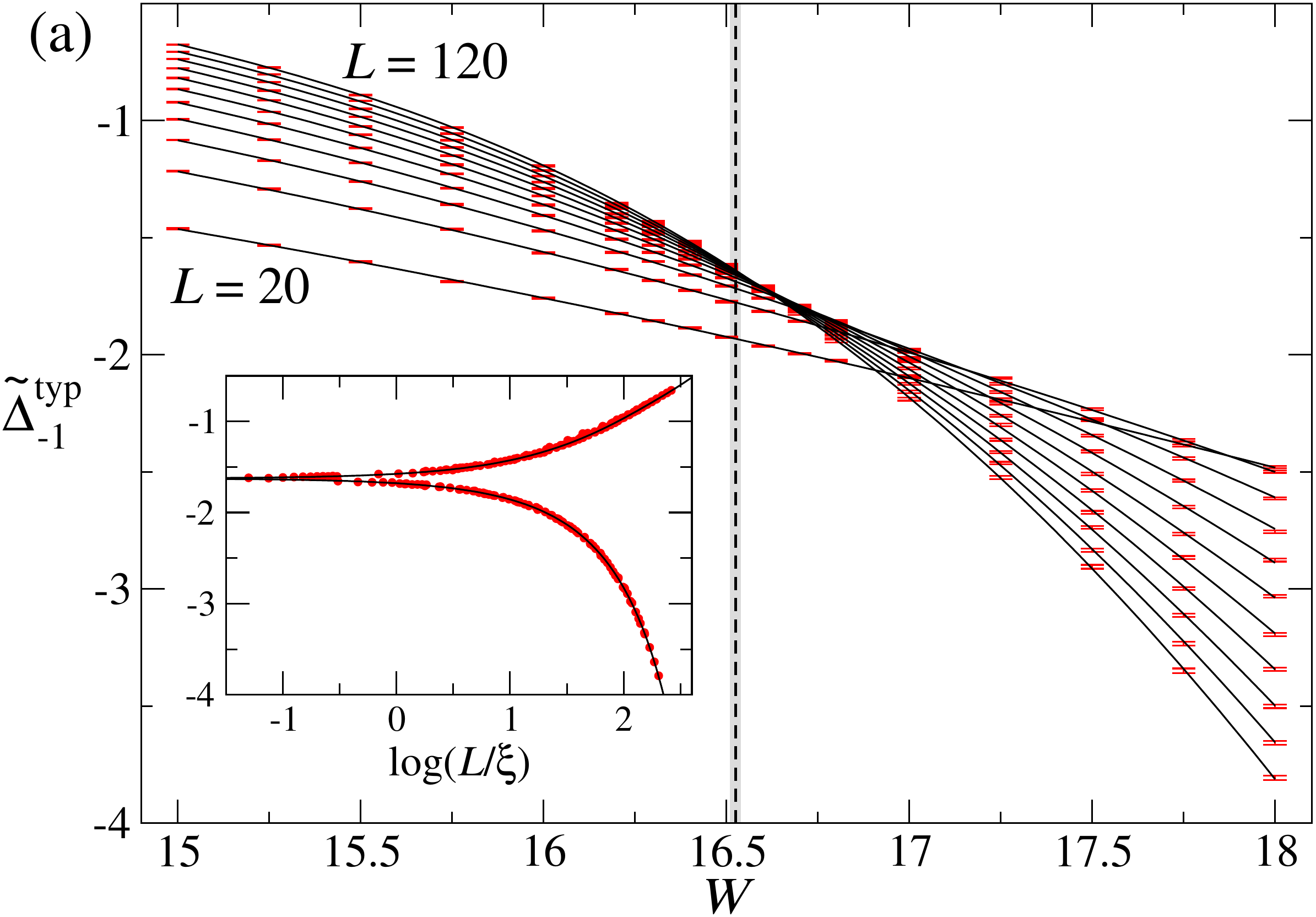}
 \includegraphics[width=.33\textwidth]{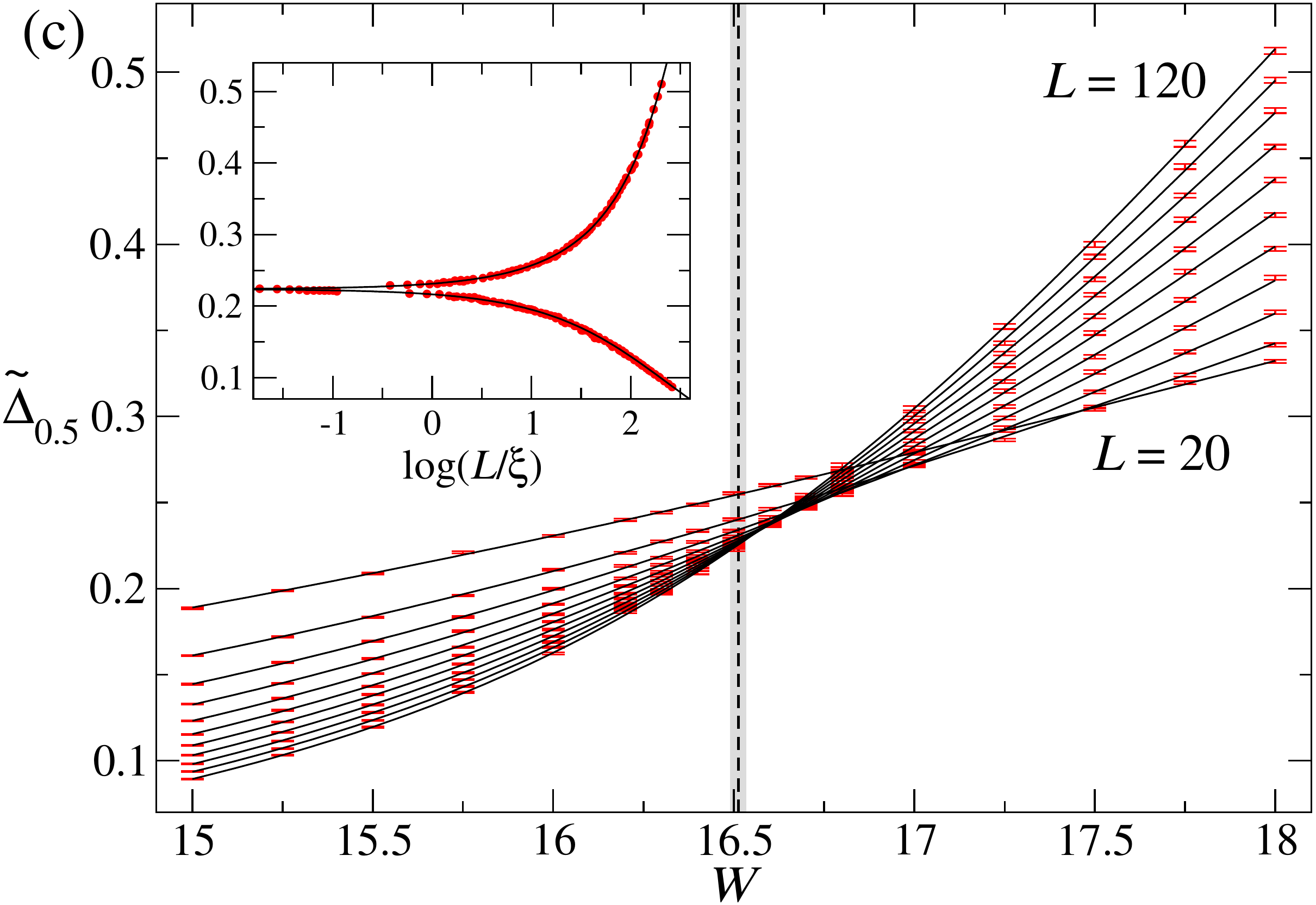}}
 \parbox{.33\textwidth}{
 \includegraphics[width=.33\textwidth]{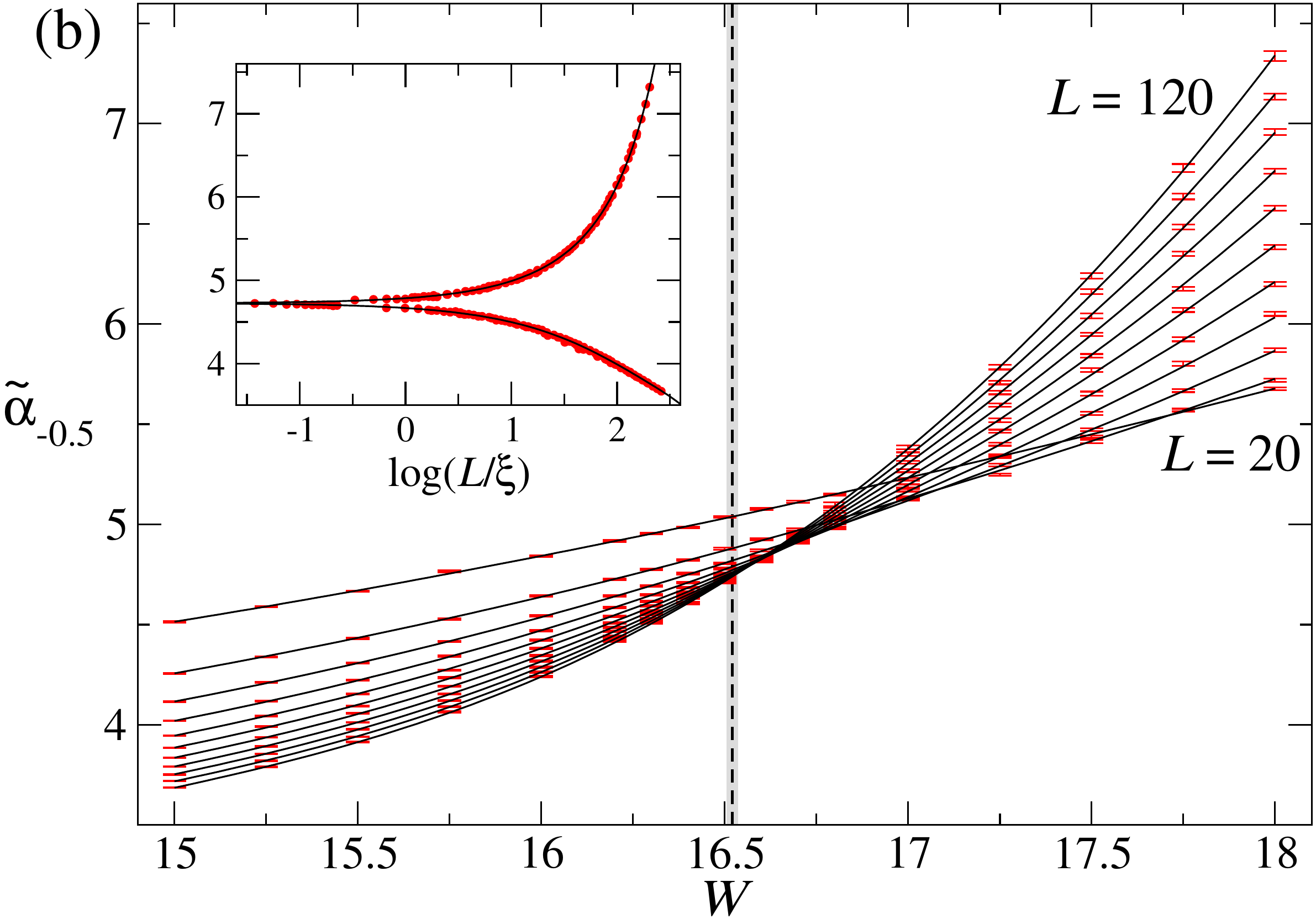}
 \includegraphics[width=.33\textwidth]{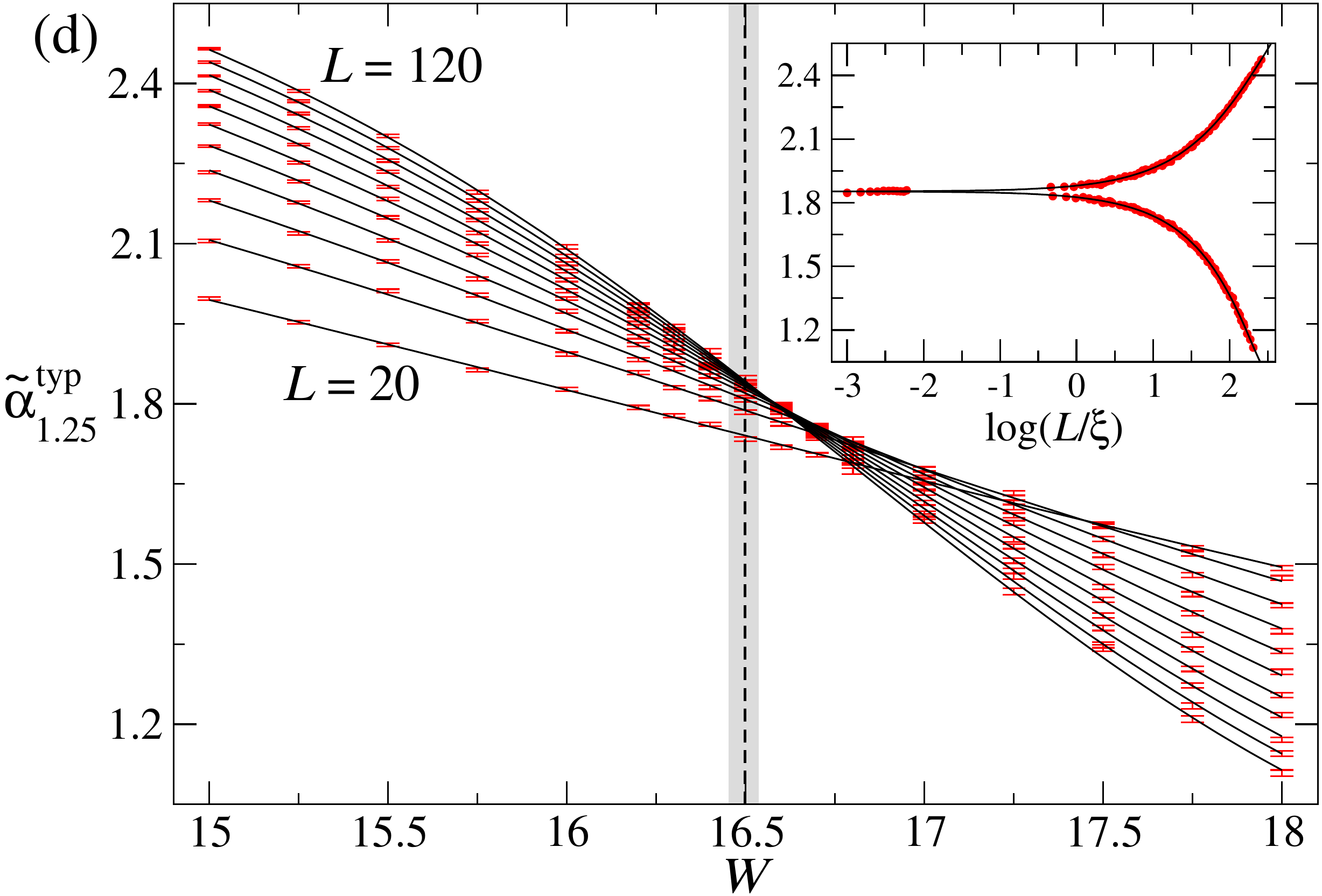}}
 \caption{(color online) Plots of several GMFEs for $\lambda=0.1$ as functions of disorder at various system sizes $L\in[20,120]$.
 The error bars are standard deviations.
 The lines are the best fits listed in Table \ref{tab-results0.1}.
 The estimated $W_c$ are shown by vertical dashed lines and 95\% confidence intervals by the shaded regions.
 The insets show the data plotted vs $L/\xi$ with the irrelevant contribution subtracted and the scaling function (solid line).}
 \label{fig-FSSplots0.1}
\end{SCfigure*}

The estimates of the critical disorder, critical exponent and irrelevant exponent, both for different values of the power $q$ and the type of average considered, are mutually consistent.
The values are also consistent with previous estimates obtained using transfer-matrix methods.\cite{SleO99a,MilRSU00,SleOK00}
The value of the irrelevant exponent is not directly comparable with previous transfer matrix studies since it is not clear that the dominant irrelevant correction should be the same for wavefunction intensity data.

In Fig.~\ref{fig-CPvsq0.1} there is a clear tendency for the error bars to increase for $q>1$. This occurs because the average precision of the data (see the inset in Fig.~\ref{fig-CPvsq0.1})
degrades quickly for $q>1$, especially for the ensemble average.
In general, for the same number of states, the precision is better when typical average is considered.
Also, for $q<1$, the precision of the data for $\widetilde{\alpha}_q$ is significantly better than that for $\widetilde{\Delta}_q$,
although this doesn't translate into smaller uncertainties for the critical parameters.

In Fig.~\ref{fig-FSSplots0.1} we show some examples of the calculated GMFEs together with the corresponding best fits.
The FSS plots can be understood from the phase diagrams for the GMFEs discussed in Section \ref{sec-gmfe} and depicted in Fig.~\ref{fig-MEphases}.
The GMFEs for $W<W_c$ ($W>W_c$) tend towards the metallic (insulating) limit as $L\rightarrow\infty$, which is located above or below the corresponding critical
value depending on $q$.
We recall that, at fixed $\lambda$, the value of the GMFE at the crossing point $W_c$ does not correspond to the scale invariant multifractal exponent, which
is recovered only as $\lambda\rightarrow 0$.

%
\subsection{The range $0<q\leqslant 1/2$}

For $0<q\leqslant 1/2$, the expected critical value of $\widetilde{\alpha}_q$ is not located between the values of the metallic and insulating limits (see Fig.~\ref{fig-MEphases}).
This anomalous behavior makes a reliable FSS analysis in this interval of $q$ very difficult.

\begin{figure}
 \includegraphics[width=.48\columnwidth]{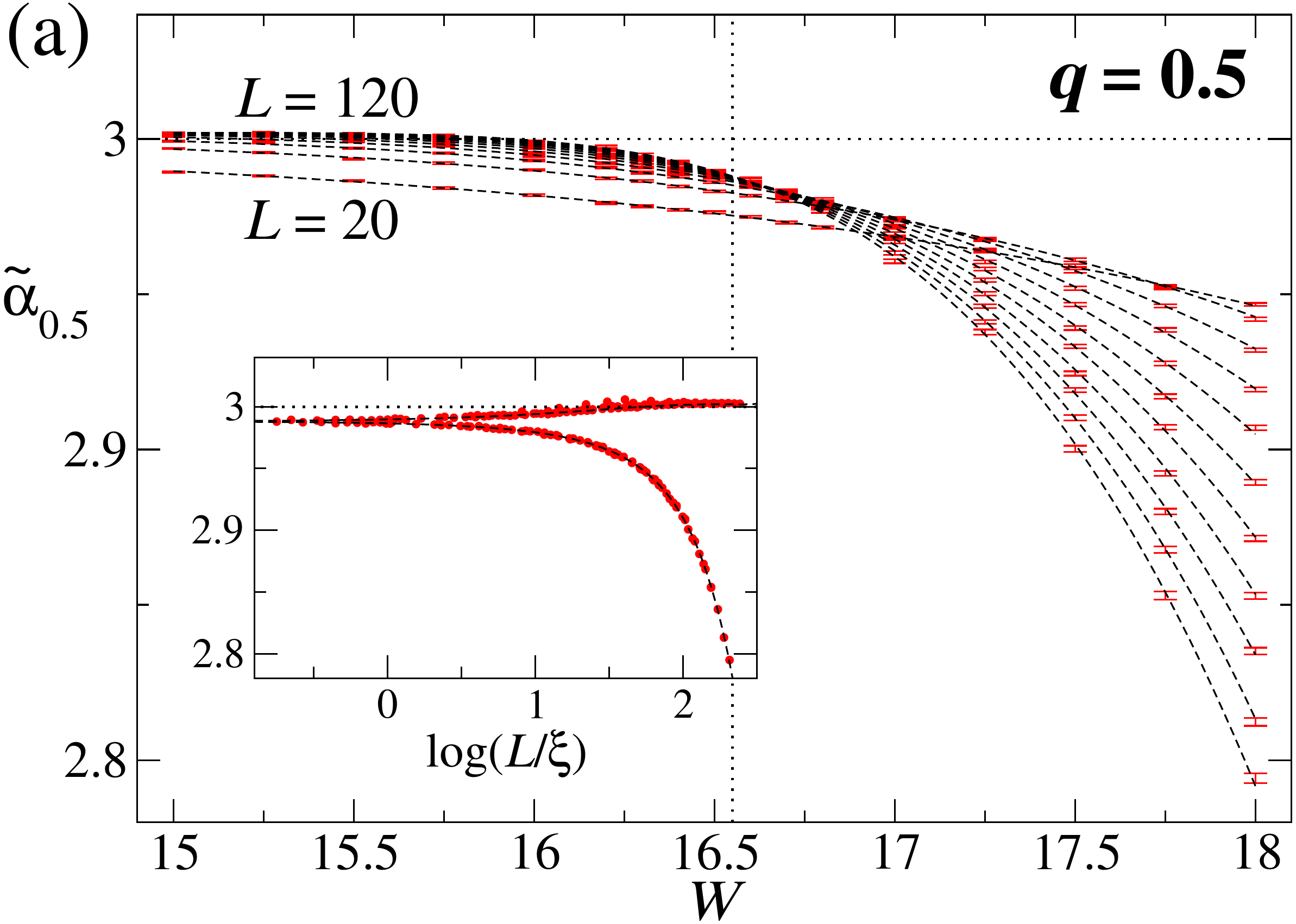}\hfill
 \includegraphics[width=.49\columnwidth]{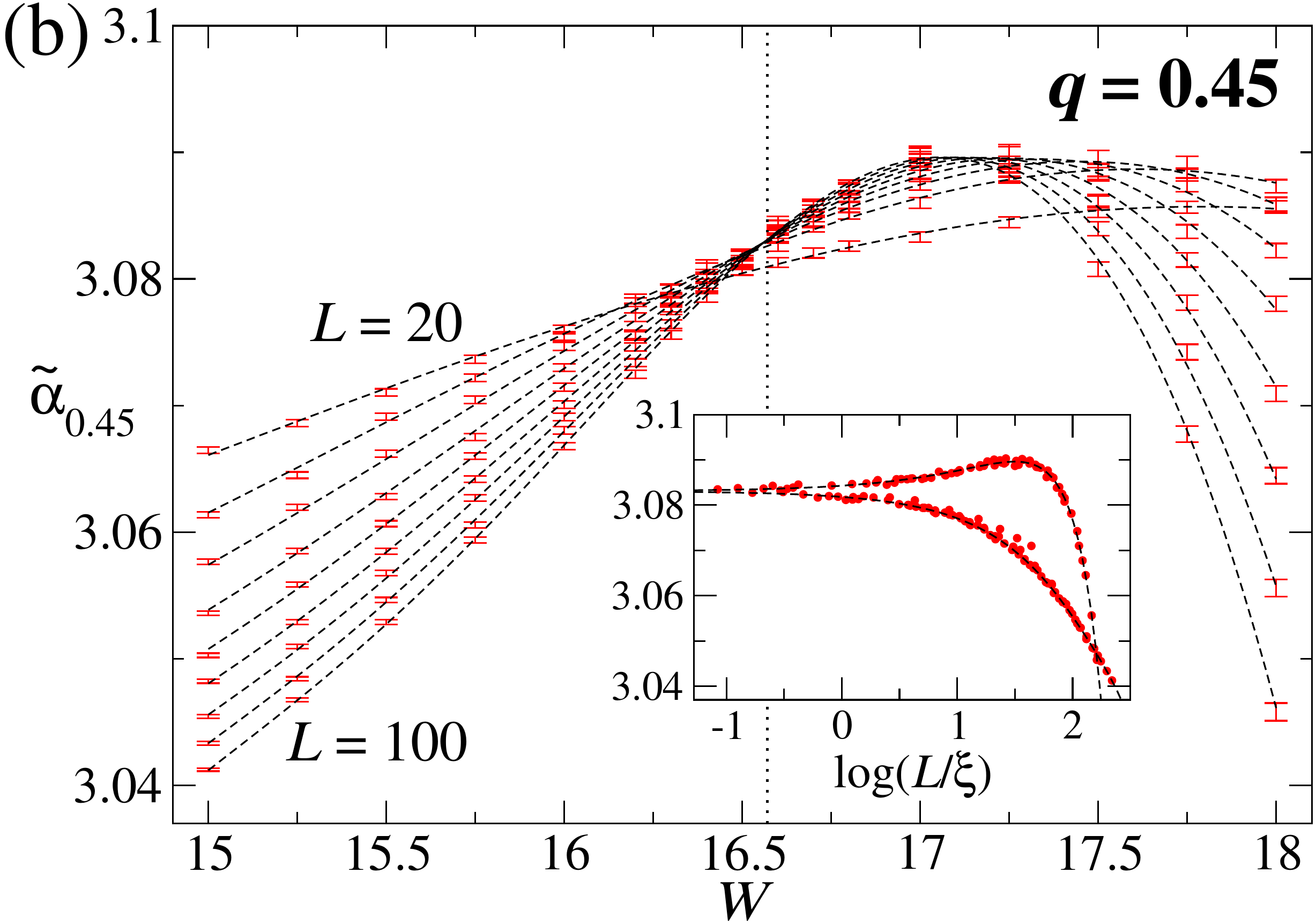}
 \caption{(color online) Behavior of (a) $\widetilde{\alpha}_{0.5}$ and (b) $\widetilde{\alpha}_{0.45}$ for $\lambda=0.1$ as functions of disorder at various system sizes.
 The dashed lines are tentative fits that serve as guides to the eye only.
 The vertical dotted lines indicate the position of $W_c$ according to these tentative fits.
 The horizontal dotted line in (a) indicates the metallic limit $\widetilde{\alpha}_{0.5}=3$.
 The insets display the corrected data and the scaling functions according to the tentative fits.}
 \label{fig-AFITalpha}
\end{figure}

For $q=1/2$ the values in the metallic limit and at the critical point are expected to be the same ($=d$), while the value in the insulating limit is expected to be zero.
As shown in Fig.~\ref{fig-AFITalpha}(a), for $q=1/2$ and $W<W_c$ the data are almost independent of both $L$ and $W$.
While, for $W>W_c$, the dependence on $L$ is very strong, as the data tend to the insulating limit.
This behavior cannot be reliably described by a power series expansion.

For $0<q<1/2$ the value of $\widetilde{\alpha}_q$ at the critical point is now expected to be larger than the values in both the metallic and insulating limits.
Data for $q=0.45$ is shown in Fig.~\ref{fig-AFITalpha}(b).
Very close to the critical point, a standard scaling behavior with opposite $L$ dependence at each side of the transition is visible.
However, for larger disorder, the $L$ dependence is again reversed so as to approach the value in the insulating limit.
We wish to emphasize that this is not a numerical artefact. Rather, analytical calculations for 3D exponentially localized states confirm the behavior.

\section{Multifractal FSS}
\label{sec-3dfss}

We now consider the scaling behavior of the GMFEs as function not only of disorder $W$ and system size $L$ but \emph{in addition} the box-size $l$
using scaling laws \eqref{eq-deltaq} and \eqref{eq-alphaq}.
This permits the \emph{simultaneous} estimation of the multifractal exponents and the critical parameters $\nu$, $W_c$, $y$.
This is a major advance over traditional multifractal analysis, where the position of the critical point must be estimated in a separate FSS
analysis before the multifractal analysis.

\subsection{ Fitting of correlated data}

Different coarse-grainings $l$ for the same disorder $W$ and system size $L$ use the \emph{same} set of eigenstates,
which induces correlations among the estimates of GMFEs for different $l$ and the same $W$ and $L$.
These correlations must be properly taken into account in MFSS.
To do so we generalize the definition of $\chi^2$ in the numerical minimization by including the full covariance matrix for the GMFEs.
We describe the calculation of the covariance matrix and the $\chi^2$-minimization procedure for correlated data in Appendices \ref{app-covmat} and \ref{app-correlatedfit}.

\subsection{ Expansion of scaling functions}

In MFSS, the scaling functions are functions of two independent variables $L/\xi$ and $l/\xi$. While these variables are independent
because the system size and the box size vary independently, they involve the same scaling variable and renormalize in the same way.
In addition, we need to allow for non-linear dependence in $W$ and for irrelevant scaling variables.
We have found that the most important irrelevant contribution is due to the box-size $l$.
Therefore, we use the expansion
\begin{multline}
  \widetilde{\Delta}_q(\varrho L^{1/\nu}, \varrho l^{1/\nu}, \eta l^{y}) = \\ \Delta_q+\frac{1}{\ln (l/L)} \sum_{k=0}^2 \left(\eta l^{y}\right)^k \mathcal{T}^k_q(\varrho L^{1/\nu}, \varrho l^{1/\nu}),
  \label{eq-expand3D}
\end{multline}
and similarly for $\widetilde{\alpha}_q$.
Here, $\varrho$ and $\eta$ are the relevant and irrelevant scaling variables, with $1/\nu$ and $y<0$ the corresponding exponents.
Note that we expand to second order in the irrelevant variable.
We find that this is necessary to fit the data reliably.
The functions $\mathcal{T}_q^k$ are expanded,
\begin{equation}
 \mathcal{T}^k_q(\varrho L^{1/\nu}, \varrho l^{1/\nu})=\sum_{i=0}^{n_L^k} \sum_{j=0}^{n_l^k} a_{kij} \varrho^{i+j} L^{i/\nu} l^{j/\nu},
\end{equation}
for $k=0,1,2$, as are the fields as described in Eq.~\eqref{eq-fieldex}.
The expansion of the scaling function is then characterized by the indices $n^0_L,n^0_l,n^1_L,n^1_l,n^2_L,n^2_l,m_\varrho,m_\eta$.
The number of free parameters is given by
\begin{equation}
N_P=\sum_{k=0}^2 (n_L^k+1)(n_l^k+1) +m_\varrho +m_\eta +3.
\end{equation}
After subtraction of irrelevant corrections we have
\begin{equation}
\widetilde{\Delta}_q^\text{corr}=\Delta_q+ \mathcal{T}^0_q(\pm (L/\xi)^{1/\nu}, \pm (l/\xi)^{1/\nu})/\ln(l/L)
\end{equation}
and the numerical data should fall on a common surface rather than a common curve as in standard FSS.

When evaluated at fixed $\lambda$, Eq.~\eqref{eq-expand3D} leads to the FSS expansions given in
Sec.\ref{sec-fixlambda}.
Since $l=\lambda L$, when performing FSS at fixed $\lambda$ the irrelevant correction is determined by the system size.
%
\subsection{Results}

We have carried out the MFSS analysis on the ensemble averaged GMFEs $\widetilde{\Delta}_q$ for different $q\in[-1,2]$, and $\widetilde{\alpha}_q$ for $q=0,1$.
The estimates of the critical parameters and the multifractal exponents, together with full details of the fits, are included in Table \ref{tab-results3D}.

We have considered different ranges of data for different $q$ trying to maximize the number of points that we could fit in a stable manner,
while keeping the value of $\lambda\leqslant 0.1$.
The minimum values of $\lambda$ occurring in our data sets are $\lambda_{\rm min}=0.017$ ($l_{\rm min}=2$) for $q\leqslant 1/2$ and $\lambda_{\rm min}=0.008$ ($l_{\rm min}=1$) for $q>1/2$. Because of the need for coarse-graining for negative $q$, as discussed in Section \ref{sec-model}, we exclude all $l=1$ data for $q\leqslant 1/2$.

\begin{table*}
\caption{The estimates of the critical parameters and multifractal exponents together with 95\% confidence intervals,
from MFSS of $\widetilde{\Delta}_q$ for $q\in[-1,2]$ and $\widetilde{\alpha}_q$ for $q=0,1$, under ensemble average.
The number of data used is $N_D$ (average percentage precision in parentheses),
the number of free parameters in the fit is $N_P$, $\chi^2$ is the value of the chi-squared statistic for the best fit, and $p$ is the goodness of fit probability. The last column specifies the orders of the expansion: $n^0_L,n^0_l,n^1_L,n^1_l,n^2_L,n^2_l,m_\varrho,m_\eta$. The system sizes considered are $L\in[20, 120]$, the range of disorder is $W\in\left[15,18\right]$, minimum box-size $l_{\rm min}=2$ $(\lambda_{\rm min}=0.017)$ for $q\leqslant1/2$ and $l_{\rm min}=1$ $(\lambda_{\rm min}=0.008)$ for $q>1/2$. The maximum values considered for $\lambda$ change from $\lambda_{\rm max}=0.063$ to $\lambda_{\rm max}=0.1$ for different $q$.%
}
\begin{tabular}{ccccccccccc}
\hline\hline
    $q$ & $\Delta_q$ ($\alpha_q$ for $q=0,1$) & $\nu$  & $W_c$ & $-y$ & $N_D$(prec.) & $N_P$ & $\chi ^2$ & $p$ & Expansion \\
\hline
 $-1$ & $-1.844(-1.854,-1.832)$ & $1.598(1.584,1.612)$ & $16.526(16.516,16.535)$ & $1.76(1.68,1.83)$ & $680\,(0.27)$ & 25 & 667 & 0.36 & 3\,2\,1\,1\,1\,1\,2\,0 \\
 $-0.75$ & $-1.252(-1.256,-1.247)$ & $1.592(1.580,1.603)$ & $16.526(16.520,16.533)$ & $1.77(1.72,1.82)$ & $680\,(0.23)$ & 25 & 667 & 0.36 & 3\,2\,1\,1\,1\,1\,2\,0 \\
 $-0.5$ & $-0.740(-0.742,-0.738)$ & $1.591(1.579,1.602)$ & $16.528(16.522,16.534)$ & $1.78(1.75,1.82)$ & $493\,(0.20)$ & 25 & 460 & 0.59 & 3\,2\,1\,1\,1\,1\,2\,0 \\
 $-0.25$ & $-0.318(-0.319,-0.317)$ & $1.594(1.583,1.606)$ & $16.527(16.520,16.534)$ & $1.77(1.72,1.81)$ & $425\,(0.19)$ & 26 & 379 & 0.76 & 4\,2\,1\,1\,0\,1\,2\,0 \\
 $0$ & $4.048(4.045,4.050)$ & $1.590(1.579,1.602)$ & $16.530(16.524,16.536)$ & $1.81(1.79,1.84)$ & $493\,(0.05)$ & 27 & 473 & 0.40 & 3\,2\,2\,1\,1\,1\,2\,0 \\
 $0.25$ & $0.1997(0.1988,0.2005)$ & $1.593(1.580,1.607)$ & $16.529(16.521,16.536)$ & $1.78(1.72,1.83)$ & $425\,(0.21)$ & 27 & 429 & 0.14 & 4\,2\,1\,1\,0\,1\,3\,0 \\
 $0.5$ & $0.2683(0.2672,0.2693)$ & $1.595(1.579,1.612)$ & $16.529(16.522,16.537)$ & $1.80(1.74,1.84)$ & $493\,(0.23)$ & 29 & 500 & 0.12 & 4\,2\,2\,1\,0\,1\,3\,0 \\
 $0.75$ & $0.1993(0.1982,0.2004)$ & $1.600(1.583,1.617)$ & $16.524(16.514,16.535)$ & $1.70(1.66,1.75)$ & $544\,(0.20)$ & 27 & 530 & 0.34 & 4\,2\,1\,1\,0\,1\,3\,0 \\
 $1$ & $1.958(1.953,1.963)$ & $1.603(1.583,1.623)$ & $16.528(16.516,16.538)$ & $1.70(1.65,1.74)$ & $612\,(0.12)$ & 27 & 597 & 0.35 & 4\,2\,1\,1\,0\,1\,3\,0 \\
 $1.25$ & $-0.317(-0.320,-0.313)$ & $1.598(1.573,1.626)$ & $16.536(16.512,16.559)$ & $1.71(1.60,1.83)$ & $544\,(0.25)$ & 23 & 515 & 0.57 & 4\,1\,1\,1\,1\,1\,2\,0 \\
 $1.5$ & $-0.730(-0.739,-0.719)$ & $1.583(1.544,1.624)$ & $16.536(16.502,16.567)$ & $1.67(1.52,1.83)$ & $544\,(0.29)$ & 23 & 531 & 0.37 & 4\,1\,1\,1\,1\,1\,2\,0 \\
 $1.75$ & $-1.217(-1.238,-1.197)$ & $1.583(1.524,1.639)$ & $16.528(16.484,16.573)$ & $1.64(1.43,1.91)$ & $544\,(0.34)$ & 23 & 549 & 0.20 & 3\,1\,2\,1\,1\,1\,2\,0 \\
 $2$ & $-1.763(-1.792,-1.727)$ & $1.622(1.555,1.691)$ & $16.529(16.468,16.582)$ & $1.62(1.34,1.95)$ & $544\,(0.41)$ & 19 & 566 & 0.11 & 3\,1\,1\,1\,0\,1\,2\,0 \\
\hline
\hline
\end{tabular}
\label{tab-results3D}
\end{table*}
\begin{figure*}[tb]
 \includegraphics[height=6.2cm]{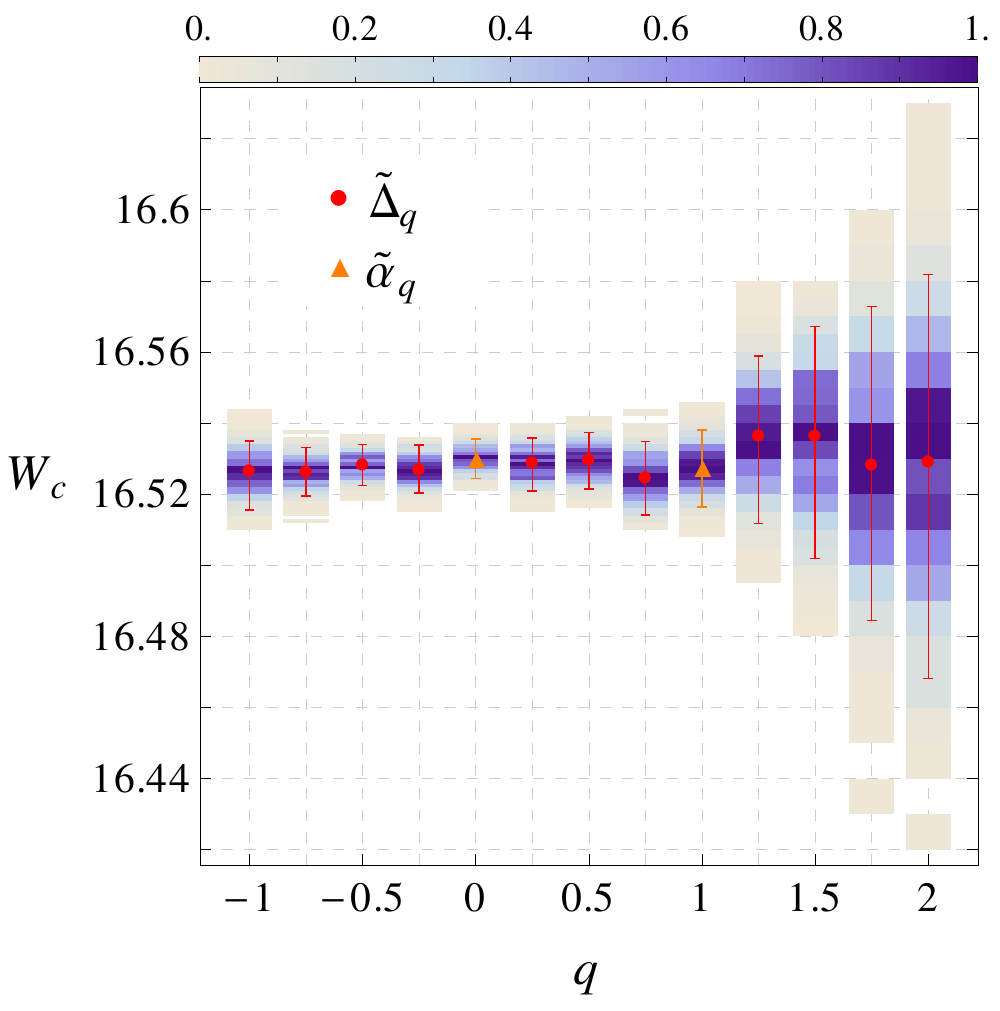}\hfill
 \includegraphics[height=6.2cm]{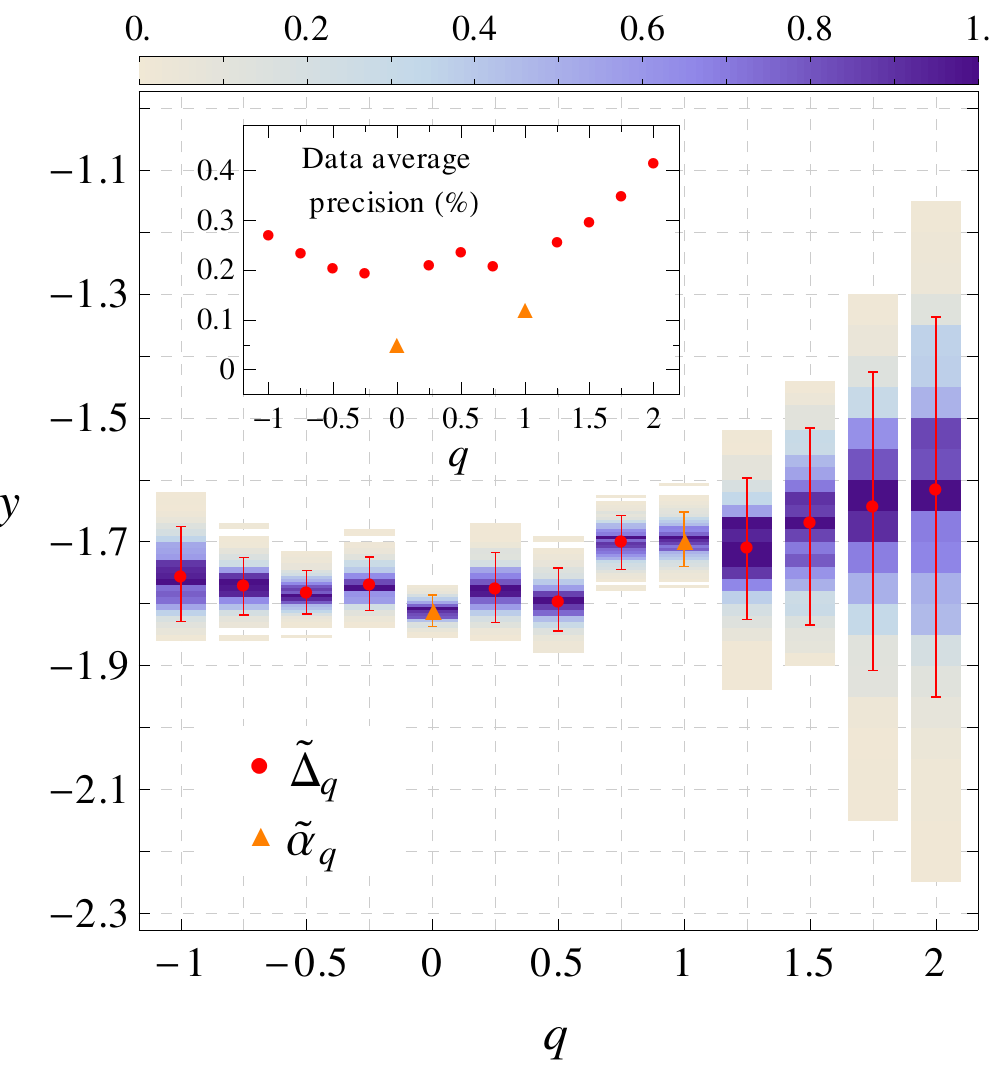}\hfill
 \includegraphics[height=6.2cm]{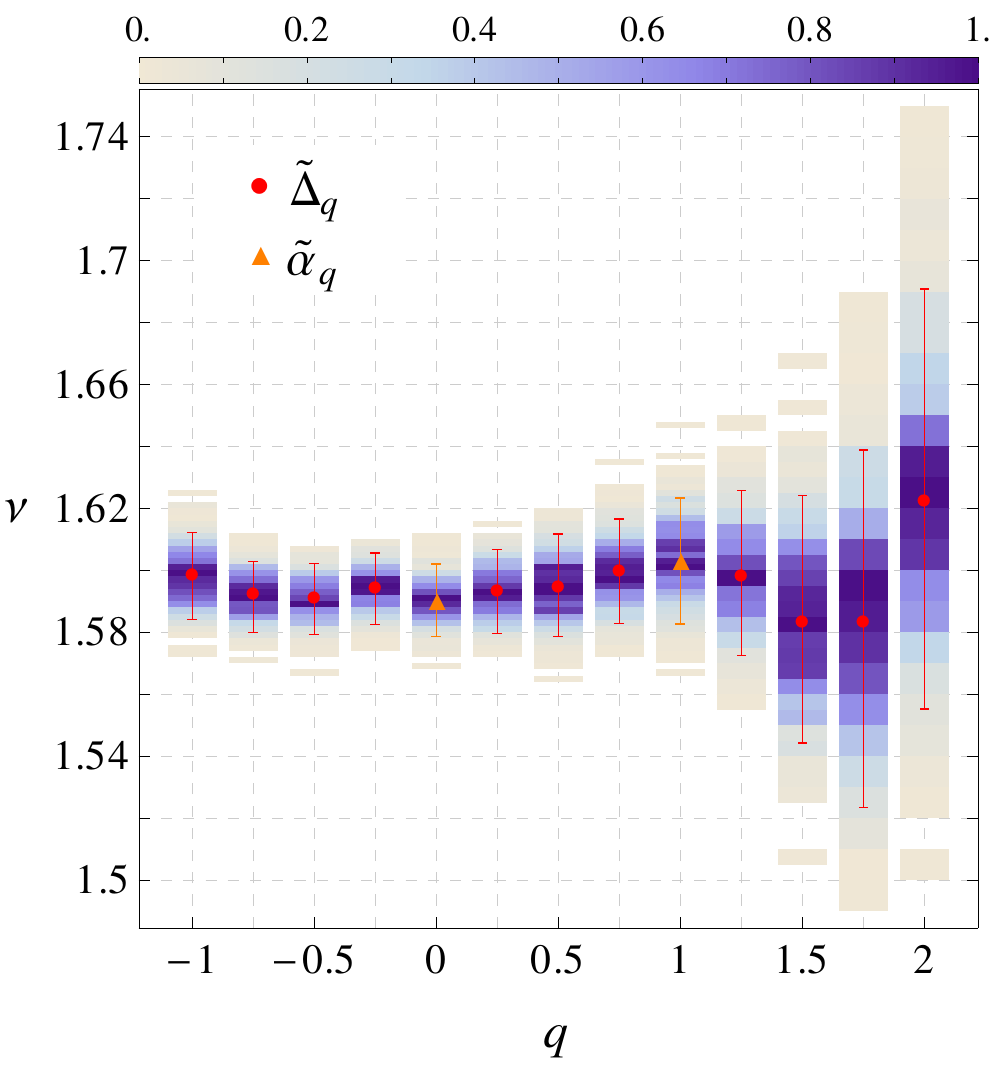}
 \caption{(color online) The estimates of the critical parameters $\nu$, $W_c$ and $y$, as functions of $q$, obtained from MFSS for $\widetilde{\Delta}_q$ and $\widetilde{\alpha}_q$ (only $q=0,1$). Error bars are 95\% confidence intervals. The corresponding values are listed in Table \ref{tab-results3D}. The inset in the centre plot shows the average data precision versus $q$ for the data set used. A density plot of the histograms obtained from the Monte Carlo simulations used to determine the uncertainty of the estimates is shown for each $q$. The color scale on top of each graph is for the density plot. The histograms are normalized so that their maximum value is unity.}
 \label{fig-CPvsq3D}
\end{figure*}
The best-fit estimates for $W_c$, $y$ and $\nu$ as functions of $q$ are shown in Fig.~\ref{fig-CPvsq3D}.
The consistency of the estimates of the critical parameters for different $q$ is remarkable.
Compared with the FSS at fixed $\lambda$ of Section \ref{sec-fixlambda}, the estimate of the irrelevant exponent $y$ is more stable.
As $q$ increases, the uncertainty for the estimates of $W_c$, $\nu$ and $y$ grow.
This is partly due to the loss of precision in the data for high $q$ (see the inset in Fig.~\ref{fig-CPvsq3D}),
but this does not fully explain the large difference in the uncertainty between sets with similar precisions, for example $q=-1$ and $q=1.5$.
We believe that the difference is caused by the amplitude of the irrelevant component in the data,
which is larger for $q<1/2$.
(We have confirmed this by examining the $q$ dependence of the coefficient $a_{100}$ in the expansion \eqref{eq-expand3D}.)
If the amplitude of the irrelevant shift in the data is small, the estimation of the irrelevant exponent becomes very ambiguous.
This in turn leads to an increase in the uncertainties of all the other parameters.
The best precision is achieved for $\widetilde{\alpha}_0$ ($0.05\%$). From the MFSS analysis of this GMFE
we find
\begin{equation}
W_c=16.530\; (16.524,16.536)
\end{equation}
and
\begin{equation}
\nu=1.590\; (1.579,1.602)
\end{equation}
where the error limits correspond to $95\%$ confidence intervals.

\subsection{Scaling surfaces}

\begin{figure*}[tb]
 \includegraphics[width=.45\textwidth]{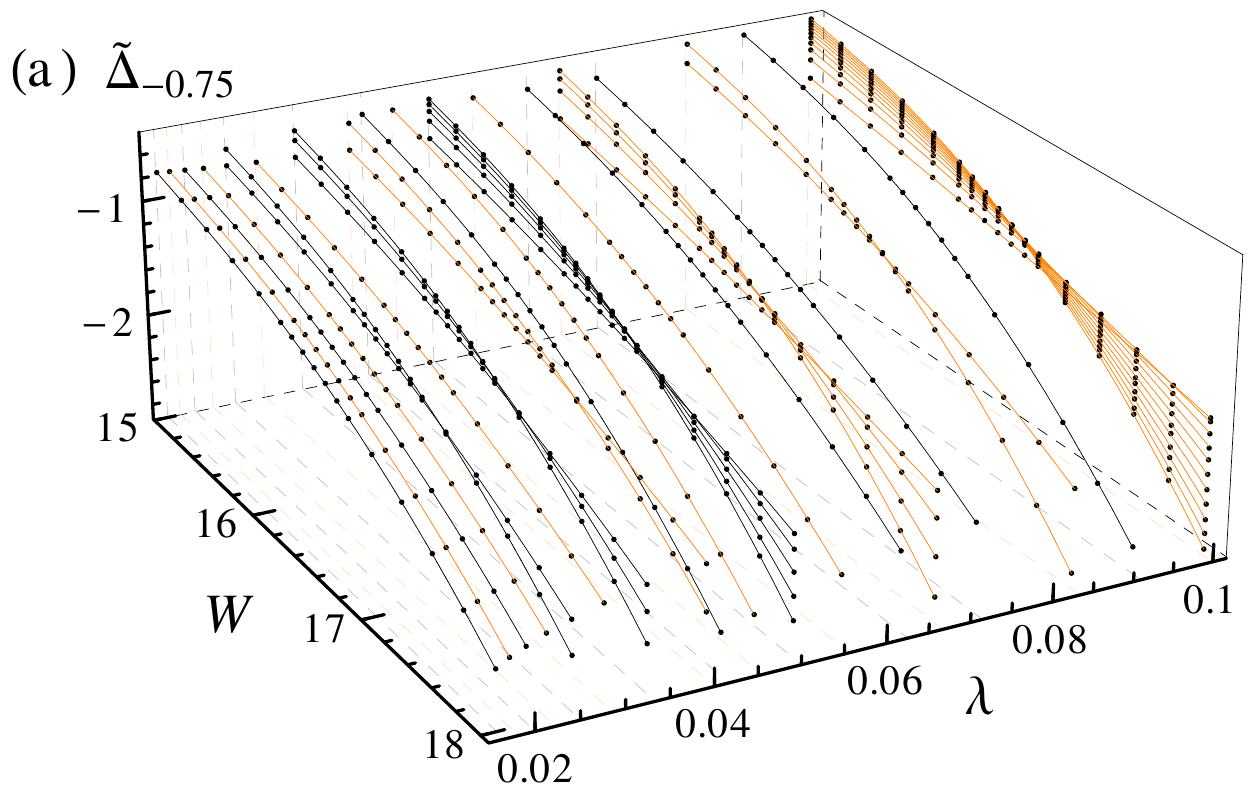}\qquad
 \includegraphics[width=.45\textwidth]{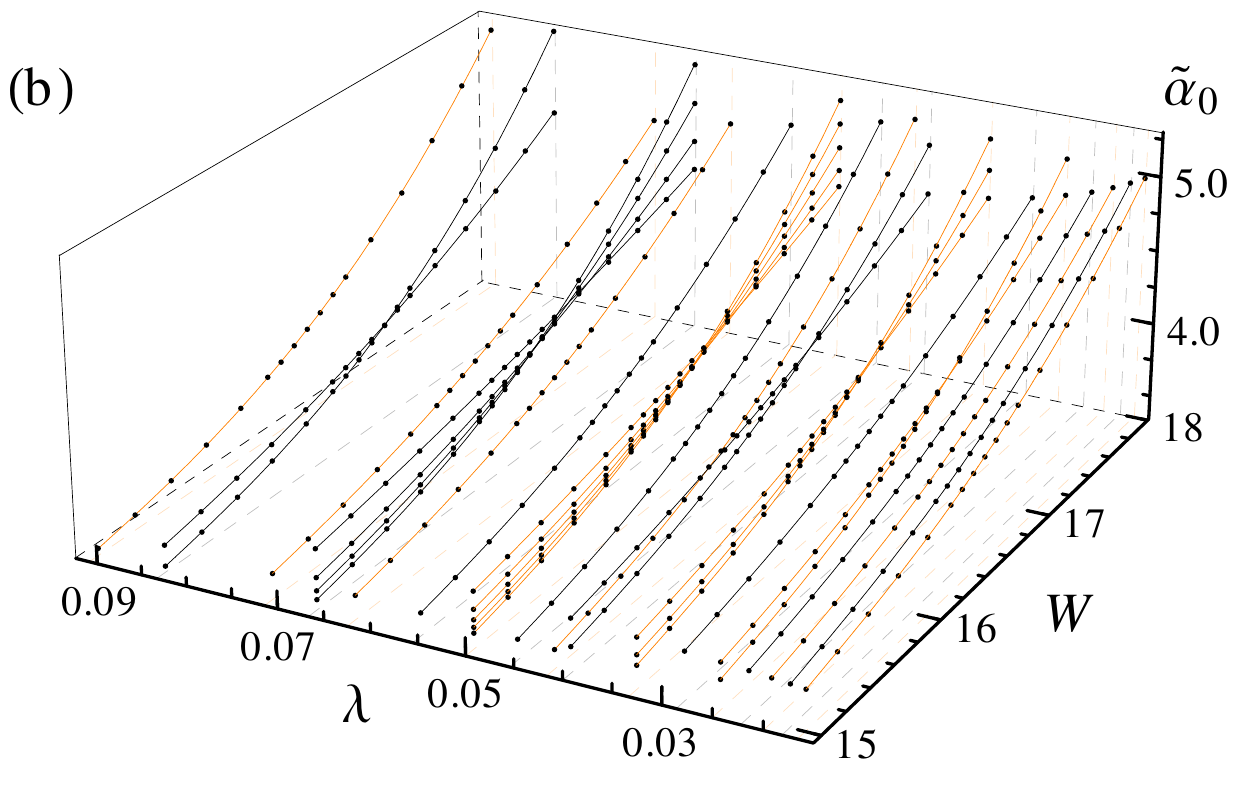}
 \includegraphics[width=.46\textwidth]{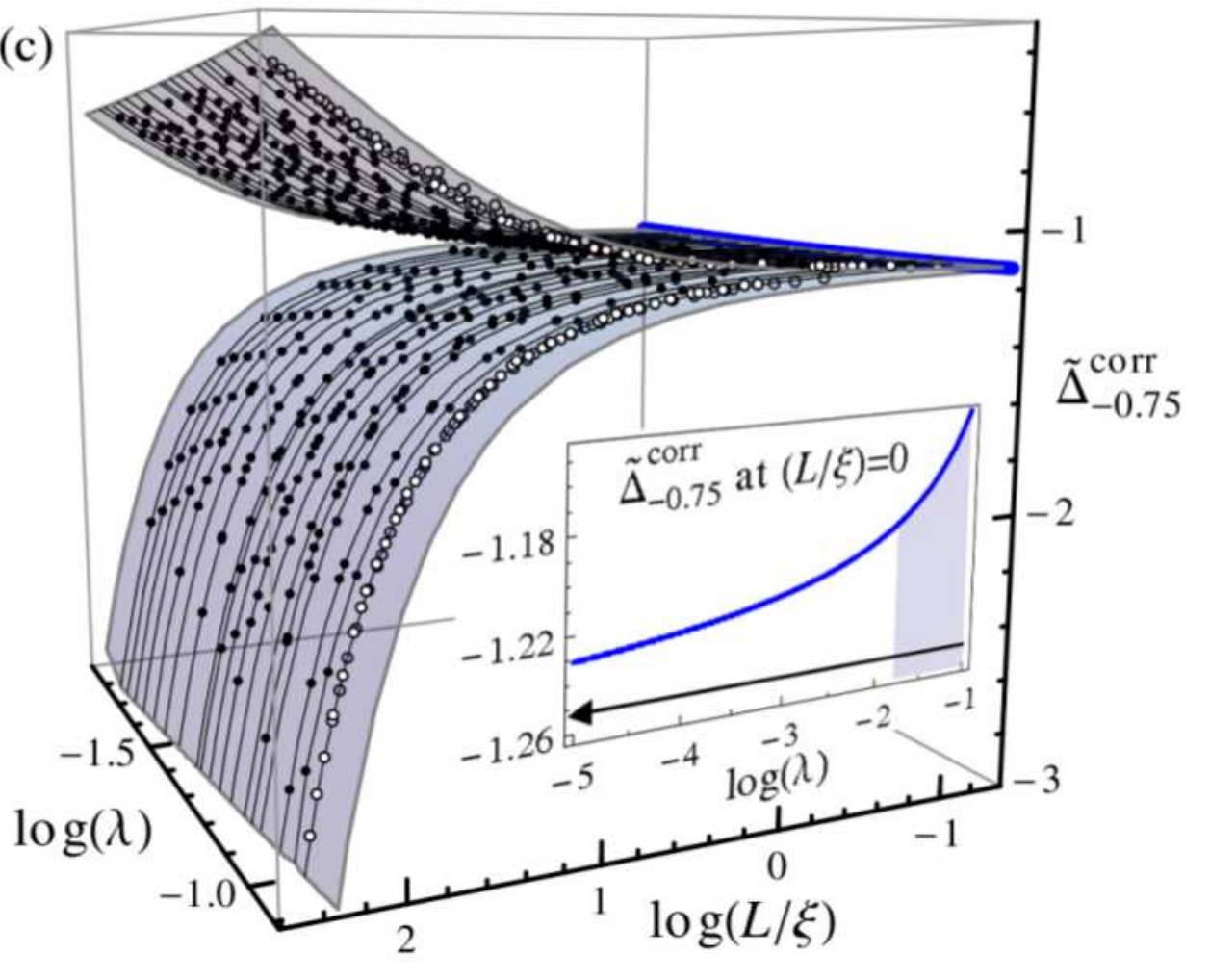}\qquad
 \includegraphics[width=.45\textwidth]{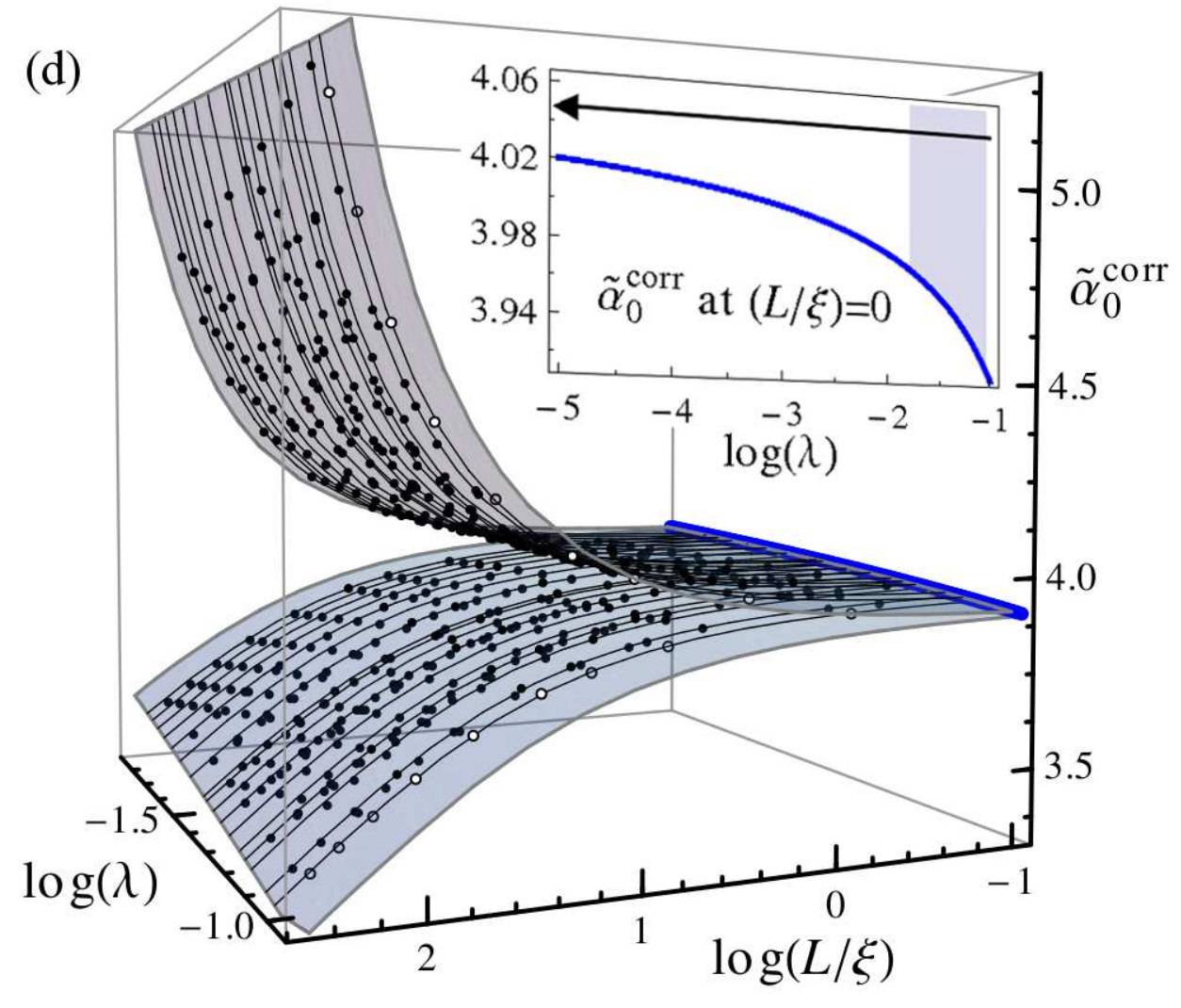}
 \caption{(color online) MFSS of $\widetilde{\Delta}_{-0.75}$ (a,c) and $\widetilde{\alpha}_0$ (b,d). (Upper plots) GMFEs (\textbullet) as functions of disorder $W$, for different $\lambda=l/L$. The solid lines are cross-sections at fixed $\lambda$ of the best fit, plotted for different $L$. Note that all points are fitted simultaneously.
 Alternating colors have been used for better visualization.
 (Lower plots) GMFEs with irrelevant contribution subtracted (\textbullet, $\circ$) and the scaling surfaces
 (symbol $\circ$ highlights the maximum value of $\lambda$).
 The insets are the scaling functions at the critical point, highlighted also in the right face of the main plot.
 The arrows indicate the multifractal exponents given by the extrapolation $\lambda \rightarrow 0$.
 The shaded regions indicate the range of $\lambda$ accessed in our simulations.
 }
 \label{fig-3Dfss}
\end{figure*}

In Fig.~\ref{fig-3Dfss} we show the best fits and the corresponding scaling surfaces for $\widetilde{\Delta}_{-0.75}$ and $\widetilde{\alpha}_0$.
The upper plots display the GMFEs and cross-sections of the global fit displayed at the different $\lambda$ values occurring in the data set.
The bottom plots show the scaling functions in terms of the two variables $L/\xi$ and $\lambda$.
(The visualization of the scaling functions is improved when $\lambda$ is chosen instead of $l/\xi$ as the second variable.)
The scale invariant multifractal exponents correspond to the asymptotic value at the critical point as $\lambda\rightarrow 0$.
This is highlighted in the insets of Fig.~\ref{fig-3Dfss}(c,d), where the behavior of the scaling function at criticality --- when the sheets of extended and localized phases meet --- is shown versus $\log(\lambda)$.

\begin{figure}[tb]
 \includegraphics[width=.95\columnwidth]{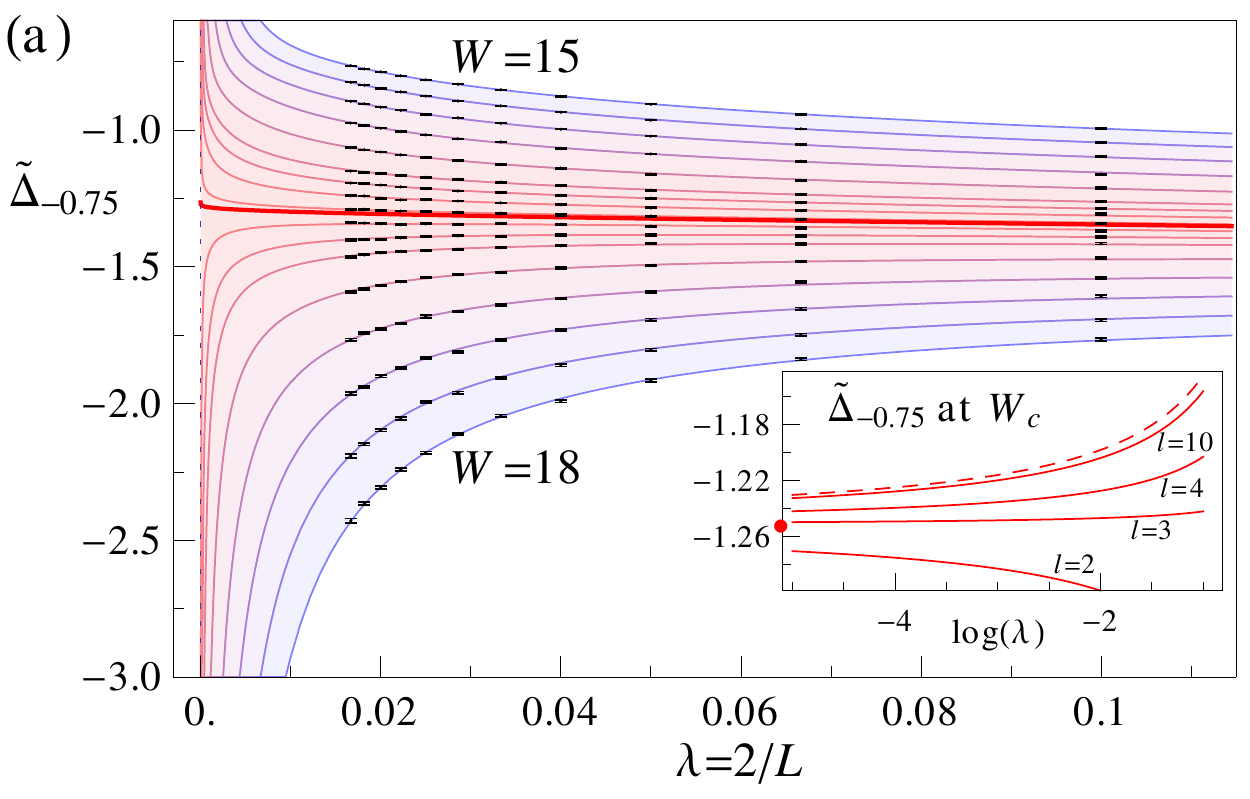}
 \includegraphics[width=.95\columnwidth]{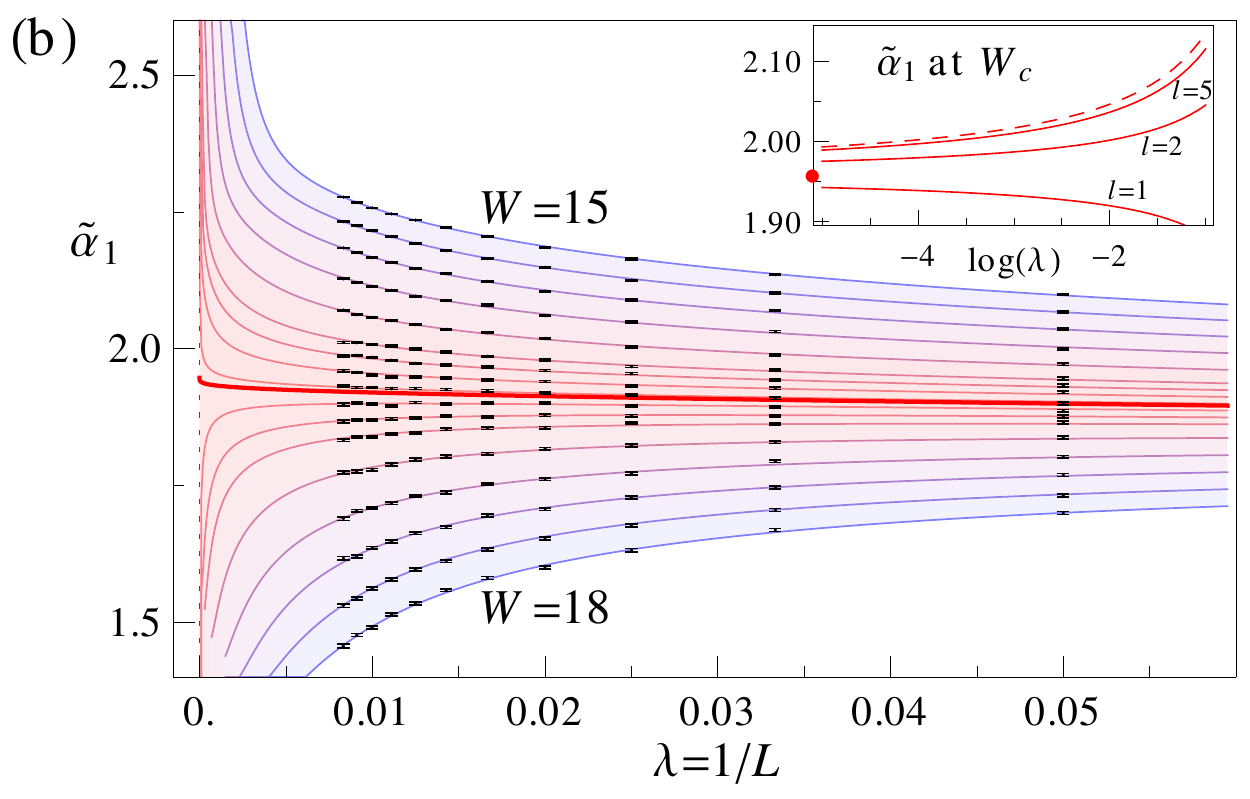}
 \caption{(color online) Plot of (a) $\widetilde{\Delta}_{-0.75}$  and (b) $\widetilde{\alpha}_1$ for fixed box-size $l$ as a function of $\lambda$ and increasing disorder values $W\in[15,18]$ from top to bottom. The lines are cross-sections at (a) $l=2$ and (b) $l=1$ of the global multifractal FSS fit. The thick lines (red) correspond to the GMFE at the critical point $W_c$ [Eq.~\eqref{eq-scalingatWc}]. The insets
 show the latter function for different values of $l$. The dashed line in the inset displays the GMFE at criticality with vanishing irrelevant corrections (effectively $l=\infty$). The circle on the inset $y$-axis marks the asymptotic ($\lambda\rightarrow 0$) multifractal exponent.}
 \label{fig-3Dfixb}
\end{figure}

The transition can also be visualized as shown in Fig.~\ref{fig-3Dfixb}
where we display a cross-section at fixed box-size of the data and the fit for $\widetilde{\Delta}_{-0.75}$ and $\widetilde{\alpha}_1$.
This provides an alternative way to monitor the Anderson transition.\cite{WafPL99,AniKPF07,KraS10}
We see the flow with increasing $L$ (decreasing $\lambda$) towards the metallic or insulating limits depending on the disorder
$W$, and at the critical point $W_c$ the convergence to the multifractal exponent.
At the critical point, the scaling law \eqref{eq-expand3D} reduces to
\begin{equation}
 \widetilde{\Delta}_q(W_c)=\Delta_q+\frac{1}{\ln\lambda}\sum_{k=0}^2l^{ky} a_{k00},
 \label{eq-scalingatWc}
\end{equation}
From this we find that this convergence differs for different $l$.
This is illustrated in the insets of Fig.~\ref{fig-3Dfixb}.

\subsection{The multifractal spectrum}

\begin{figure}[tb]
 \includegraphics[width=.85\columnwidth]{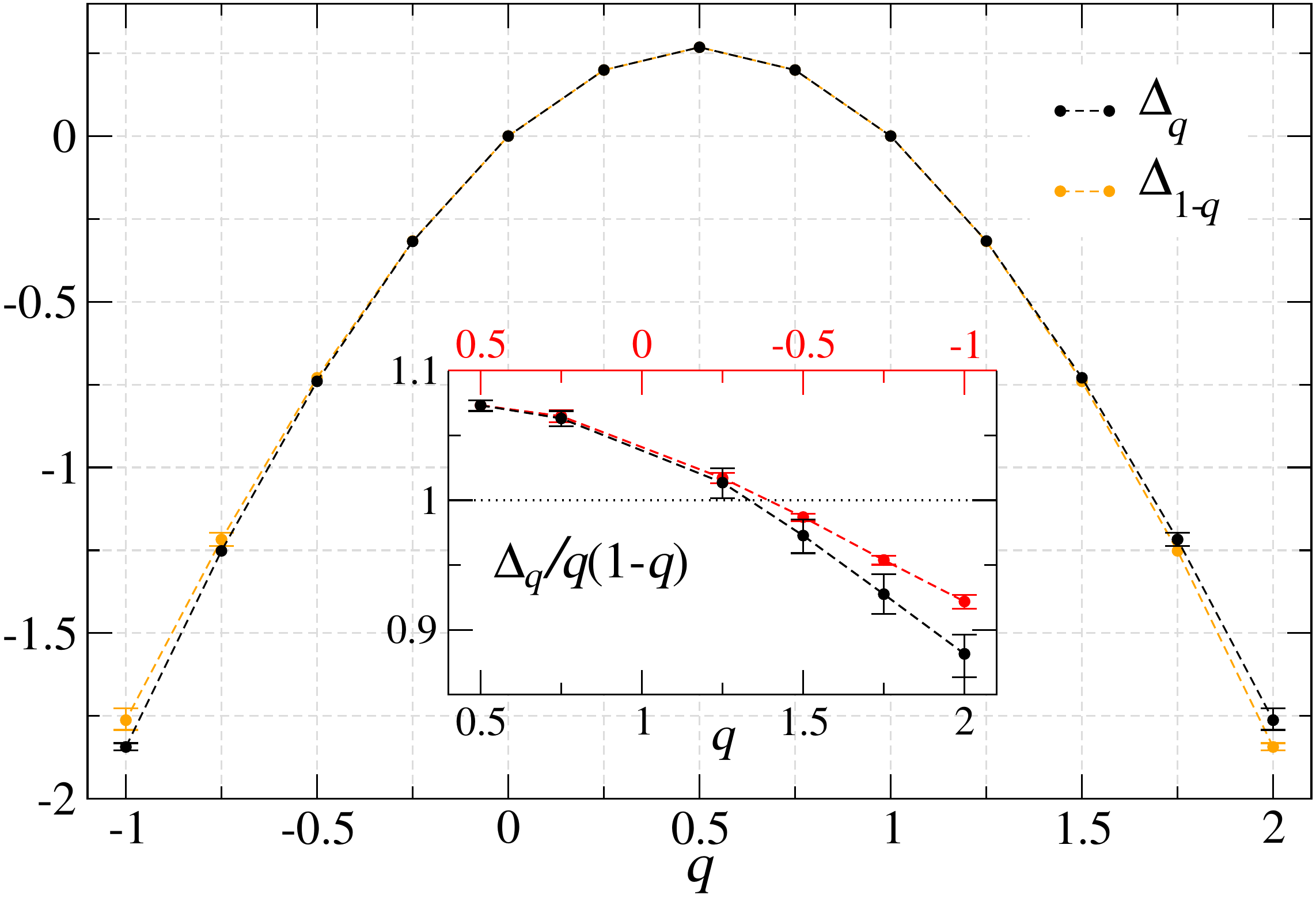}
 \caption{(color online) Multifractal exponents $\Delta_q$ obtained from MFSS. The numerical values are listed in Table \ref{tab-results3D}. Error bars denote 95\% confidence intervals, and are contained within symbol
 size whenever not shown. Note that $\Delta_0=\Delta_1=0$ by definition. The inset shows the reduced multifractal exponents $\Delta_q/q(1-q)$. The horizontal dotted line corresponds to Wegner's parabolic approximation.\cite{Weg89}}
 \label{fig-symmetry}
\end{figure}

Previous work by some of the authors, using standard multifractal analysis at $W=16.5\approx W_{c}$, suggested that the symmetry relation \eqref{deltaq_symrel} for $\Delta_q$ holds for the 3D Anderson model.\cite{VasRR08,RodVR08,RodVR09}
Our improved analysis given here confirms this at $W_{c}$.
As shown in Fig.~\ref{fig-symmetry}, the exponents $\Delta_q$ satisfy the symmetry relation for the range of $q$ values considered.
The symmetry relation \eqref{alphaq_symrel} for $\alpha_q$, is also satisfied by our estimates $\alpha_0=4.048 (4.045, 4.050)$ and $\alpha_1=1.958  (1.953, 1.963)$.
A more careful analysis in terms of the reduced exponents $\Delta_q/q(1-q)$ reveals a slight violation of the symmetry beyond $q=1.75$.
We suspect that this reflects minor limitations in our $\lambda\rightarrow 0$ extrapolation,
rather than a genuine violation of the symmetry relation.
(Particulary for $q<0$ where the range of $\lambda$ is more restricted because of the absence of data for $l=1$.)
We note that the validity of the symmetry relation for the whole $q$-range depends on the absence of termination points for the ensemble average
multifractal spectrum,\cite{EveM08} which is still an open question for the 3D Anderson transition.\cite{RodVR09}

Our data in Fig.~\ref{fig-symmetry} (inset) clearly show that the multifractal spectrum is not parabolic, i.e.~that the reduced exponents depend on $q$.
This is in agreement with previous results at the 3D Anderson transition, where non-parabolicity has been directly observed in $f(\alpha)$,\cite{RodVR08} and in the non-Gaussian nature of the PDF, $\mathcal{P}(\alpha)$.\cite{RodVR09,HuaW10}
A non-parabolic multifractal spectrum implies that the distribution of wavefunction intensities at the transition, and consequently also of the LDOS, is not a log-normal distribution, in contrast to recent claims.\cite{SchSBF10}

\section{Conclusions}
\label{sec-conclusions}

We have shown how to exploit the persistence of multifractal fluctuations away from the critical point on scales below the correlation length to perform a multifractal finite size scaling analysis.
We demonstrated the potential of this approach by applying it to the Anderson localization-delocalization transition in the 3D
Anderson model.
We validated our proposed scaling laws for the generalized multifractal exponents and we estimated the critical disorder at the band center $W_c=16.530(16.524,16.536)$
and the critical exponent describing the divergence of the localisation length $\nu=1.590(1.579,1.602)$.
A remarkable consistency of these estimated critical parameters for different averages (ensemble and typical) and different generalized multifractal exponents (different $q$) was seen.
We also found that the multifractal spectrum exhibits the predicted symmetry and confirmed its non-parabolic nature.

Recently, the value of $\nu$ for the Anderson transition has been estimated for a variety of systems, including disordered phonons
\cite{PinR11}, electrons in non-conventional lattices \cite{EilFR08} or with topological disorder, \cite{KriA11} a disordered Lenard-Jones fluid, \cite{HuaW09} and cold-atom quantum kicked-rotors, \cite{ChaLGD08,LemCSG09} finding agreement with our results.
The universality of the critical exponent for the Anderson transition is, therefore, well established and
a value for $\nu\sim 1.6$ is widely accepted by the community.

In our approach, scaling versus system size yields the critical exponent, while scaling versus box size yields the multifractal exponents. Thus, these exponents
 do not appear to be intrinsically related.
Nevertheless, such a connection has been suggested by Kramer \cite{Kra93} and Janssen,\cite{Jan94a} who discussed a lower bound for $\nu$ in terms of the multifractal exponents, namely
\begin{equation}
\nu> 2/\tau_2.
\end{equation}
From our estimate of $\Delta_2$, the confidence interval for $2/\tau_2$ is $(1.571,1.656)$. Remarkably, all our estimates for $\nu$ fall within this interval.
Although this bound has not been widely discussed in the literature, we find the agreement intriguing.

Multifractal fluctuations have been observed in the local density of states near the metal-insulator transition in semiconductors\cite{MorKMG02,HasSWI08,RicRMZ10} and in the
intensity fluctuations of ultrasound near the Anderson transition in a random elastic network.\cite{HuSPS08,FaeSPL09}
We believe that multifractal finite size scaling as demonstrated here has potential in the quantitative analysis of this sort of experimental data.\cite{KraCWC10,MilKRR10,BilJZB08,RoaDFF08,ChaLGD08,LemCSG09,LemLDS10}

\begin{acknowledgments}%
 The authors gratefully acknowledge EPSRC (EP/F32323/1, EP/C007042/1, EP/D065135/1) for financial support. A.R.\ acknowledges financial support from the Spanish government (FIS2009-07880).
\end{acknowledgments}

\appendix

\section{Fitting procedure}
\label{app-fss}

\subsection{Table of fits.}

The first step is to generate a table of fits by varying the expansion indices of the model over a reasonable range of values.
Since the number of combinations of the indices increases exponentially, we impose sensible restrictions.
Let us consider, for example, the single-parameter FSS of Section \ref{sec-fixlambda}, where the expansion \eqref{eq-Gexpansion} is characterized by the indices $n_0,\,n_1,\,m_\varrho,\,m_\eta$.
We expect the irrelevant components to be less important than the relevant part of the scaling function.
Therefore we impose the restriction $n_1\leqslant n_0$ and $m_\eta\leqslant m_\varrho$.
For the single-parameter FSS we generate tables for each $q$ containing usually a few hundred index-combinations.
For the MFSS of Section \ref{sec-3dfss} we generate initial tables of $\sim2000$ index-combinations for each $q$ and for different sets of data.

We decide whether or not a fit is acceptable based on the goodness-of-fit or $p$-value. This is the probability that for $k$ degrees of freedom
we would observe a value of $\chi^2$ larger than the one obtained for the best fit.
To calculate this probability we use the approximation
\begin{equation}
 p=\frac{\Gamma(k/2,\chi^2/2)}{\Gamma(k/2)}.
\end{equation}
Here, $\Gamma(x)$ is the Euler gamma function, $\Gamma(a,x)$ is the upper incomplete gamma function, and $k$ denotes the number of degrees of freedom in the model, i.e.\  the number of data minus the number of free parameters in the fit.
We use $p\geqslant 0.1$ as a criterion for an acceptable fit.

After excluding unacceptable fits we order the table according to the number of free parameters.

\subsection{Stability of fits.}
A good fit must not only give an acceptable $p$ value, it must also be stable.
By stable we mean that the estimates of the fit parameters, and specially of the critical parameters,
must not change significantly when the expansion orders are increased.
In order to check the stability of a particular expansion, for example $n_0,\,n_1,\,m_\varrho,\,m_\eta$ in fixed-$\lambda$ FSS,
we consider all expansions where each index is separately increased by $1$, and also the case where all indices are increased by $1$ at the same time.
If the critical parameters from all these fits lie within the uncertainty interval obtained for the estimates of the initial expansion, then we regard that fit as stable.
The search for a stable fit proceeds by considering acceptable fits from the initial table in increasing order with the number of free parameters.
As a general rule we try to find the simplest stable fit.
We give examples of this procedure in Tables \ref{tab-stability} and \ref{tab-stability3D}.
While the criterion for a stable fit can certainly be debated we think that our criterion is sensible and helps avoid
the ambiguity that may arise from an arbitrary choice of a particular fit with an acceptable $p$ value.

\begin{table}
 \caption{Table of fits for FSS of $\widetilde{\Delta}_{-0.5}^\text{typ}$ at fixed $\lambda=0.1$.
  Our choice of fit is highlighted in bold. In subsequent rows the expansion indices are progressively increased.
  Dashes indicate the expansions used to check if our chosen fit is stable. The stability of our choice is shown in Fig.~\ref{fig-montecarlo}.}
 \begin{tabular}{ccccccc}
 \hline\hline
    $n_0\,n_1\,m_{\rho }\,m_{\eta }$ & $N_P$ & $\chi^2$ & $p$ & $W_c$ & $y$ & $\nu$ \\
 \hline
\textbf{3\,1\,2\,0} & \textbf{10} & \textbf{178} & \textbf{0.47} & \textbf{16.521} & \textbf{1.726} & \textbf{1.610} \\
- 3\,1\,2\,1 - & 11 & 172 & 0.58 & 16.512 & 1.701 & 1.602 \\
- 3\,1\,3\,0 - & 11 & 174 & 0.53 & 16.521 & 1.718 & 1.609 \\
3\,1\,3\,1 & 12 & 171 & 0.57 & 16.516 & 1.691 & 1.600 \\
- 3\,2\,2\,0 - & 11 & 172 & 0.58 & 16.520 & 1.712 & 1.613 \\
3\,2\,2\,1 & 12 & 172 & 0.56 & 16.519 & 1.702 & 1.603 \\
3\,2\,3\,0 & 12 & 171 & 0.57 & 16.517 & 1.704 & 1.618 \\
3\,2\,3\,1 & 13 & 171 & 0.55 & 16.516 & 1.697 & 1.610 \\
- 4\,1\,2\,0 - & 11 & 173 & 0.55 & 16.523 & 1.723 & 1.619 \\
4\,1\,2\,1 & 12 & 172 & 0.56 & 16.519 & 1.704 & 1.604 \\
4\,1\,3\,0 & 12 & 173 & 0.53 & 16.523 & 1.722 & 1.618 \\
4\,1\,3\,1 & 13 & 171 & 0.55 & 16.517 & 1.696 & 1.605 \\
4\,2\,2\,0 & 12 & 172 & 0.56 & 16.521 & 1.714 & 1.615 \\
4\,2\,2\,1 & 13 & 172 & 0.54 & 16.521 & 1.716 & 1.618 \\
4\,2\,3\,0 & 13 & 170 & 0.56 & 16.519 & 1.709 & 1.624 \\
- 4\,2\,3\,1 - & 14 & 170 & 0.54 & 16.518 & 1.704 & 1.619 \\
 \hline
 \end{tabular}
 \label{tab-stability}
\end{table}
%

\begin{table}
 \caption{Table of fits for MFSS of $\widetilde{\alpha}_0$. Our choice of fit is highlighted in bold.
The stability of our choice is shown in Fig.~\ref{fig-montecarlo3D}.}
 \begin{tabular}{cccccccc}
 \hline\hline
    Expansion & $N_P$ & $\chi^2$ & $p$ & $W_c$ & $y$ & $\alpha_0$ & $\nu$ \\
 \hline
    \textbf{3\,2\,2\,1\,1\,1\,2\,0} &  \textbf{27} & \textbf{473} & \textbf{0.40} &  \textbf{16.530} & \textbf{1.811} & \textbf{4.048} & \textbf{1.590} \\
  4\,2\,2\,1\,1\,1\,2\,0 & 30 & 466 & 0.46 & 16.531 & 1.811 & 4.048 & 1.593 \\
  3\,3\,2\,1\,1\,1\,2\,0 & 31 & 467 & 0.43 & 16.530 & 1.810 & 4.048 & 1.592 \\
  3\,2\,3\,1\,1\,1\,2\,0 & 29 & 466 & 0.46 & 16.531 & 1.813 & 4.048 & 1.593 \\
  3\,2\,2\,2\,1\,1\,2\,0 & 30 & 466 & 0.46 & 16.532 & 1.813 & 4.048 & 1.591 \\
  3\,2\,2\,1\,2\,1\,2\,0 & 29 & 472 & 0.39 & 16.530 & 1.806 & 4.047 & 1.591 \\
  3\,2\,2\,1\,1\,2\,2\,0 & 29 & 469 & 0.42 & 16.530 & 1.813 & 4.048 & 1.590 \\
  3\,2\,2\,1\,1\,1\,3\,0 & 28 & 471 & 0.42 & 16.531 & 1.812 & 4.048 & 1.589 \\
  3\,2\,2\,1\,1\,1\,2\,1 & 28 & 467 & 0.46 & 16.530 & 1.809 & 4.047 & 1.581 \\
  4\,3\,3\,2\,2\,2\,3\,1 & 48 & 435 & 0.62 & 16.528 & 1.795 & 4.046 & 1.597 \\
 \hline
 \end{tabular}
 \label{tab-stability3D}
\end{table}
%

\subsection{Uncertainties for fitting parameters.}
To determine the uncertainties (confidence intervals) of the estimates of the fit parameters we use a Monte Carlo method.

Given a candidate expansion for a stable fit, we generate a perfect data set from that fit.
We then generate at least $1000$ \emph{synthetic} data sets by sampling randomly from an appropriate (see below) multivariate normal distribution centered on the perfect data set.
We then fit these data sets, using the same expansion of the scaling functions, and build histograms of the estimates for the critical parameters.
From these histograms, the corresponding, possibly asymmetric, $95\%$ confidence interval is obtained by removing $2.5\%$ of the events at each end of the distribution.

\begin{figure}[tb]
\includegraphics[width=.49\columnwidth]{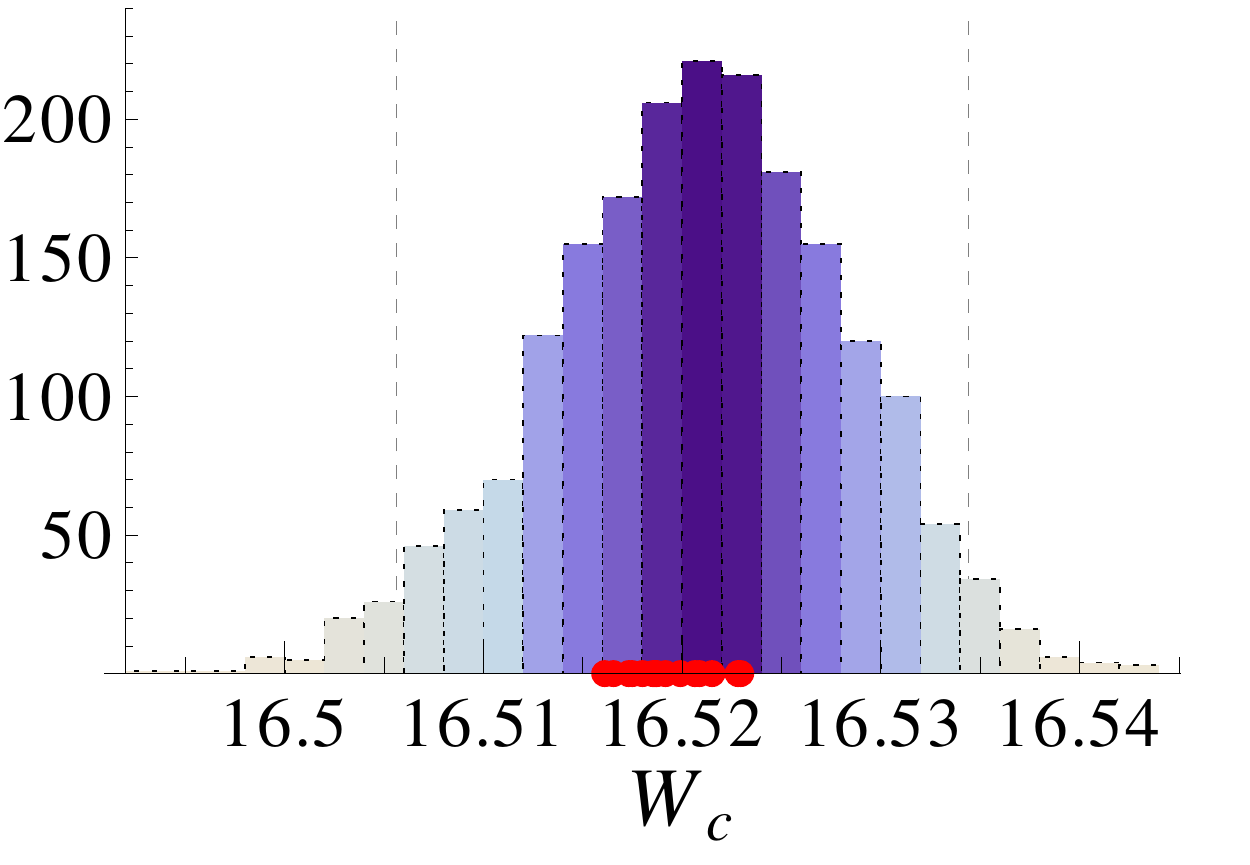}
\includegraphics[width=.49\columnwidth]{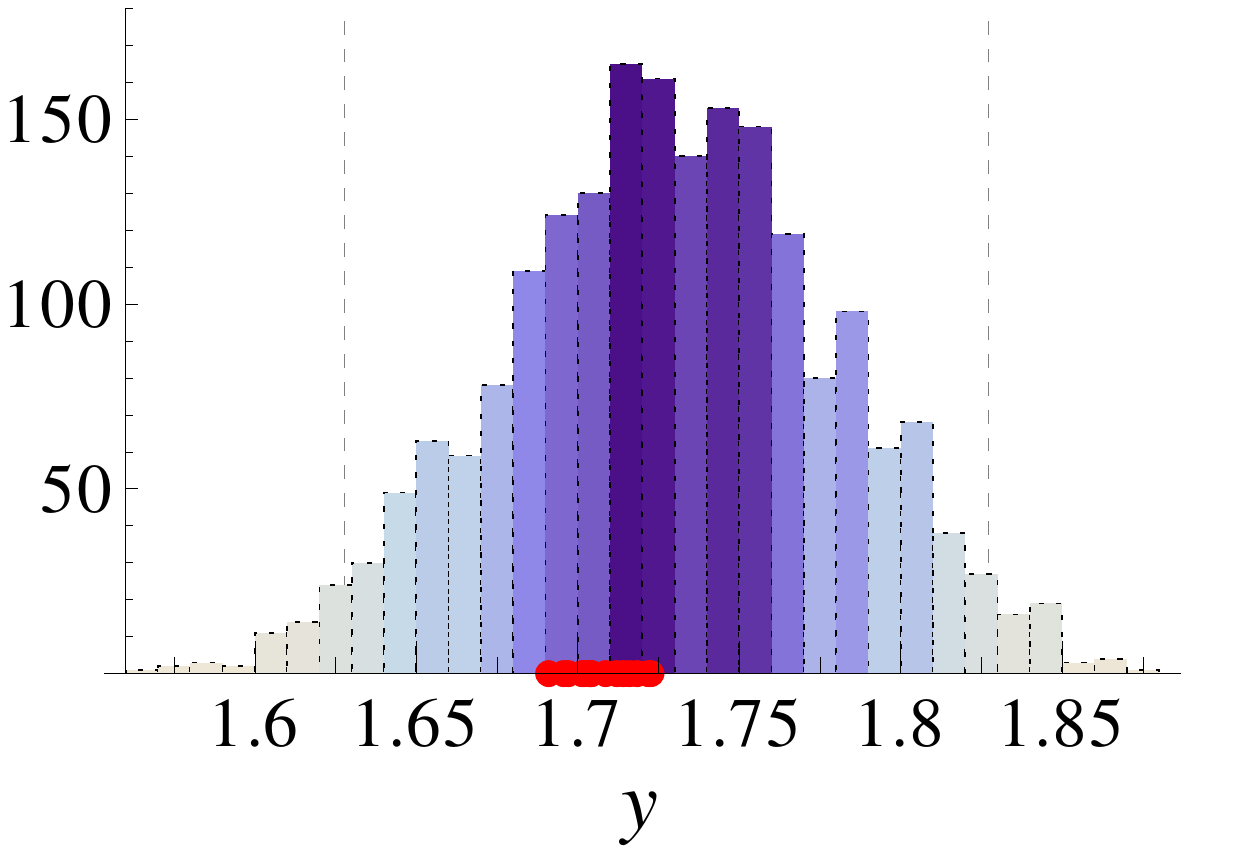}
\includegraphics[width=.49\columnwidth]{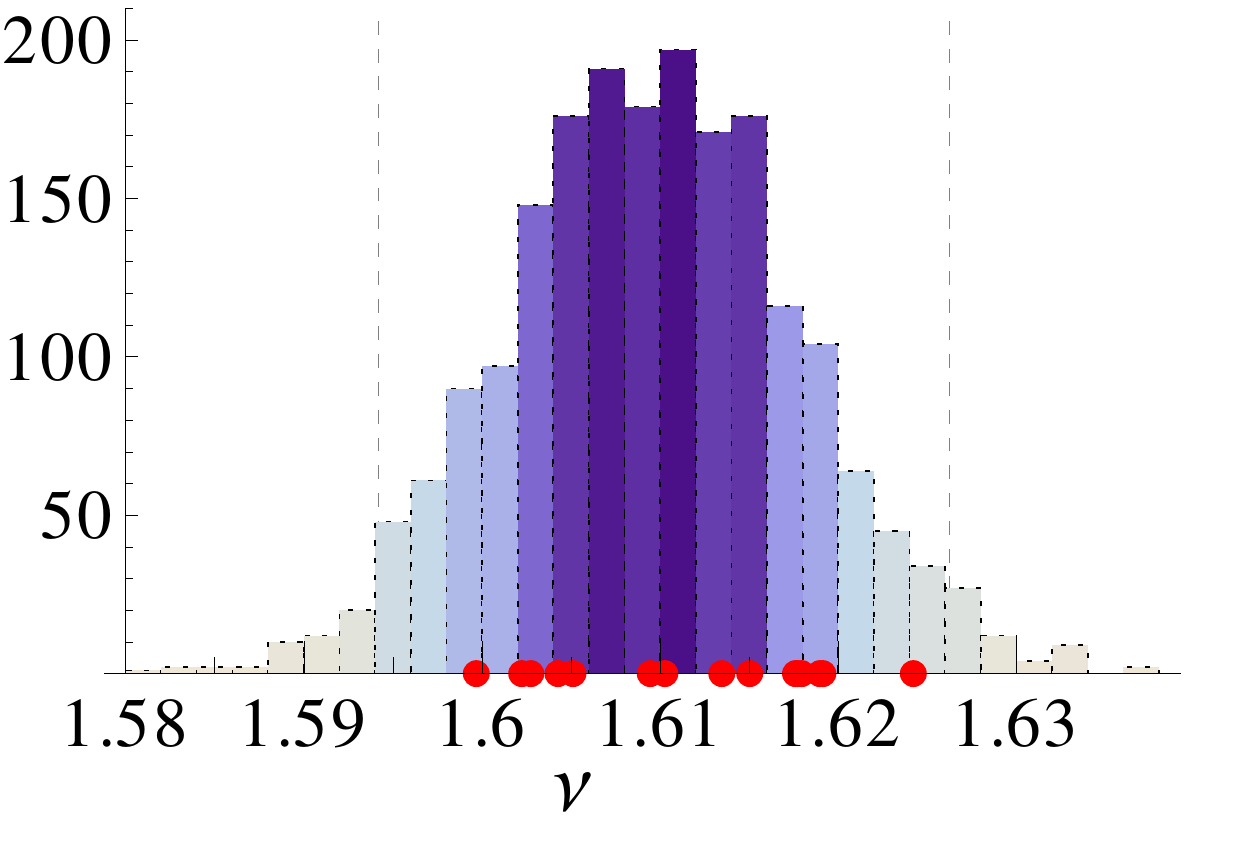}
\caption{(color online) Histograms of the critical parameters for expansion (3\,1\,2\,0) for FSS at fixed $\lambda=0.1$ of $\widetilde{\Delta}^\text{typ}_{-0.5}$.
 The histograms are obtained from fits of $2000$ synthetic data sets. The $y$-axis indicates the number of events. The vertical dashed lines denote the 95\% confidence interval. The circles on the $x$-axis mark the position of the values obtained from the stability analysis shown in Table \ref{tab-stability}. The color scale is the same as that used for the histogram density plots in Fig.~\ref{fig-CPvsq3D}.}
\label{fig-montecarlo}
\end{figure}

For single-parameter FSS at fixed $\lambda$,
the simulation data are uncorrelated and it is sufficient to use a product of independent normal distributions centered on the perfect data set with
the same standard deviations as the simulation data to generate the synthetic data sets.
In Fig.~\ref{fig-montecarlo} we show the resulting histograms for the fit of $\widetilde{\Delta}^\text{typ}_{-0.5}$ at fixed $\lambda=0.1$.

For MFSS the simulation data are correlated and these correlations must be taken into account when generating the synthetic data sets.
In this case, the appropriate distribution is a multivariate normal distribution centered on the perfect data set and with the covariance matrix
estimated from the simulation data as described in Appendix \ref{app-covmat}.
A simple way to generate the necessary correlated data is to use the Cholesky decomposition of the covariance matrix
to define an auxiliary set of independent normally distributed random variables with unit variance (see Appendix \ref{app-correlatedfit}).
In Fig.~\ref{fig-montecarlo3D} we show the histograms for critical parameters obtained from the MFSS of $\widetilde{\alpha}_0$.

\begin{figure}[tb]
\includegraphics[width=.49\columnwidth]{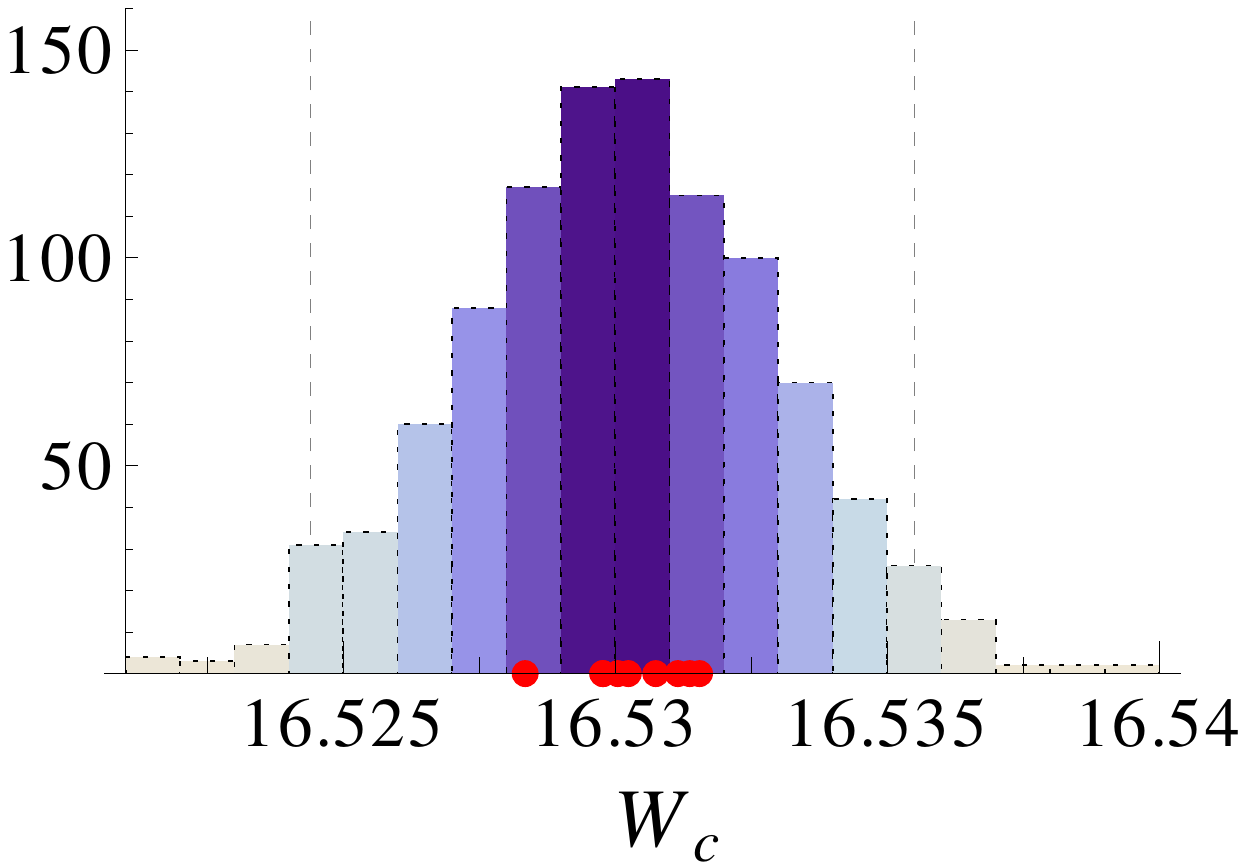}
\includegraphics[width=.49\columnwidth]{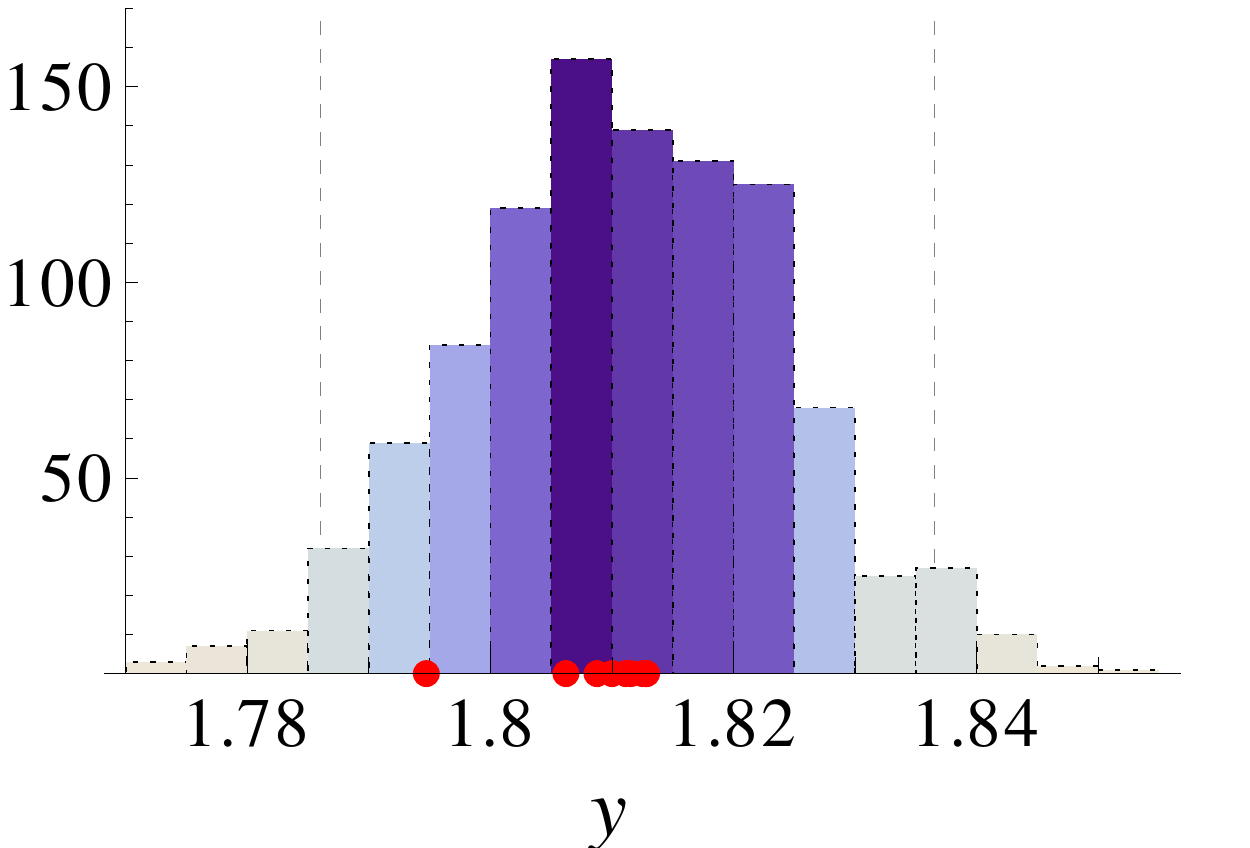}
\includegraphics[width=.49\columnwidth]{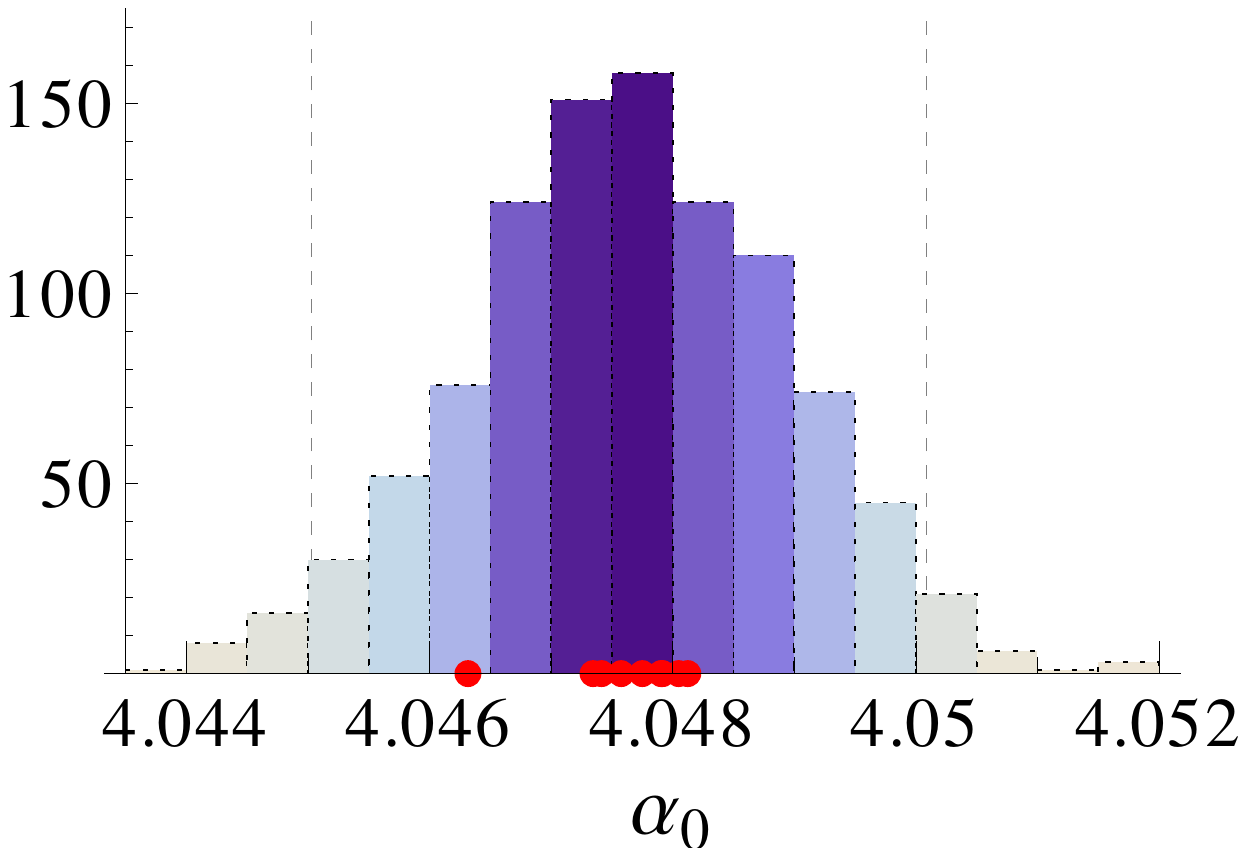}
\includegraphics[width=.49\columnwidth]{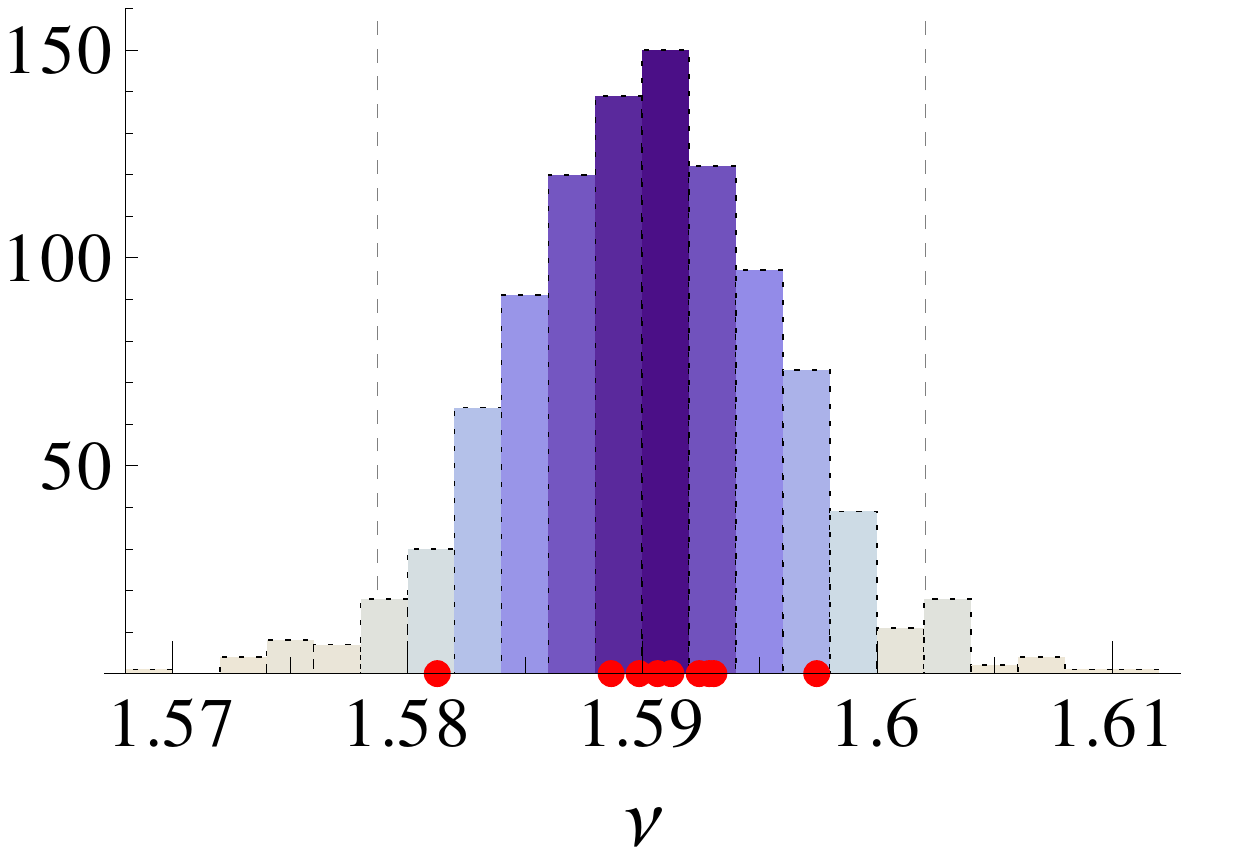}
\caption{(color online) Histograms of the critical parameters obtained for expansion (3\,2\,2\,1\,1\,1\,2\,0) for MFSS of $\widetilde{\alpha}_0$. The histograms are obtained from fits of $1000$ synthetic data sets. The $y$-axis shows the number of events. The vertical dashed lines indicate the 95\% confidence interval. The circles on the $x$-axis mark the position of the values obtained from the stability analysis shown in Table \ref{tab-stability3D}. The color scale is the same as that used for the histogram density shown in Fig.~\ref{fig-CPvsq3D}.}
\label{fig-montecarlo3D}
\end{figure}

\section{Calculation of the covariance matrix for correlated data}
\label{app-covmat}
In this section only we distinguish between the expectation value or population mean indicated by angular brackets,
and the sample mean indicated by an overline.
To avoid a cumbersome notation we do not make this distinction elsewhere in the paper.
In practical calculations, the population mean is always estimated using the sample mean.

The correlation between two random variables $X$ and $Y$ can be characterized in terms of the covariance, defined as
\begin{equation}\label{eq-covdef}
\text{cov}(X,Y)\equiv\langle XY \rangle -\langle X\rangle \langle Y\rangle.
\end{equation}
Note that $\text{cov}(X,X)=\text{var}(X)\equiv \sigma^2_X$, gives the variance for the variable $X$.
From $n$ samples $(X_1,Y_1),\ldots, (X_n,Y_n)$,
of the pair of random variables, the covariance can be estimated as,\cite{Davidson03}
\begin{equation}
 \textrm{cov}(X,Y)=\frac{1}{n-1}\sum_{i=1}^n \left( X_i - \overline{X} \right)\left( Y_i - \overline{Y}\right).
 \label{eq-covestimate}
\end{equation}
Here, the overline indicates the sample mean
\begin{equation}
 \overline{X} \equiv \frac{1}{n}\sum_{i=1}^n X_i,
 \label{eq-mean}
\end{equation}
and similarly for $Y$.
We also need the covariance of the sample means. This follows from (\ref{eq-covdef})
\begin{equation}
 \textrm{cov}(\overline{X}, \overline{Y}) =  \frac{1}{n} \textrm{cov}(X,Y).
\end{equation}

Furthermore, we need to consider random variables
$Y^{(1)}, \ldots, Y^{(s)}$ that are in turn functions of random variables $X^{(1)}, \ldots, X^{(r)}$ such that
\begin{equation}
    Y^{(i)} = Y^{(i)} \left( X^{(1)}, \ldots , X^{(r)} \right),
\end{equation}
where $i=1,\ldots ,s$.
Assuming that it is reasonable to linearize the functions in the range of interest, the covariance matrix $C$
of the random variables $X^{(1)}, \ldots, X^{(r)}$,
\begin{equation}
C_{ij} = \textrm{cov}\left( X^{(i)}, X^{(j)} \right),
\end{equation}
is related to the covariance matrix $C^{\prime}$ of the variables $Y^{(1)}, \ldots, Y^{(s)}$ by
\begin{equation}
    C^{\prime} = J^{T} C J.
\end{equation}
Here, $J$ is the matrix of derivatives
\begin{equation}
J_{ij} \equiv \frac{ \partial Y^{(j)} } {\partial X^{(i)}},
\end{equation}
evaluated at $\langle X^{(1)}\rangle, \ldots, \langle X^{(r)}\rangle$.

We now consider the application of the formulae above to the determination of the uncertainties of the GMFE $\widetilde{\Delta}_q$.
Correlations only arise between values of this exponent calculated for different coarse-grainings of the same set of samples.
Therefore, the covariance matrix for this quantity is block diagonal in disorder $W$ and system size $L$.
Consider two different coarse-grainings $l_i$ and $l_j$, and corresponding estimates, $\widetilde{\Delta}_q^{(i)}$ and $\widetilde{\Delta}_q^{(j)}$, of the values of this
GMFE for these coarse-grainings.
Application of the formulae above is then straightforward because $\widetilde{\Delta}_q$ is a function of $\overline{ R_q }$ only (see Table \ref{tab-gme}).
The result is
 \begin{equation}
 \textrm{cov}\left(\widetilde{\Delta}^{(i)}_q, \widetilde{\Delta}^{(j)}_q\right)=
 \frac{ \textrm{cov}\left( \overline{ R_q^{(i)} }, \overline{ R_q^{(j)} } \right)} { \langle R_q^{(i)} \rangle \; \langle R_q^{(j)} \rangle \ln \lambda^{(i)} \ln \lambda^{(j)} }.
\end{equation}

Application of the formulae above to the determination of the uncertainties of the GMFE $\widetilde{\alpha}_q$ is more complicated because it
is a function of $\overline{ R_q }$ and $\overline{ S_q }$.
The result is
\begin{multline}
 \textrm{cov}\left(\widetilde{\alpha}_q^{(i)},\widetilde{\alpha}_q^{(j)}\right)=
 \bigg[ \frac{\langle S_q^{(i)}\rangle\langle S_q^{(j)}\rangle}{\langle R_q^{(i)}\rangle^2 \langle R_q^{(j)}\rangle^2}
 \textrm{cov}\left(\overline{ R_q^{(i)}},\overline{R_q^{(j)}}\right) \\
   - \frac{\langle S_q^{(j)}\rangle \textrm{cov}\left(\overline{S_q^{(i)}},\overline{R_q^{(j)}}\right)}{\langle R_q^{(i)}\rangle \langle R_q^{(j)}\rangle^2}
   - \frac{\langle S_q^{(i)}\rangle \textrm{cov}\left(\overline{S_q^{(j)}},\overline{R_q^{(i)}}\right)}{\langle R_q^{(i)}\rangle^2 \langle R_q^{(j)}\rangle} \\
   + \frac{\textrm{cov}\left(\overline{S_q^{(j)}},\overline{S_q^{(i)}}\right)}{\langle R_q^{(i)}\rangle \langle R_q^{(j)}\rangle} \bigg]/ \left(\ln\lambda^{(i)}\ln\lambda^{(j)}\right).
\end{multline}
For $q=0,1$ the formulae are considerably simpler,
\begin{align}
  \textrm{cov}\left(\widetilde{\alpha}_0^{(i)},\widetilde{\alpha}_0^{(j)}\right) &= \frac{\left[\lambda^{(i)}\lambda^{(j)}\right]^d  \textrm{cov}\left(\overline{S_0^{(i)}},\overline{S_0^{(j)}}\right)} { \ln \lambda^{(i)} \ln \lambda^{(j)}}, \\
  \textrm{cov}\left(\widetilde{\alpha}_1^{(i)},\widetilde{\alpha}_1^{(j)}\right) &= \frac{\textrm{cov}\left(\overline{S_1^{(i)}},\overline{S_1^{(j)}}\right)} { \ln \lambda^{(i)} \ln \lambda^{(j)}}.
\end{align}

\section{$\chi^2$ minimization for correlated data}
\label{app-correlatedfit}
Let $Y_i$ be the result of the $i$-th simulation performed for simulation parameters $X_i$. The uncertainties and correlations amongst the simulation
data are described by a covariance matrix $C\equiv\{\textrm{cov}(Y_i,Y_j)\}$.
We fit a model $F_i\equiv F(X_i,\{a\})$ to the simulation data
by varying the parameters $\{a\}$ of the model so as to minimize the chi-squared statistic
\begin{equation}
 \chi^2 = \sum_{ij} \left( Y_i - F_i \right)  \left( C^{-1} \right)_{ij}  \left( Y_j - F_j\right).
 \label{eq-chi2corr}
\end{equation}
In practice it is convenient to perform a Cholesky factorization, $C^{-1}=R^{T} R$, where $R$ is an upper triangular matrix,
so that $\chi^2$ can be expressed as the square norm of a vector $\chi^2= ||R (Y-F)||^2$.

In MFSS, $Y_i$ are the values of the GMFE under consideration and $X_i=\{W_i,L_i,l_i\}$.
The covariance matrix is calculated as described in Appendix \ref{app-covmat}.

%

\begin{thebibliography}{10}
\bibitem{note-gmfelimit}
For 3D exponentially localized states, with a finite localization length $\xi>1$, an analytical calculation of the insulating limit for the GMFEs for $l=1$ reveals the following behavior.
For $q>0$, the moments $R_q$ and $S_q$ saturate and become $L$-independent as $L\rightarrow\infty$. This implies
\begin{displaymath}\quad
\lim_{\lambda\rightarrow 0}\widetilde{\tau}_q=-\lim_{L\rightarrow\infty} \frac{\ln R_q}{\ln L} =0,\;
\lim_{\lambda\rightarrow 0}\widetilde{\alpha}_q=-\lim_{L\rightarrow\infty} \frac{S_q/R_q}{\ln L} =0.
\end{displaymath}
For $q<0$ we find $R_q\sim L^2 \exp(|q|L/\xi)$ and $S_q/R_q\sim -L$ asymptotically, which lead to
\[
\lim_{\lambda\rightarrow 0}\widetilde{\tau}_q=-\infty, \quad
\lim_{\lambda\rightarrow 0}\widetilde{\alpha}_q=\infty.
\]
The latter limit is also obtained for $\widetilde{\alpha}_0$.
\bibitem{note-scalinglaw}
It must be stressed that the scaling laws given here are different from those suggested in the past. In particular for $\widetilde{\alpha}_0$, Eq.~\eqref{eq-alphaq} for a fixed box size $l=1$ implies a relation of the form
\[
\widetilde{\alpha}_0(L,W) = \alpha_0 +  \frac{1}{\ln L} \mathcal{A}_0\left(\frac{L}{\xi}, \frac{1}{\xi}\right),
\]
This differs from the scaling law proposed in Refs.~\onlinecite{Jan94a,HucS92}
\[
\widetilde{\alpha}_0(L,W) = \alpha_0 +  F\left(\frac{L}{\xi}\right).
\]
Moreover, this latter scaling law is in error because the renormalization of the lattice constant (or box size) has been neglected. In fact,
for the 3D Anderson model, the latter scaling law produces estimates for $\nu$ which violate the universality of the transition (Ref.\ \onlinecite{VasSRR09}).
\bibitem{note-fits}
The fits at fixed $\lambda=0.1$ for $\widetilde{\Delta}_q$, $\widetilde{\Delta}_q^\text{typ}$ for $q=0.75,1.25,1.5,1.75$, and for $\widetilde{\alpha}_q$ for $q=1.25$ do not fully satisfy our stability requirements, as one of the expansions in the stability test gives estimates slightly outside the corresponding confidence intervals. We suspect this is due to the numerical $\chi^2$-minimization not being able to resolve two neighboring minima. If we reduce the data set by restricting slightly the range of $W$ or the range of $L$, we quickly find stability for the same expansion, giving very similar confidence intervals. Therefore, we include the results for the fits using the whole data set.
\end{thebibliography}

\end{document}